\documentclass[11pt]{article}

\usepackage[margin=1in]{geometry}
\usepackage{mathtools}
\usepackage{amssymb}
\usepackage{amsthm}
\usepackage{bm}

\usepackage{booktabs,array,microtype}
\usepackage[colorlinks=true,linkcolor=blue,urlcolor=blue]{hyperref}
\usepackage[x11names]{xcolor}

\hypersetup{
  citecolor=SpringGreen4
}

\usepackage{enumitem}
\usepackage{needspace}

\usepackage{mathtools}
\mathtoolsset{showonlyrefs=true,showmanualtags=true}

\usepackage[bottom]{footmisc}
\usepackage{tablefootnote}\usepackage{diagbox}
\usepackage{cellspace} 
\usepackage{tikz}
\usetikzlibrary{calc}
\usetikzlibrary{decorations.pathreplacing}

\usepackage[dvipsnames]{xcolor}

\newtheorem{theorem}{Theorem}[section]
\newtheorem{lemma}[theorem]{Lemma}

\newtheorem{proposition}[theorem]{Proposition}

\theoremstyle{definition}
\newtheorem{definition}[theorem]{Definition}

\theoremstyle{remark}
\newtheorem{remark}[theorem]{Remark}

\newcommand{\RR}{\mathbb{R}}

\newcommand{\bc}{\bm{c}}
\newcommand{\be}{\bm{e}}
\newcommand{\bg}{\bm{g}}
\newcommand{\bh}{\bm{h}}
\newcommand{\bn}{\bm{n}}
\newcommand{\bo}{\bm{o}}
\newcommand{\bt}{\bm{t}}
\newcommand{\bu}{\bm{u}}
\newcommand{\bv}{\bm{v}}
\newcommand{\bw}{\bm{w}}
\newcommand{\bx}{\bm{x}}
\newcommand{\by}{\bm{y}}
\newcommand{\bz}{\bm{z}}

\newcommand{\bone}{\mathbf{1}}

\newcommand{\EQ}{\;=\;}
\newcommand{\LE}{\;\le\;}

\newcommand{\GE}{\;\ge\;}
\newcommand{\GT}{\;>\;}

\newcommand{\supp}{\mathrm{supp}}
\newcommand{\conv}{\mathrm{conv}}

\newcommand{\OPT}{\mathrm{OPT}}
\newcommand{\LP}{\mathrm{LP}}
\newcommand{\cost}{\mathrm{cost}}

\newcommand{\cD}{\mathcal{D}}

\newcommand{\cU}{\mathcal{U}}

\newcommand{\dir}{\mathrm{Dir}}

\DeclareMathOperator{\Div}{div}

\newcommand{\LPvertex}{\LP_{\mathrm{vertex}}}
\newcommand{\LPedge}{\LP_{\mathrm{edge}}}
\newcommand{\LPCKR}{\LP_{\mathrm{CKR}}}

\newcommand{\OPTno}{\OPT^{\mathrm{no}}}
\newcommand{\LPCKRno}{\LPCKR^{\mathrm{no}}}
\newcommand{\LPvertexno}{\LP_{\mathrm{vertex}}^{\mathrm{no}}}
\newcommand{\LPedgeno}{\LP_{\mathrm{edge}}^{\mathrm{no}}}

\newcommand{\Omegano}{\Omega^{\mathrm{no}}}
\newcommand{\Lno}{L^{\mathrm{no}}}
\newcommand{\bestNum}{1.23003}

\newcommand{\val}{\mathrm{val}}

\title{Improved lower bounds on the approximability of \\ Multiway Cut using Sherali--Adams relaxations}

\author{Joshua Brakensiek\thanks{University of California, Berkeley. Email: \texttt{josh.brakensiek@berkeley.edu}. Supported in part by Venkatesan Guruswami's Simons Investigator award and NSF grants CCF-2211972 and DMS-2503280.}\and Neng Huang\thanks{TU Dresden. Work was partly done when NH was affiliated with the University of Michigan. NH has been funded by the European Union (Project POCOCOP, ERC Synergy Grant 101071674). Views and opinions expressed are however those of the author(s) only and do not necessarily reflect those of the European Union or the European Research Council Executive Agency. Neither the European Union nor the granting authority can be held responsible for them. Email: \texttt{neng.huang@tu-dresden.de}}\and Aaron Potechin\thanks{University of Chicago. Supported in part by NSF grant CCF-2611927. Email: \texttt{potechin@uchicago.edu}}\and Uri Zwick\thanks{Blavatnik School of Computer Science and AI, Tel Aviv University, Israel. Email: \texttt{zwick@tau.ac.il}. Work supported by ISF grant 2735/2025.}}

\begin{document}

\maketitle

\null\vspace*{-0.75cm}
\begin{abstract}
    The Multiway Cut problem is among the most well-studied problems in the area of combinatorial approximation algorithms. The currently best approximation ratio for the problem is $1.2787$, obtained recently by the authors (2026). The best lower bound on the approximation ratio that can be achieved for the problem in polynomial time, under the Unique Games Conjecture (UGC), is $1.20016$, obtained by Bérczi, Chandrasekaran, Király, and Madan (2020), building on and slightly improving a bound of $\frac65=1.2$, obtained by Angelidakis, Makarychev, and Manurangsi (2017) by introducing the important framework of \emph{non-opposite} multiway cuts. 

    We obtain a substantially improved lower bound of $1.22232$ on the approximation ratio of the Multiway Cut problem.  Manokaran, Naor, Raghavendra, and Schwartz (2008) showed that under UGC, the approximation ratio of Multiway Cut is equal to the integrality ratio of a simple LP relaxation, known as the CKR relaxation, introduced by C{\u a}linescu, Karloff, and Rabani (2000). Thus, stronger LP relaxations are not required for obtaining an optimal approximation algorithm for the problem. We show, however, that relying on the framework of Angelidakis, Makarychev, and Manurangsi (2017), stronger LP relaxations from the Sherali--Adams hierarchy for non-opposite multiway cuts can be used to obtain improved integrality ratios for the CKR relaxation, and hence improved lower bounds on the approximation ratio of the Multiway Cut problem. One of the novel features of our analysis is considering a continuous limit of discrete LPs defined on a simplex.

    Our proof of the new $1.22232$ bound is completely analytic. We also present strong numerical evidence that suggests that our techniques can be used to improve the lower bound to at least $\bestNum$.
\end{abstract}

\paragraph{\footnotesize The Unique Games Conjecture} {\footnotesize A proof of the Unique Games Conjecture was recently announced by Open AI~\cite{OAI:UGC}. To the best of our knowledge it has not been independently verified yet. Thus, we continue to document which of our results depend on UGC. A proof of the Unique Games Conjecture would make our results unconditional, and thus stronger.}

\paragraph{\footnotesize AI Disclosure} {\footnotesize See page~\pageref{para:ai} for a detailed AI Disclosure.}

\pagebreak

\section{Introduction}

In the field of combinatorial approximation algorithms, the (Minimum) Multiway Cut problem has served for over three decades as a central benchmark for designing algorithms and establishing lower bounds. The input to the (Minimum) Multiway Cut problem is a weighted undirected graph $G = (V,E,w)$, where $w:E\to\RR^+$, with~$k$ designated vertices called \emph{terminals}. The objective is to find a partition of $V$ into $k$ parts, each containing exactly one terminal, so as to minimize the total weight of edges whose endpoints lie in different parts of the partition. For $k=2$ this is the classical min $s$-$t$ cut problem which admits polynomial-time algorithms. For $k\ge 3$, however, the problem is APX-hard~\cite{DJPSY94}. This motivated a long series of papers, see the left side of Table~\ref{T-approx}, that design approximation algorithms for the problem, with gradually improving approximation ratios. The currently best approximation algorithm for arbitrary~$k$, achieves a ratio of $1.2787$~\cite{brakensiek2026improved}. \footnote{The final journal version of the paper is expected to present an upper bound below $1.275$.}

All the approximation algorithms mentioned in Table~\ref{T-approx}, except the first one, use a simple linear programming (LP) relaxation of the problem introduced by C{\u a}linescu, Karloff, and Rabani~\cite{CKR00}. We refer to this relaxation as the CKR relaxation. Crucially, Manokaran, Naor, Raghavendra, and Schwartz \cite{MNRS08} showed that, under the Unique Games Conjecture (UGC)~\cite{Khot02}, the best approximation ratio that can be obtained for the Multiway Cut problem in polynomial time is equal to the integrality ratio of the CKR relaxation, the maximum possible ratio (over all instances) between the optimal solution and the objective value of the CKR relaxation. (See also \cite{KKSTY04,Raghavendra08}.) 

Since the works of \cite{CKR00,MNRS08}, the central challenge in the study of Multiway Cut is to establish rigorous bounds on the integrality ratio of the CKR relaxation. Although a rich line of works has established increasingly better upper bounds for the CKR relaxation, the focus of this work is on better understanding lower bounds for the CKR relaxation. The first such works, by Cheung, Cunningham, and Tang~\cite{CT99,CCT06} and Karger, Klein, Stein, Thorup, and Young~\cite{KKSTY04} established an exact integrality ratio of $\frac{12}{11} \simeq 1.090909$ for the CKR relaxation when $k=3$. See the right side of Table~\ref{T-approx} for a list of lower bounds established on the integrality ratio of the CKR relaxation for general $k$ (i.e., $k \to \infty$). Using the result of \cite{MNRS08} in combination with the Unique Games Conjecture, these lower bounds translate into lower bounds on the approximation ratio of the Multiway Cut problem.

The first general lower bound for Multiway Cut was $\frac87 \simeq 1.14285$ due to Freund and Karloff~\cite{FK00}, which stood for over a decade. In a major innovation, Angelidakis, Makarychev, and Manurangsi~\cite{AMM17} pushed the integrality ratio all the way to $\frac65 = 1.2$ via a new method involving \emph{non-opposite} multiway cuts, which we soon describe. More recently, Bérczi, Chandrasekaran, Király, and Madan \cite{BCKM20} used the non-opposite cut framework to slightly improve the best lower bound to $1.20016$. Our primary contribution is a substantially improved lower bound on the integrality gap for the CKR relaxation.

\begin{theorem}\label{thm:main}
    Assuming the Unique Games Conjecture, Multiway Cut is NP-hard to approximate within a factor of 1.22232.
\end{theorem}

This lower bound is obtained by rigorously bounding a certain continuous optimization problem closely related to the CKR relaxation. In fact, through numerical estimates (based on solving huge linear programs) our method establishes a stronger lower bound of \bestNum; see the right side of Table~\ref{T-approx}.

\begin{table}[t]
\begin{center}
\begin{tabular}{c@{\hspace{3em}}c}
Upper bound & Paper\\[3pt]
\hline
\noalign{\vskip 3pt}\noalign{\vskip 3pt}
2 & \cite{DJPSY94} \\
1.5 & \cite{CKR00} \\
1.3438 & \cite{KKSTY04} \\
1.3239 & \cite{BNS18} \\
1.3022 & \cite{SV14} \\
1.2970 & \cite{BSW21} \\
1.2965 & \cite{SV14} \\
1.2787 & \cite{brakensiek2026improved}\\
\end{tabular}
\hspace*{1.5cm}
\begin{tabular}{c@{\hspace{3em}}c}
Lower bound & Paper\\[3pt]
\hline
\noalign{\vskip 3pt}\noalign{\vskip 3pt}
$\frac{12}{11} \simeq 1.09091$ & \cite{CCT06,KKSTY04}\\[2pt]
$\frac87 \simeq 1.14285$  & \cite{FK00} \\[2pt]
$\frac65 = 1.2$ & \cite{AMM17} \\[2pt]
$1.20016$ & \cite{BCKM20} \\[2pt]
$1.22232$ & This work (Theorem \ref{thm:main})\\[2pt]
$\bestNum$ & This work (numerical)
\end{tabular}
\end{center}
\caption{Upper and lower bounds on the approximation ratio of the Multiway Cut problem with an arbitrary number of terminals. The lower bounds are on the integrality ratio of the CKR relaxation, which translate to lower bounds on the approximation ratio under the Unique Games Conjecture.}\label{T-approx}
\end{table}

\subsection{CKR, Simplex Partitions and Non-opposite Cuts}
Before we describe our contributions in more detail, we first give some background and notation for the CKR relaxation and the non-opposite cut framework. For a weighted graph $G = (V, E, w)$ with~$k$ terminals $[k] \subset V$ \footnote{Throughout the paper, we let $[k]=\{1,2,\ldots,k\}$. We also assume for simplicity that $V=[n]$, where $n=|V|$.}, where $w:E\to \mathbb{R}_{\geq 0}$, its CKR relaxation is defined as follows:

\begin{equation}
\tag{$\LP_{\mathrm{CKR}}$}
\label{LP:CKR}
\begin{array}{llll}
 \min & \multicolumn{3}{l}{\displaystyle \sum_{e = \{i,j\}\in E} w_e \cdot \frac{\|\bu_i-\bu_j\|_1}{2}} \\[5pt]
 \text{s.t.} & \bu_i\in \Delta_k &,& \qquad i\in V,\\
             & \bu_i=\be_i &,& \qquad i\in[k]. \\
\end{array}
\end{equation}

Here $\Delta_k = \{\bx \in \left(\mathbb{R}_{\geq 0}\right)^k\mid \sum_{i = 1}^kx_i = 1\}$ is the unit simplex with $k$ vertices, and $\be_i$ denotes the $i$-th standard basis vector. Geometrically, the CKR relaxation embeds the graph into $\Delta_k$ such that terminal~$i$ is mapped to $\be_i$. Observe that $\frac12{\|\be_i - \be_j\|_1} = 1$ for $i \neq j$, so when all vertices in $G$ are mapped to simplex vertices, the LP objective corresponds exactly to the total weight of cut edges. A fractional solution allows mapping to arbitrary simplex points in $\Delta_k$, which can be thought of as a probability distribution over the $k$ terminals. $ \frac12{\|\bu_i-\bu_j\|_1}$ then gives the \emph{earth mover's distance} between the distributions given by $\bu_i$ and $\bu_j$.

In the many algorithms for Multiway Cut based on the CKR relaxation, one first computes this vector embedding $\{\bu_i \mid i \in V\}$ and then applies a suitable (randomized) \emph{rounding function} $f : \Delta_k \to [k],$ where we think of $[k]$ as a \emph{label set} corresponding to the terminals. In particular, for any edge $\{i,j\} \in E$, the rounding function cuts the edge if and only if $f(\bu_i) \neq f(\bu_j)$. The construction of distributions of these rounding functions can be quite complex; we refer the reader to \cite{brakensiek2026improved} for an overview.

If we let $\OPT(G)$ denote the minimum weight of a multiway cut of $G$, and let $\LPCKR(G)$ denote the value of the CKR relaxation, then the \emph{integrality ratio} (or \emph{integrality gap}) with respect to $G$ is defined to be $\frac{\OPT(G)}{\LPCKR(G)}$. The maximum value of this integrality ratio over all graphs $G$ is the integrality ratio of the CKR relaxation. As previously discussed, Manokaran, Naor, Raghavendra, and Schwartz~\cite{MNRS08} showed that, if there is a Multiway Cut instance with some \emph{integrality gap} for the CKR relaxation, then the same gap can be translated into hardness of approximation under the Unique Games Conjecture with an arbitrarily small additive loss. 

The first lower bound papers for Multiway Cut~\cite{CCT06,KKSTY04,FK00} directly constructed instances $G$ which have the prescribed integrality ratios. In fact, nearly without loss of generality, one can assume the graph $(V, E)$ of the underlying instance is itself a discretization of the simplex. For a positive integer $n$, let $\Delta_{k, n} = \left\{\bx \in \left(\frac{1}{n}\mathbb{Z}_{\geq 0}\right)^k\mid \sum_{i = 1}^kx_i = 1\right\}$ be the points in $\Delta_k$ whose coordinates are multiples of $\frac1n$, and let $E_{k, n}$ be the set of all pairs of points in $\Delta_{k, n}$ with $\ell_1$ distance $\frac2n$. Given a weight function $w: E_{k, n} \to \mathbb{R}_{\geq 0}$, observe that if
\begin{align}
    \sum_{e \in E_{k,n}} w_e \EQ n \label{eq:w-normalized}
\end{align}
then we must have that $\LPCKR(w) \le 1$. This can be achieved by setting $\bu_{\bx} = \bx$ for all $\bx \in \Delta_{k,n}$ in \eqref{LP:CKR}. A weight function $w$ satisfying \eqref{eq:w-normalized} is said to be \emph{normalized}. For a normalized $w$, $\OPT(w)$ serves as a lower bound on the integrality ratio.

In the low-dimensional setting of $k=3$, the works~\cite{CCT06,KKSTY04} show that it is quite feasible to construct by hand a suitable normalized weight function $w : E_{3,n} \to \mathbb {R}_{\geq 0}$ for which $\OPT(w) \approx \frac{12}{11}$, see also Appendix~\ref{app:k=3}. However for larger values of $k$ (in particular as $k \to \infty$) getting a grasp on the structure of gap-inducing weight functions is much more unwieldy.

A major step toward mitigating this problem was the introduction of the \emph{non-opposite cut} framework by Angelidakis, Makarychev, and Manurangsi~\cite{AMM17}. The key insight is that strong integrality gaps on $\Delta_{k',n}$ can be built by studying the structure of $\Delta_{k,n}$ for some $k$ much smaller than $k'$. To be more precise, a \emph{non-opposite} cut on $\Delta_{k, n}$ is a function $P: \Delta_{k, n} \to [k+1]$ satisfying $P(\be_i) = i$ for every $i \in [k]$, and $P(\bx) \in \{k+1\} \cup \supp(\bx)$ \footnote{We use $\supp(\bx) \coloneqq \{i \mid x_i \neq 0\}$ to denote the support of the vector $\bx$.} for every $\bx \in \Delta_{k, n}$. Note that there are now $k+1$ labels with the final label corresponding to an assignment of an additional \emph{neutral} label not associated with any of the~$k$ vertices of $\Delta_k$. The condition $P(\bx) \in \{k+1\} \cup \supp(\bx)$ can be viewed as a variant of Sperner colorings in combinatorial geometry (e.g., \cite{MV15}).

A non-opposite cut instance is simply a (normalized) weight function $w: E_{k, n} \to \mathbb{R}_{\geq 0}$. We say that a non-opposite cut $P$ cuts an edge $e = \{\bu, \bv\} \in E_{k, n}$ if $P(\bu) \neq P(\bv)$, and for any $w$ we define $\OPTno(w)$ to be the minimum total weight of edges cut by a non-opposite cut. The CKR relaxation, as we show below in \eqref{LP:CKRno}, can be easily adapted to the non-opposite cut setting. Furthermore, for any normalized weight function $w$, we still have that $\LPCKRno(w) \le 1$.  Angelidakis, Makarychev, and Manurangsi~\cite{AMM17} showed that, for any $\epsilon > 0$, any normalized non-opposite cut instance $w$ can be turned into a Multiway Cut instance on $k' \gg k$ terminals, whose CKR integrality gap is at least $\OPTno(w) - \epsilon$, see Theorem~\ref{thm:amm}. 

The crucial gain of this framework is that we can study the performance of the CKR relaxation on high-dimensional simplices by understanding non-opposite cut gaps in much smaller dimensions. In particular, Angelidakis, Makarychev, and Manurangsi~\cite{AMM17} adapt the ``traditional'' integrality gaps for $k=3$~\cite{CCT06,KKSTY04} to the non-opposite cut setting, leading to an incredible boost in the integrality ratio of Multiway Cut to $\frac65=1.2$. Bérczi, Chandrasekaran, Király, and Madan \cite{BCKM20} made the first progress on understanding non-opposite cuts when $k=4$, leading to a small quantitative improvement of $1.20016$.

\subsection{Our Contributions}\label{subsec:results}

We prove Theorem~\ref{thm:main} by giving a new method of constructing and rigorously analyzing non-opposite cut instances, particularly when $k=4$. The primary vehicle for our strengthened analysis is a detailed investigation of \textbf{Sherali--Adams relaxations} of Multiway Cut.

At first glance, such an investigation seems essentially pointless. Since in principle rounding the CKR relaxation yields the optimal approximation algorithm for Multiway Cut, assuming the Unique Games Conjecture~\cite{KKSTY04,MNRS08,Raghavendra08}, there is no \emph{worst-case} benefit of using any strengthened LP relaxations for the Multiway Cut problem. Nevertheless, for a \emph{specific instance}, a strengthened LP can greatly improve the estimate of the true optimum. In particular, for a set of non-opposite cut weights $w : E_{k,n} \to \mathbb R_{\ge 0}$, we have the following chain of inequalities.
\[
    \LPCKR^{\mathrm{no}}(w) \LE \LP^{\mathrm{no}}_{\mathrm{strengthened}}(w) \LE \OPTno(w).
\]
Recall the ultimate goal of the non-opposite cut framework is to find normalized instances $w$ which maximize $\OPTno(w)$. Since $\LP_{\mathrm{strengthened}}^{\mathrm{no}}(w) \le \OPTno(w)$, we can use the strengthened LP relaxation as a proxy for OPT! See Section~\ref{subsec:related} for a discussion of other works in the approximation algorithms literature which have made similar observations.

This insight raises a new question: how do we find normalized instances $w$ for which we have that $\LP_{\mathrm{strengthened}}^{\mathrm{no}}(w)$ is large? For a fixed vertex set $\Delta_{k,n}$ and a fixed strengthened LP relaxation, the maximizing $w$ can be found by solving a linear program only slightly more complicated than the strengthened LP itself.  See \eqref{LP:dual-vertex-w} and \eqref{LP:dual-edge-w} below for examples of progressively stronger Sherali--Adams-based relaxations. Numerical solutions of these linear programs, although not fully rigorous, are able to estimate a lower bound on the approximation ratio of at least $1.218$ when $k=4$ and $1.225$ when $k=5$. See Tables~\ref{table:LP_vertex} and \ref{table:LP_edge} for more details.

Our second main contribution is a \emph{rigorous analysis} of this technique when $k=4$ by casting these strengthened LP formulations as \emph{multi-commodity flow} problems. To get some intuition, recall that the $k=2$ case of Multiway Cut is equivalent to the min $s$-$t$ cut problem, whose answer is dual to the classic single-source single-sink maximum flow problem. In an analogous manner, one can construct a flow-like formulation of non-opposite cuts for general $k$ by considering a multi-commodity flow from the corners of $\Delta_{k,n}$ to the opposite faces. See \eqref{eq:flow_lp} below for a precise implementation. This flow formulation is again considerably stronger than the CKR relaxation, and thus can be used to construct integrality gaps for Multiway Cut, leading to the best numerical lower bound we can get for the integrality gap of \bestNum. See Table~\ref{table:flow} and the last paragraph of Section~\ref{sub-lift} for a detailed summary of the results.

To get a rigorous analysis in the $k=4$ setting, we consider a \emph{continuous limit} of these discrete flow problems, leading to a continuous linear program. We construct an explicit (although slightly suboptimal) solution to this continuous LP, leading to our claimed lower bound of $1.22232$ on the approximation ratio. We refer the reader to the technical overview in Section~\ref{subsec:tech-overview} for a more thorough discussion of this continuous LP and its analysis.

\subsection{Technical Overview}\label{subsec:tech-overview}

We now give an overview of the main technical details leading to our proof of Theorem~\ref{thm:main}. Our exposition of these techniques focuses on the non-opposite cut framework of Angelidakis, Makarychev, and Manurangsi~\cite{AMM17}, although our techniques can also be used to obtain improved integrality gaps for standard cuts.\footnote{We use the term \emph{standard} (multiway) cuts to refer to cuts in $\Delta_{k,n}$ without the non-opposite condition. Namely, all functions $P : \Delta_{k, n} \to [k]$ such that $P(\be_i) = i$ for every $i \in [k]$. Such standard cuts can be used to construct integrality gaps for the CKR relaxation for a fixed value of $k$. See Appendix~\ref{app:standard-cuts} for further discussion.} 

\paragraph{Strengthened LP relaxations.} Our approach is to use strengthened LP relaxations for non-opposite cuts whose optimal values serve as certified lower bounds for $\OPTno(w)$. First, note that the CKR relaxation extends naturally to the non-opposite cut setting. For a non-opposite cut instance $w: E_{k, n} \to \mathbb{R}_{\geq 0}$, we can write:
\begin{equation}
\tag{$\LPCKRno$}
\label{LP:CKRno}
    \begin{array}{llll}
 \min & \multicolumn{3}{l}{\displaystyle \sum_{e = \{\bu,\bv\}\in E_{k, n}} w_e \cdot \frac{\|\bx(\bu)-\bx(\bv)\|_1}{2}} \\[5pt]
 \text{s.t.} & \bx(\bu)\in \Delta_{k+1} &,& \qquad\bu \in \Delta_{k, n},\\
             & x_i(\bu) = 0 &,& \qquad\bu \in \Delta_{k, n},\, i \in [k] \text{ such that } u_i = 0, \\
             & \bx(\be_i)=\be_i &,& \qquad i\in[k]. \\
\end{array}
\end{equation}

This is a valid LP relaxation for computing $\OPTno(w)$. As previously mentioned, for every $\bu \in \Delta_{k, n}$, we can obtain a feasible solution $\bx(\bu)$ by appending a zero to $\bu$. For normalized $w$ satisfying \eqref{eq:w-normalized}, this choice gives a \emph{canonical} LP solution with objective value equal to $1$. Analogous to the standard CKR relaxation, any point $\bx(\bu) \in \Delta_{k+1}$ can be thought of as a distribution over the valid non-opposite cut labels for $\bu$, and for any edge connecting~$\bu$ and~$\bv$, $\frac12{\| \bx(\bu) - \bx(\bv) \|_1}$ is the minimum possible probability that two jointly distributed labels differ, subject to their marginals being $\bx(\bu)$ and $\bx(\bv)$ respectively. We may think of this as choosing a pairwise distribution over the edge so that the marginals agree with what the two endpoints prescribe. Our first observation is that for the canonical LP solution, the optimal edge distributions associated with the identity embedding need not be jointly consistent, and this can already be detected if we add \emph{local distribution} constraints involving more than 2 vertices to the LP. 

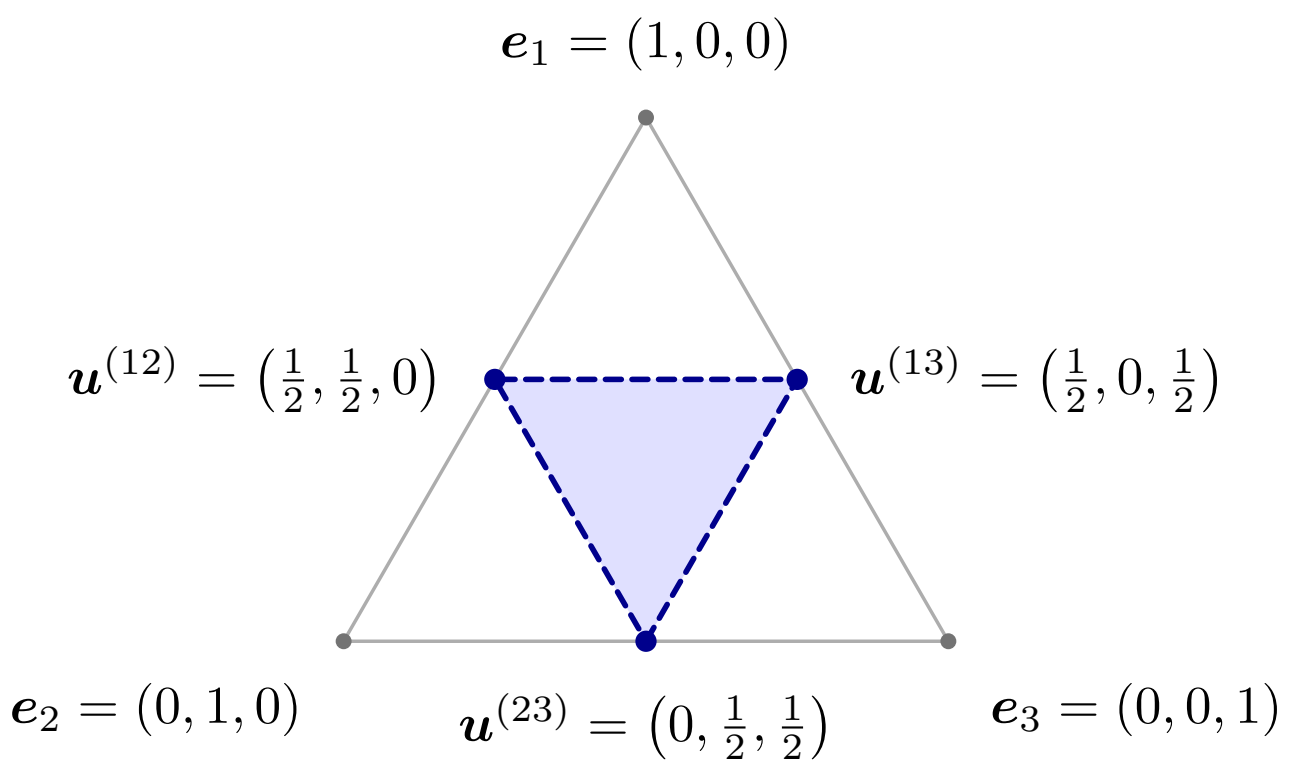
\begin{figure}
\centering
\begin{tikzpicture}[
  line cap=round,
  line join=round,
  ambient/.style={draw=black!32,line width=.65pt},cell edge/.style={
  draw=blue!55!black,
  line width=1.05pt,
  dash pattern=on 3pt off 2pt
},
  every node/.style={font=\small}
]
  \coordinate (A1) at (0,{2*sqrt(3)});
  \coordinate (A2) at (-2,0);
  \coordinate (A3) at (2,0);

  \coordinate (X1) at ($(A2)!.5!(A3)$);
  \coordinate (X2) at ($(A1)!.5!(A3)$);
  \coordinate (X3) at ($(A1)!.5!(A2)$);

  \fill[blue!12] (X1)--(X2)--(X3)--cycle;
  \draw[ambient] (A1)--(A2)--(A3)--cycle;
  \draw[cell edge] (X1)--(X2)--(X3)--cycle;

  \foreach \i in {1,2,3} {
    \fill[black!55] (A\i) circle[radius=1.5pt];
    \fill[blue!55!black] (X\i) circle[radius=2pt];
  }

  \node[above=5pt]       at (A1) {$\be_1 = (1,0,0)$};
  \node[below left=4pt]  at (A2) {$\be_2 = (0,1,0)$};
  \node[below right=4pt] at (A3) {$\be_3 = (0,0,1)$};

  \node[below=6pt] at (X1)
    {$\bu^{(23)} = \bigl(0,\tfrac12,\tfrac12\bigr)$};
  \node[right=6pt] at (X2)
    {$\bu^{(13)} = \bigl(\tfrac12,0,\tfrac12\bigr)$};
  \node[left=6pt] at (X3)
    {$\bu^{(12)} = \bigl(\tfrac12,\tfrac12,0\bigr)$};
\end{tikzpicture}
\caption{An illustration of $\Delta_{3,2}$. The triangle with the dashed boundary is the \emph{downward triangle}.}
\label{Fig:Delta_3_2}
\end{figure}

Take $\Delta_{3, 2}$, shown in Figure~\ref{Fig:Delta_3_2}, as a concrete example. For the edge $\{\bu^{(12)}, \bu^{(13)}\}$, the minimum disagreement probability in the canonical LP solution is equal to $\frac12{\| \bu^{(12)} - \bu^{(13)} \|_1} = \frac12$, realized by assigning probability $\frac12$ to $(1, 1)$ and $\frac12$ to $(2, 3)$. It can be easily seen that a similar construction works for any edge in $E_{3, 2}$ (and $E_{k, n}$ in general). The key observation here is that the optima for the three edges $\{\bu^{(12)}, \bu^{(13)}\}$, $\{\bu^{(12)}, \bu^{(23)}\}$, $\{\bu^{(13)}, \bu^{(23)}\}$ cannot be attained simultaneously by any \emph{joint} distribution over the labels of $(\bu^{(12)}, \bu^{(13)}, \bu^{(23)})$ with the prescribed marginals. Indeed, for any label, at least one of the three vertices gets that label with probability 0, and therefore the three vertices never receive the same label. Consequently, with probability one, at least two of the three edges are cut, whereas the sum of their LP contributions is only $\frac32$. 

This observation motivates our \emph{vertex-consistency} LP relaxation. The three points $\bu^{(12)}$, $\bu^{(13)}$, and $\bu^{(23)}$ from the preceding example form a downward-pointing triangle, which we call a \emph{downward cell}. To account for the grid edges on the boundary of $\Delta_{3,n}$, we also treat each of them as a degenerate downward cell of two vertices, so that every edge in $E_{3,n}$ is contained in exactly one downward cell. For $k \geq 4$, we will define downward cells in a similar manner so that every edge $e$ is contained in a unique downward cell $D(e)$ (see Section~\ref{subsec:vertex_lp} for more detail). We let $\cD_{k,n}$ be the collection of all downward cells in $\Delta_{k,n}$. Our vertex-consistency relaxation maintains a joint distribution $\mu_C$ over the set of valid non-opposite cut labelings $\Omegano_C$ for each downward cell $C \in \cD_{k,n}$, such that whenever two cells share a vertex $\bu$, their marginals on $\bu$ must be consistent (we use $\lambda_{\bu}$ to denote this marginal on $\bu$). Formally, we may formulate our vertex-consistency relaxation as follows:

\begin{equation}
\tag{$\LPvertexno$}
\begin{aligned}
\text{min}\quad & \sum_{e\in E_{k, n}} w_e p_e \\
\text{s.t.}\quad & \sum_{\sigma \in \Omegano_C} \mu_{C, \sigma} \EQ 1\,, && \qquad C \in \cD_{k, n}\,, \\
    & \sum_{\substack{\sigma \in \Omegano_{C} \\ \sigma(\bu) = a}} \mu_{C, \sigma} \EQ \lambda_{\bu, a}\,, &&  \qquad C \in \cD_{k, n},\, \bu \in C,\,  a \in \Lno(\bu)\,,\\
    & p_e \EQ \sum_{\substack{\sigma\in \Omegano_{D(e)}\\ \sigma(\bu)\ne\sigma(\bv)}} \mu_{D(e),\sigma}\,, && \qquad e = \{\bu, \bv\}\in E_{k, n}\,, \\
    & \mu_{C, \sigma} \GE 0 \,, && \qquad C \in \cD_{k, n}, \, \sigma \in \Omegano_C\,.
\end{aligned}
\end{equation}

Here we use $\Lno(\bu) \subseteq [k + 1]$ to denote the set of valid non-opposite cut labels for $\bu$. The objective value $\sum_{e\in E_{k, n}} w_e p_e $ of this LP is equal to the expected total weight of cut edges, where the expectation is taken in the local distributions $\mu_C$. For any $w$, we denote the optimum objective value by $\LPvertexno(w)$. Using a standard argument, we can prove that $\LPvertexno(w) \leq \OPTno(w)$ (see Lemma~\ref{lem:vertex_LP}), so any lower bound on $\LPvertexno(w)$ is also a lower bound on $\OPTno(w)$. It turns out that this relaxation is already strong enough to recover the optimal $k = 3$ lower bound for non-opposite cuts obtained by Angelidakis, Makarychev, and Manurangsi~\cite{AMM17}, see Appendix~\ref{app:k=3} for more details. In this paper, for $k=4$, we present a normalized instance $w$ which attains the bound claimed in our main theorem.

It is natural to consider more local distributions in the LP and impose even stronger consistency requirements. In fact, our vertex-consistency LP can be seen as a weaker version of some level in the Sherali--Adams LP hierarchy~\cite{sherali1990hierarchy}, which considers local distributions of all subsets of variables up to some size and imposes marginal consistency for all intersections, whereas we only considered geometrically adjacent vertices and marginal consistency for intersections of size one. We discuss this further in Section~\ref{subsec:edge}. For now, we only remark that numerically these even stronger relaxations offer only slightly stronger bounds, but their analysis seems to be much more involved. We therefore base our hardness construction on the vertex-consistency LP relaxation.

\paragraph{Searching for gap instances.} So far we are only using the strengthened LP relaxations to certify lower bounds of $\OPTno(w)$ for some given weights $w$. We now show how to search for these weights simultaneously while solving the LP. The key tool here is \emph{LP duality}. We may write down the dual LP to the vertex-consistency relaxation as follows, where we remind the reader that $\Lno$ is the set of valid labels for a vertex and $\Omegano$ is the set of valid labelings for a downward cell.

\begin{align*}
\text{max}\quad & \sum_{C \in \cD_{k, n}} y_C \\
\text{s.t.}\quad & y_C + \sum_{\bv \in C} z_{C, \bv, \sigma(\bv)} \LE \sum_{\substack{\{\bu, \bv\} \subseteq C\\ \bu\ne\bv}}w_{\{\bu, \bv\}} \cdot \bone[\sigma(\bu) \neq \sigma(\bv)], &&\qquad C \in \cD_{k, n},  \sigma \in \Omegano_C, \tag{$\ast$} \label{eq:dual_cell}\\
    & \sum_{C: \bv \in C} z_{C, \bv, a} \EQ 0, && \qquad \bv \in \Delta_{k, n}, a \in \Lno(\bv),\\
    & y_C, z_{C, \bv, a} \in \mathbb{R}. && \\
\tag{dual-$\LPvertexno$}
\end{align*}

Here we have a dual variable $y_C$ for each downward cell $C$, corresponding to the normalization constraint for the local distribution for $C$ in the primal. We also have a variable $z_{C, \bu, a}$, corresponding to each vertex marginal consistency constraint. See Section~\ref{subsec:vertex_lp} for more discussion on the construction of the dual and an explanation of the dual constraints. The key observation here is that the dual LP is also linear in the variables $w_e$, so we may treat them as LP variables and jointly optimize over them as well. We can also add the normalization constraint $\sum_{e \in E_{k, n}}w_e = n$ to the dual LP. By optimizing this augmented dual LP, we obtain a normalized instance $w$ with $\LPvertexno(w)$ maximized. We remark that the same approach works for stronger relaxations of the Sherali--Adams hierarchy. Numerically, this strategy already produces much better integrality gap ratios for Multiway Cut (see Table~\ref{table:LP_vertex} for a compilation of numerical ratios), and in particular this leads to a ratio above $1.225$.

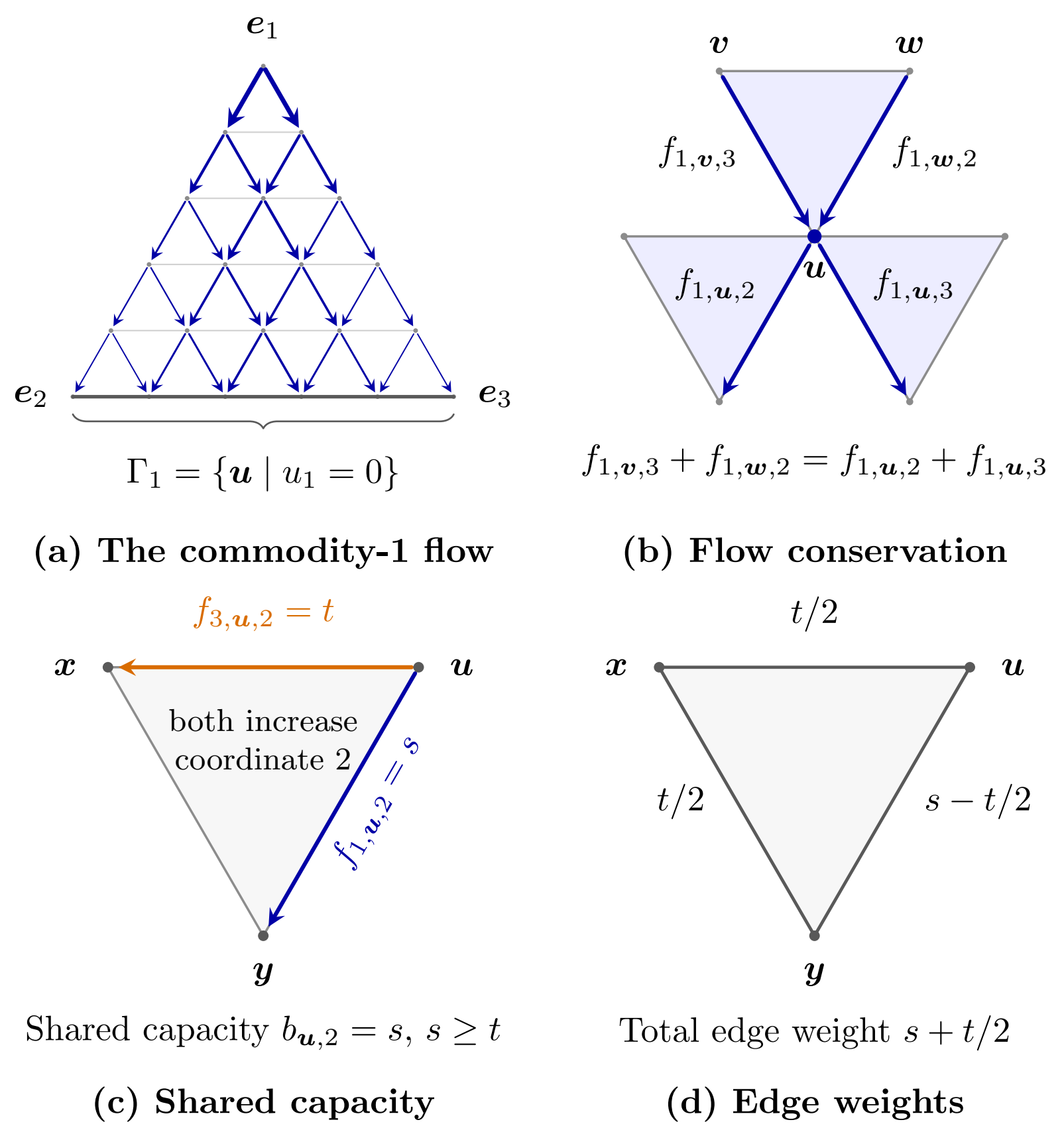
\begin{figure}[t]
\centering
\begin{tikzpicture}[
  x=1cm,y=1cm,>=stealth,
  line cap=round,line join=round,
  every node/.style={font=\small},
  grid edge/.style={draw=black!18,line width=.4pt},
  cell edge/.style={draw=black!45,line width=.65pt},
  one flow/.style={draw=blue!65!black,->,line width=1.15pt},
  three flow/.style={draw=orange!85!black,->,line width=1.15pt},
  subplot caption/.style={font=\small\bfseries,anchor=north}
]
  \begin{scope}
    \foreach \r in {0,...,5} {
      \foreach \j in {0,...,\r} {
        \coordinate (G\r-\j) at
          ({2.05+1.9*(2*\j-\r)/5},{3.3*(1-\r/5)});
      }
    }
    \foreach \r in {1,...,5} {
      \pgfmathtruncatemacro{\last}{\r-1}
      \foreach \j in {0,...,\last} {
        \pgfmathtruncatemacro{\next}{\j+1}
        \draw[grid edge] (G\r-\j)--(G\r-\next);
      }
    }
    \foreach \r in {0,...,4} {
      \pgfmathtruncatemacro{\nextrow}{\r+1}
      \foreach \j in {0,...,\r} {
        \pgfmathtruncatemacro{\next}{\j+1}
        \pgfmathsetmacro{\mass}
          {factorial(\r)/(factorial(\j)*factorial(\r-\j)*2^(\r+1))}
        \pgfmathsetmacro{\flowwidth}{.3+2*\mass}
        \draw[one flow,line width=\flowwidth pt,
          shorten <=1.1pt,shorten >=1.4pt]
          (G\r-\j)--(G\nextrow-\j);
        \draw[one flow,line width=\flowwidth pt,
          shorten <=1.1pt,shorten >=1.4pt]
          (G\r-\j)--(G\nextrow-\next);
      }
    }
    \foreach \r in {0,...,5}
      \foreach \j in {0,...,\r}
        \fill[black!45] (G\r-\j) circle[radius=.65pt];

    \draw[black!65,line width=1pt] (G5-0)--(G5-5);
    \node[above=4pt] at (G0-0) {$\be_1$};
    \node[left=3pt] at (G5-0) {$\be_2$};
    \node[right=3pt] at (G5-5) {$\be_3$};

    \draw[decorate,decoration={brace,mirror,amplitude=4pt},
      draw=black!65,line width=.5pt]
      ([yshift=-5pt]G5-0)
      --node[midway,below=8pt]
        {$\Gamma_1=\{\bu\mid u_1=0\}$}
      ([yshift=-5pt]G5-5);

    \node[subplot caption] at (2.05,-1.25)
      {(a) The commodity-1 flow};
  \end{scope}

  \begin{scope}[shift={(5.5,0)}]
    \coordinate (X)  at (2.05,1.6);
    \coordinate (TL) at (1.1,3.25);
    \coordinate (TR) at (3,3.25);
    \coordinate (L)  at (.15,1.6);
    \coordinate (R)  at (3.95,1.6);
    \coordinate (BL) at (1.1,-.05);
    \coordinate (BR) at (3,-.05);

    \fill[blue!7] (TL)--(TR)--(X)--cycle;
    \fill[blue!7] (L)--(X)--(BL)--cycle;
    \fill[blue!7] (X)--(R)--(BR)--cycle;
    \draw[cell edge] (TL)--(TR)--(X)--cycle;
    \draw[cell edge] (L)--(X)--(BL)--cycle;
    \draw[cell edge] (X)--(R)--(BR)--cycle;

    \draw[one flow,shorten <=2pt,shorten >=3pt]
      (TL)--node[left=4pt,pos=.5]
        {$f_{1,\bv,3}$} (X);
    \draw[one flow,shorten <=2pt,shorten >=3pt]
      (TR)--node[right=4pt,pos=.5]
        {$f_{1,\bw,2}$} (X);
    \draw[one flow,shorten <=3pt,shorten >=2pt]
      (X)--node[left=4pt,pos=.3]
        {$f_{1,\bu,2}$} (BL);
    \draw[one flow,shorten <=3pt,shorten >=2pt]
      (X)--node[right=4pt,pos=.3]
        {$f_{1,\bu,3}$} (BR);

    \foreach \pt in {TL,TR,L,R,BL,BR}
      \fill[black!45] (\pt) circle[radius=1pt];

    \node[above=4pt,inner sep=1pt] at (TL) {$\bv$};
    \node[above=4pt,inner sep=1pt] at (TR) {$\bw$};
    \fill[blue!65!black] (X) circle[radius=2pt];
    \node[below=6pt,inner sep=1pt] at (X) {$\bu$};

    \node at (2.05,-.65)
      {$f_{1,\bv,3}+f_{1,\bw,2}
        =f_{1,\bu,2}+f_{1,\bu,3}$};

    \node[subplot caption] at (2.05,-1.25)
      {(b) Flow conservation};
  \end{scope}

  \begin{scope}[shift={(0,-5.5)}]
    \coordinate (U) at (.5,2.8);
    \coordinate (V) at (3.6,2.8);
    \coordinate (W) at (2.05,.12);

    \fill[black!3] (U)--(V)--(W)--cycle;
    \draw[cell edge] (U)--(V)--(W)--cycle;

    \draw[one flow,shorten <=2pt,shorten >=3pt]
      (V)--(W);
    \draw[three flow,shorten <=2pt,shorten >=3pt]
      (V)--node[above=6pt,text=orange!85!black]
        {$f_{3,\bu,2}=t$} (U);
    \node[blue!65!black,rotate=60] at (3.2,1.45)
      {$f_{1,\bu,2}=s$};

    \foreach \pt in {U,V,W}
      \fill[black!65] (\pt) circle[radius=1.5pt];

    \node[left=5pt] at (U) {$\bx$};
    \node[right=5pt] at (V) {$\bu$};
    \node[below=4pt] at (W) {$\by$};

    \node[align=center,font=\footnotesize] at (2.05,2.08)
      {both increase\\coordinate 2};
    \node[align=center] at (2.05,-.85)
      {Shared capacity $b_{\bu,2}=s$, $s\geq t$};

    \node[subplot caption] at (2.05,-1.25)
      {(c) Shared capacity};
  \end{scope}

  \begin{scope}[shift={(5.5,-5.5)}]
    \coordinate (U) at (.5,2.8);
    \coordinate (V) at (3.6,2.8);
    \coordinate (W) at (2.05,.12);

    \fill[black!3] (U)--(V)--(W)--cycle;
    \draw[draw=black!65,line width=.9pt]
      (U)--node[above=6pt] {$t/2$} (V)
         --node[right=5pt,pos=.5] {$s-t/2$} (W)
         --node[left=5pt,pos=.5] {$t/2$} cycle;

    \foreach \pt in {U,V,W}
      \fill[black!65] (\pt) circle[radius=1.5pt];

    \node[left=5pt] at (U) {$\bx$};
    \node[right=5pt] at (V) {$\bu$};
    \node[below=4pt] at (W) {$\by$};

    \node[align=center] at (2.05,-.85)
      {Total edge weight $s+t/2$};

    \node[subplot caption] at (2.05,-1.25)
      {(d) Edge weights};
  \end{scope}
\end{tikzpicture}
\caption{An illustration for the Flow LP for $k=3$. (a) The commodity-$1$ flow travels from $\be_1$ to its opposite face $\Gamma_1$. (b) Conservation of commodity-$1$ flow at an interior vertex $\bu$. (c) Two flows both increasing coordinate $2$ and sharing capacity $b_{\bu, 2} = s$. (d) The corresponding edge weights, totaling $s+t/2$.}
\label{fig:flow_lp_overview}
\end{figure}

\paragraph{The Flow LP.} We now seek a way to explicitly construct feasible solutions to the augmented dual vertex-consistency LP. To do so, we introduce an auxiliary LP which we call the \emph{Flow LP}. Observe that to obtain a non-opposite cut, we require that each terminal $\be_a$ is separated from its opposite face $\Gamma_a \coloneqq \{\bu \mid u_a = 0\}$. This suggests a \emph{multi-commodity flow} formulation, where for each terminal~$a$ we send one unit of commodity-$a$ flow from $\be_a$ to its opposite face $\Gamma_a$. For simplicity, we assume that the flow is \emph{monotone}, i.e., as it travels the $a$-th coordinate monotonically decreases. For every vertex $\bu \in \Delta_{k, n} \setminus \Gamma_a$ and terminal $i \in [k] \setminus \{a\}$, we let $f_{a, \bu, i} \geq 0$ denote the amount of commodity-$a$ flow that travels from $\bu$ to $\bu - \frac{1}{n}\be_a + \frac{1}{n}\be_i$. Naturally, the flows must satisfy the flow conservation constraint, namely, at any vertex (other than a sink or a source) the amount of incoming flow is equal to the amount of outgoing flow. See Figure~\ref{fig:flow_lp_overview}(a) and (b) for an illustration of the commodity-1 flow and its conservation when $k = 3$.

To connect these flow variables $f_{a, \bu, i}$ to the dual vertex-consistency LP, let $C$ be the downward cell that contains both $\bu$ and $\bv = \bu - \frac{1}{n}\be_a + \frac{1}{n}\be_i$, then we can think of the flow represented by $f_{a, \bu, i}$ as entering $C$ through $\bu$ and leaving $C$ through $\bv$, so we add $f_{a, \bu, i}$ to $z_{C, \bu, a}$ and subtract the same amount from $z_{C, \bv, a}$. Then, when $\bu$ is not the source $\be_a$, the dual vertex-consistency LP constraint $\sum_{C: \bu \in C}z_{C, \bu, a} = 0$ is precisely the flow conservation at $\bu$. With this construction, for any cell $C$, we also get $\sum_{\bw \in C}z_{C, \bw, a} = 0$ by flow conservation for a cell $C$. When $\bu = \be_a$ is the source, we need an additional adjustment: we include each terminal as a singleton cell, and we set $z_{\{\be_a\}, \be_a, a} = -1$ to account for the unit commodity-$a$ flow starting at $\be_a$, so we satisfy $\sum_{C: \bu \in C}z_{C, \bu, a} = 0$ at $\bu = \be_a$. 

We now choose the $y_C$ variables. For a terminal singleton cell, the dual LP constraint \eqref{eq:dual_cell} implies that $y_{\{\be_a\}} + z_{\{\be_a\}, \be_a, a} \leq 0$, and therefore $y_{\{\be_a\}} \leq 1$, and we choose $y_{\{\be_a\}} = 1$ in order to maximize the dual vertex-consistency LP objective. Similarly, for all other non-singleton cells, \eqref{eq:dual_cell} with a feasible constant labeling $\sigma$ implies that $y_C \leq 0$, and we set $y_C = 0$. This gives the dual LP objective~$k$.

It remains to choose the edge weights to satisfy \eqref{eq:dual_cell} for all feasible labelings, which also motivates how we define \emph{capacities} in the Flow LP. Consider a flow variable $f_{a, \bu, i} = t$ from $\bu$ to $\bv = \bu - \frac{1}{n}\be_a + \frac{1}{n}\be_i$, and let $C$ be the cell containing both $\bu$ and $\bv$. The contribution of $f_{a, \bu, i}$ to the left hand side of \eqref{eq:dual_cell} is equal to
\[
t \cdot (\bone[\sigma(\bu) = a] - \bone[\sigma(\bv) = a]),
\]
while its contribution to the right hand side is equal to
\[
w_{\{\bu, \bv\}} \cdot \bone[\sigma(\bu) \neq \sigma(\bv)].
\]

An easy way to satisfy \eqref{eq:dual_cell} is to set $w_{\{\bu, \bv\}} = t$. However, this would give the total edge weights~$kn$ (since every path from $\be_a$ to $\Gamma_a$ has $n$ edges), and after normalization we obtain a feasible solution with $\sum_e{w_e} = n$ and a dual certificate of value only 1, which is not useful for proving new lower bounds. We improve over this naïve construction by considering multiple flows together. Take a $k = 3$ example where we have two flows that start at the same vertex $\bu$ and both increase coordinate~$2$, with $f_{1, \bu, 2} = s$ and $f_{3, \bu, 2} = t$, as illustrated in Figure~\ref{fig:flow_lp_overview}(c). Without loss of generality, assume $s \geq t$. We may split $f_{1, \bu, 2} = s$ into two parts, $t$ and $s - t$. The first part, together with $f_{3, \bu, 2} = t$, can be dealt with by adding $t/2$ to all three edges in the triangle $\{\bu, \bx, \by\}$: this is because their contribution to the left hand side of \eqref{eq:dual_cell} is equal to
\[
t \cdot (\bone[\sigma(\bu) \in \{1, 3\}] - \bone[\sigma(\bx) = 3] - \bone[\sigma(\by) = 1]) \LE t,
\]
whereas their contribution to the right hand side is at least $t$ if the three vertices do not all receive the same label, in which case at least two of the edges are cut. For the second part with flow amount $s - t$, we proceed as in the naïve construction and add $s - t$ to the weight of $\{\bu, \by\}$. This gives a total edge weight of 
\[
s - t + 3 \cdot \frac{t}{2} \EQ \frac{s}{2} + \frac{s + t}{2} \EQ \frac{\max_{a\neq 2} f_{a, \bu, 2}}{2} + \frac{\sum_{a\neq 2} f_{a, \bu, 2}}{2}.
\]
If we sum the second term over the entire simplex, we obtain a fixed quantity $kn / 2$. This motivates us to define $b_{\bu, i} = \max_{a\neq i} f_{a, \bu, i}$, as the \emph{shared capacity} for flows of every commodity at vertex $\bu$ going in the direction of increasing the $i$-th coordinate. As a special case, when $s = t$, the new construction uses total edge weight $3t/2$, while the naïve construction uses $2t$. We are now ready to state the full Flow LP as the following simple linear program that minimizes the sum of shared capacities:
\begin{equation}
\begin{aligned}
 \min \quad & q_{k, n} \coloneqq \sum_{\bx \in \Delta_{k,n}}\sum_{i=1}^{k} b_{\bx,i} \\
\text{s.t.}\quad & f_{a,\bx,i} \LE b_{\bx,i} && \quad \bx\in \Delta_{k,n}\;,\; i \in [k]\;,\; a \in \supp(\bx) \setminus \{i\} \\
& \sum_{i\ne a} f_{a,\bx,i}-\sum_{\substack{i\ne a\\ i \in \supp(\bx)}} f_{a,\,\bx + \frac{\be_a}n - \frac{\be_i}n,\,i}\EQ \bone[\bx = \be_a] && \quad \bx\in \Delta_{k,n}\;,\; a \in \supp(\bx) \\
& f_{a,\bx,i} \GE 0,\qquad b_{\bx,i} \GE 0 .
\end{aligned}
\end{equation}

The constraints are transparent: we require that the flows respect the shared capacities, and we enforce flow conservation at every vertex (with some extra care for the terminals). The preceding discussion may be generalized to give a construction that turns any Flow LP feasible solution with value $q$ into a normalized non-opposite cut instance $w$ such that
\begin{equation}\label{eq:flow_to_vertex}
\LPvertexno(w) \GE \frac{2kn}{kn + q},
\end{equation} 
(see Theorem~\ref{thm:flow_to_vertex_LP}). Computationally, Flow LP is cheaper to solve than the dual vertex-consistency LP, which allows us to obtain even better numerical bounds (see Table~\ref{table:flow} for a compilation of numerical results), although we remark that for the same $k$ and $n$ the Flow LP generally gives a slightly weaker lower bound than the dual vertex-consistency LP, especially when $k > 4$.

\paragraph{The continuous Flow LP and a feasible solution when $k = 4$.}

From Table~\ref{table:flow} it can be seen that increasing $n$ gives better lower bounds. To analyze the Flow LP with large $n$, we take the $n \to \infty$ limit and obtain a continuous formulation of the Flow LP:
\begin{align*}
 \inf \quad & Q_{k} \coloneqq \sum_{i=1}^{k} B_i(\Delta_k) \\
\text{s.t.}\quad & 0 \LE F_{a,i} \LE B_{i}, && \qquad a\ne i,\\
& \sum_{i\ne a} \int_{\Delta_k} \nabla \varphi \cdot (\be_a - \be_i) \,dF_{a, i} \EQ \varphi(\be_a) - \int_{\Gamma_a}\varphi \,d\nu_a, && \qquad a \in [k] \;, \;\varphi \in C^{(1)}(\Delta_k). \tag{$\dagger$}\label{eq:cont_conserv}
\end{align*}

Here $F_{a,i}$ and $B_{i}$ are measures on $\Delta_k$, and $\nu_a$ is a probability measure over $\Gamma_a$ describing how the commodity exits. The first constraint is the straightforward continuous limit of the shared capacity constraint, while the second constraint is the continuous analogue of the discrete flow conservation constraint. We explain the meaning of this formulation in more detail in Section~\ref{sec:flow_lp}, where we also prove in Theorem~\ref{thm:flow_lp_limit} that for every $k$, 
\[
Q_k \EQ \lim_{n \to \infty} \frac{q_{k, n}}{n},
\]
which shows that constructing a continuous feasible solution with small total capacity gives discrete feasible solutions with asymptotically the same normalized objective value.

Finally, we describe a feasible continuous solution for $k = 4$, which implies our main theorem, the new lower bound for Multiway Cut. To reduce the total capacity, we would like to have different commodities share capacity wherever possible. For $k = 4$, numerical experiments suggest that we partition $\Delta_4$ into the following regions:
\begin{itemize}
    \item \emph{Corner} regions, where for each terminal $a$ we have the corner $K_a = \{\bt \in \Delta_4 \mid t_a \geq \alpha\}$.
    \item \emph{Tube} regions, where for each pair of terminals $a$ and $b$ we have the tube \[T_{ab} \EQ \left\{\bt\in\Delta_4 \mid \max_{\ell \in [4]} t_\ell\le \alpha \text{ and } t_i+t_j\le\beta\;,\; \{i, j\} = [4] \setminus \{a, b\}\right\}.\]
    \item \emph{The central region} 
    \[
    Z \EQ \{\bt\in\Delta_4 \mid \forall\ell\in[4], t_\ell\le \alpha \text{ and }   \forall i\neq j \in [4]\;,\; t_i+t_j\ge\beta\}.
    \]
\end{itemize}

\begin{figure}
    \begin{center}
\scalebox{.6}{
\providecommand{\QFourH}{1.147861560477988}
\begingroup
\pgfmathsetmacro{\qden}{2*\QFourH*\QFourH+12*\QFourH+3}
\pgfmathsetmacro{\qrho}{2*\QFourH/\qden}
\pgfmathsetmacro{\qdelta}{0.5-\QFourH*\qrho}
\pgfmathsetmacro{\qL}{1-\qdelta}
\pgfmathsetmacro{\qm}{\qdelta-\qrho}
\definecolor{qcorner}{HTML}{94BDD9}
\definecolor{qtube}{HTML}{EDB778}
\definecolor{qcentral}{HTML}{B7D2B3}
\definecolor{qline}{HTML}{394A51}
\definecolor{qcut}{HTML}{76548C}
\definecolor{qrhoboundary}{HTML}{A15B20}
\begin{tikzpicture}[x=1cm,y=1cm,font=\small,
  boundary/.style={draw=qline,line width=.4pt,line join=round},
  regionlabel/.style={font=\small,inner sep=1pt,text=qline},
  vertexlabel/.style={font=\small,inner sep=2pt,text=qline}]
\coordinate (E1) at (0,4.6);
\coordinate (E2) at (-3,0);
\coordinate (E3) at (3,0);
\coordinate (E4) at (.45,1.3);
\node[font=\bfseries] at (0,5.55) {The tetrahedron $\Delta_4$};

\foreach \a/\b/\c/\shade in {1/2/4/93,1/4/3/78,2/3/4/100}{
  \foreach \u/\v/\w in {\a/\b/\c,\b/\c/\a,\c/\a/\b}{
    \path[boundary,fill=qcorner!\shade!white]
      (E\u) -- ($(E\u)!\qdelta!(E\v)$)
      -- ($(E\u)!\qdelta!(E\w)$) -- cycle;
  }
  \foreach \u/\v/\w in {\a/\b/\c,\b/\c/\a,\c/\a/\b}{
    \path[boundary,fill=qtube!\shade!white]
      (barycentric cs:E\u=\qL,E\v=\qdelta)
      -- (barycentric cs:E\u=\qdelta,E\v=\qL)
      -- (barycentric cs:E\u=\qm,E\v=\qL,E\w=\qrho)
      -- (barycentric cs:E\u=\qL,E\v=\qm,E\w=\qrho)
      -- cycle;
  }
  \path[boundary,fill=qcentral!\shade!white]
    (barycentric cs:E\a=\qL,E\b=\qm,E\c=\qrho)
    -- (barycentric cs:E\a=\qm,E\b=\qL,E\c=\qrho)
    -- (barycentric cs:E\a=\qrho,E\b=\qL,E\c=\qm)
    -- (barycentric cs:E\a=\qrho,E\b=\qm,E\c=\qL)
    -- (barycentric cs:E\a=\qm,E\b=\qrho,E\c=\qL)
    -- (barycentric cs:E\a=\qL,E\b=\qrho,E\c=\qm)
    -- cycle;
}
\draw[qline,line width=.85pt,line join=round] (E1)--(E2)--(E3)--cycle;
\draw[qline,line width=.65pt] (E4)--(E1) (E4)--(E2) (E4)--(E3);
\node[vertexlabel,above] at (E1) {$\be_1$};
\node[vertexlabel,below left] at (E2) {$\be_2$};
\node[vertexlabel,below right] at (E3) {$\be_3$};
\fill[qline] (E4) circle (1pt);
\node[vertexlabel,below=2pt] at (E4) {$\be_4$};
\node[regionlabel] at (-.24,3.77) {$K_1$};
\node[regionlabel] at (-2.23,.15) {$K_2$};
\node[regionlabel] at (2.28,.15) {$K_3$};
\node[regionlabel] at (.73,1.55) {$K_4$};
\node[regionlabel] at (-.64,2.18) {$Z$};
\node[regionlabel] at (1.24,1.83) {$Z$};
\node[regionlabel] at (0,.43) {$Z$};
\node[regionlabel,rotate=57] at (-1.27,2.34) {$T_{12}$};
\node[regionlabel,rotate=-57] at (1.28,2.38) {$T_{13}$};
\node[regionlabel,rotate=-82] at (.39,2.84) {$T_{14}$};
\node[regionlabel,rotate=21] at (-1.20,.84) {$T_{24}$};
\node[regionlabel,rotate=-27] at (1.61,.88) {$T_{34}$};
\draw[qline,line width=.4pt] (0,.075)--(0,-.26);
\node[regionlabel,below] at (0,-.26) {$T_{23}$};

\coordinate (Cut12) at ($(E1)!\qdelta!(E2)$);
\coordinate (Cut13) at ($(E1)!\qdelta!(E3)$);
\coordinate (Cut14) at ($(E1)!\qdelta!(E4)$);
\draw[qcut,line width=1.15pt,densely dashed] (Cut12)--(Cut13);
\draw[qcut,line width=1.15pt] (Cut12)--(Cut14)--(Cut13);
\draw[qcut,line width=.55pt] (Cut13)--(1.65,3.52)--(2.05,3.52);
\node[anchor=west,text=qcut,align=left,inner sep=2pt]
  at (2.05,3.52) {$t_1=\alpha$};
\coordinate (R12a) at (barycentric cs:E1=\qL,E2=\qm,E4=\qrho);
\coordinate (R12b) at (barycentric cs:E1=\qm,E2=\qL,E4=\qrho);
\draw[qrhoboundary,line width=1.05pt] (R12a)--(R12b);
\draw[qrhoboundary,line width=1.7pt] (Cut12)--(R12a);
\fill[qrhoboundary] (Cut12) circle (1.05pt);
\fill[qrhoboundary] (R12a) circle (1.05pt);
\draw[qrhoboundary,line width=.4pt]
  (Cut12)--($(Cut12)+(-.095,.25)$)
  (R12a)--($(R12a)+(-.095,.25)$);
\draw[qrhoboundary,<->,>=stealth,line width=.55pt]
  ($(Cut12)+(-.076,.20)$)--node[above=1pt,sloped,inner sep=1pt] {$\beta$}
  ($(R12a)+(-.076,.20)$);

\begin{scope}[xshift=7.6cm]
\coordinate (Y2) at (-3,0);
\coordinate (Y3) at (3,0);
\coordinate (Y4) at (0,4.6);
\pgfmathsetmacro{\qfraction}{\qrho/\qdelta}
\node[font=\bfseries] at (0,5.55) {The cross section $t_1=\alpha$ (enlarged)};
\node[text=qline,font=\footnotesize] at (0,5.14)
  {Showing regions from $t_1<\alpha$};
\path[boundary,fill=qcentral] (Y2)--(Y3)--(Y4)--cycle;
\foreach \a/\b/\c in {2/3/4,3/4/2,4/2/3}{
  \path[boundary,fill=qtube]
    (Y\a)--($(Y\a)!\qfraction!(Y\b)$)
    --($(Y\a)!\qfraction!(Y\c)$)--cycle;
}
\draw[qcut,line width=1.15pt,line join=round] (Y2)--(Y3)--(Y4)--cycle;
\node[vertexlabel,below left] at (Y2) {$\by_{12}$};
\node[vertexlabel,below right] at (Y3) {$\by_{13}$};
\node[vertexlabel,above] at (Y4) {$\by_{14}$};
\node[regionlabel] at (-2.14,.42) {$T_{12}$};
\node[regionlabel] at (2.14,.42) {$T_{13}$};
\node[regionlabel] at (0,3.66) {$T_{14}$};
\node[regionlabel,font=\footnotesize] at (0,1.48) {$Z\cap\{t_1=\alpha\}$};
\node[text=qline,font=\footnotesize] at (0,-.50)
  {$\by_{1b}=\alpha\be_1+(1 - \alpha) \be_b$};
\end{scope}

\begin{scope}[xshift=15.2cm]
\coordinate (U2) at (-3,0);
\coordinate (U3) at (3,0);
\coordinate (U4) at (0,4.6);
\node[font=\bfseries] at (0,5.55) {The opposite face $t_1=0$};
\foreach \a/\b/\c in {2/3/4,3/4/2,4/2/3}{
  \path[boundary,fill=qcorner]
    (U\a)--($(U\a)!\qdelta!(U\b)$)
    --($(U\a)!\qdelta!(U\c)$)--cycle;
  \path[boundary,fill=qtube]
    (barycentric cs:U\a=\qL,U\b=\qdelta)
    --(barycentric cs:U\a=\qdelta,U\b=\qL)
    --(barycentric cs:U\a=\qm,U\b=\qL,U\c=\qrho)
    --(barycentric cs:U\a=\qL,U\b=\qm,U\c=\qrho)--cycle;
}
\path[boundary,fill=qcentral]
  (barycentric cs:U2=\qL,U3=\qm,U4=\qrho)
  --(barycentric cs:U2=\qm,U3=\qL,U4=\qrho)
  --(barycentric cs:U2=\qrho,U3=\qL,U4=\qm)
  --(barycentric cs:U2=\qrho,U3=\qm,U4=\qL)
  --(barycentric cs:U2=\qm,U3=\qrho,U4=\qL)
  --(barycentric cs:U2=\qL,U3=\qrho,U4=\qm)--cycle;
\draw[qline,line width=.85pt,line join=round] (U2)--(U3)--(U4)--cycle;
\node[vertexlabel,below left] at (U2) {$\be_2$};
\node[vertexlabel,below right] at (U3) {$\be_3$};
\node[vertexlabel,above] at (U4) {$\be_4$};
\node[regionlabel] at (-2.12,.44) {$K_2$};
\node[regionlabel] at (2.12,.44) {$K_3$};
\node[regionlabel] at (0,3.69) {$K_4$};
\node[regionlabel] at (0,.27) {$T_{23}$};
\node[regionlabel,rotate=57] at (-1.22,2.17) {$T_{24}$};
\node[regionlabel,rotate=-57] at (1.22,2.17) {$T_{34}$};
\node[regionlabel,font=\footnotesize] at (0,1.43) {$Z\cap\{t_1=0\}$};
\node[text=qline,font=\footnotesize] at (0,-.50)
  {$S_1:\quad \beta\le t_2,t_3,t_4\le \alpha$};
\end{scope}

\foreach \xx/\col/\txt in {2.9/qcorner/{Corners $K_a$},7.35/qtube/{Tubes $T_{ab}$},11.5/qcentral/{Central region $Z$}}{
  \path[draw=qline,line width=.35pt,fill=\col] (\xx,-1.12) rectangle ++(.28,.22);
  \node[anchor=west,text=qline] at ({\xx+.39},-1.01) {\txt};
}
\end{tikzpicture}
\endgroup}
    \end{center}
    \caption{The Corner/Tube/Central Partition of $\Delta_4$. The first panel shows an overview of the partition. The second panel shows the cross section at $t_1 = \alpha$, and the third shows the facet $t_1 = 0$.}
    \label{fig:partition}
\end{figure}
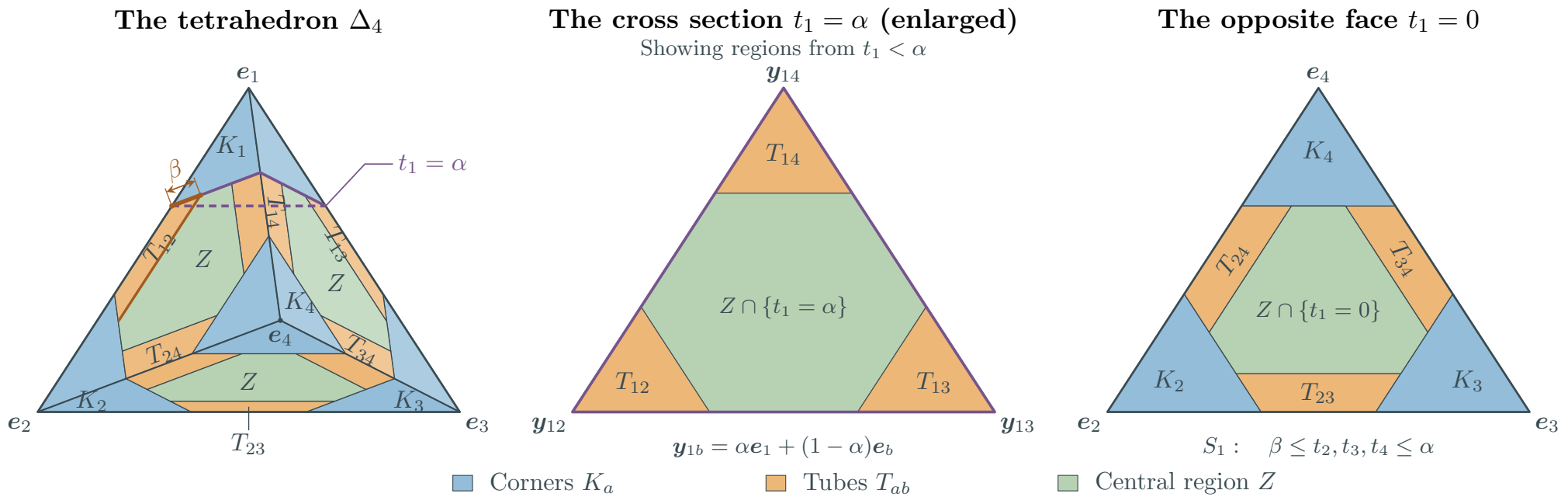

Here $\alpha, \beta > 0$ are some parameters controlling the sizes of these regions. See Figure~\ref{fig:partition} for an illustration. In this partition, the commodity-$a$ flow travels alone in the corner $K_a$, only the commodity-$a$ and commodity-$b$ flows travel and share the capacity in the tube $T_{ab}$, and in the central region all capacities are fully utilized (i.e., $F_{a, i} = B_i$ for every $a \neq i$). The flow solution is defined piecewise on these regions, in a way that the constraint $0 \leq F_{a, i} \leq B_i$ is satisfied directly by definition. The difficulty lies in proving the continuous flow conservation equation \eqref{eq:cont_conserv}, and for this proof the main tool is \emph{Gauss's  divergence theorem} (see Theorem~\ref{thm:divergence}), which enables us to reduce the integral on the left hand side of \eqref{eq:cont_conserv} in some region to some other integral on the surface of that region. We can then assemble these integrals region by region and using a careful construction make sure the contributions from the internal surfaces cancel, leaving just the right hand side of \eqref{eq:cont_conserv}. In the end, the total capacity of our construction can be expressed as a univariate rational function, and by choosing the optimal parameter we prove that $Q_4 \leq 2.54492029$. By \eqref{eq:flow_to_vertex}, this gives the desired lower bound of $1.22232$ for Multiway Cut.

\subsection{Related Work}\label{subsec:related}

In this section, we survey a few directions of research related to our investigation, particularly concerning the study of the Sherali--Adams hierarchy~\cite{sherali1990hierarchy} and related hierarchies such as the Sum of Squares (SoS) / Lasserre hierarchy~\cite{Lasserre01}.

\paragraph{Comparison of LP formulations.} For classical optimization problems such as the Traveling Salesperson Problem (TSP), the standard LP relaxation has served as a benchmark~\cite{HeldK70,HeldK71,ShmoysW90,Wolsey09,KarlinKG24}.  The work of Goemans~\cite{Goemans95} studies LP formulations of TSP of varying levels of strength and compares to what degree these formulations approximate each other. However, such comparisons do not lead to improved integrality gaps for the classical LP.

For another example, in the Maximum Agreement Forest (MAF) problem~\cite{HeinJWZ96,Wu09,OlverSSSZ23}, one seeks to compute the similarity between two phylogenetic trees. Here, linear programming is again useful for approximation algorithms, but it is less clear what the ``best'' linear programming algorithm should be. For a while, an LP formulation by Wu~\cite{Wu09} was the state of the art, but Olver, Schalekamp, van der Ster, Stougie, and van Zuylen~\cite{OlverSSSZ23} constructed a better LP formulation by showing their LP has an integrality gap of at most $2$, while Wu's LP has an integrality gap of at least $16/5$~\cite[Appendix B]{OlverSSSZ23}. As pointed out to us by ChatGPT 6 Astra, the proof of the integrality gap can be viewed as a discrepancy between the new and old LP formulations. However, unlike in our results, no corresponding computational hardness results are implied by this gap (since \cite{OlverSSSZ23} constructed a strictly superior algorithm).

\paragraph{Past evidence for the Unique Games Conjecture.} In light of Raghavendra's theorem~\cite{Raghavendra08,RaghavendraS09,Brown-CohenR16} it is known that, assuming the Unique Games Conjecture, the optimal approximation algorithm for every finite-template\footnote{In particular, this theory applies to Multiway Cut when the number of terminals is a fixed constant, although in that case the work of Manokaran, Naor, Raghavendra, and Schwartz~\cite{MNRS08} shows the much weaker CKR relaxation is sufficient for worst-case analysis. } CSP can be achieved by rounding its associated Basic Semidefinite Program (SDP). 

Implicitly, such a result implies that gap instances should exist for various strengthened algorithms such as the Sherali--Adams and Lasserre hierarchies. As a result, many works have sought to locate such integrality gaps for a variety of optimization problems, including Vertex Cover and MAX~CUT~\cite{SchoenebeckTT07,Schoenebeck08,CharikarMM09,KhotV15}. Such investigations have led to some surprising results, such as a result of O'Donnell and Schramm~\cite{ODonnellS21} that a (mild) super-constant number of rounds of Sherali--Adams can outperform the Basic LP relaxation of MAX CUT.

Of particular interest is the work of O'Donnell and Zhou~\cite{ODonnellZ13} who use degree 4 Lasserre to (partially) certify a MAX CUT integrality gap of Khot-Vishnoi~\cite{KhotV15}. Such certificates were automated more generally by Kauers, O’Donnell, Tan, and Zhou~\cite{KauersOTZ14}; see also the work of De, Mossel, and Neeman~\cite{DeMN16}. However, like in the TSP example, such methods are not being used to automate the construction of \emph{improved} integrality gaps.

\paragraph{Universal CSP algorithms.} In the study of exactly solving constraint satisfaction problems, the proofs of the CSP dichotomy theorem~\cite{Bulatov17,Zhuk20} give algorithms for solving every constraint satisfaction problem in the complexity class $\mathsf{P}$. However, such algorithms are highly complex, leading to an increasing desire to find simpler algorithms which imply the CSP dichotomy. A particularly popular topic of interest is to investigate variants of the Sherali--Adams and Lasserre hierarchies to solve various CSPs~\cite{BrakensiekGWZ20,CiardoZ23,CiardoZ25a,LichterP25,Zhuk25,CiardoZ26,BartoHZ26}. Some of the resulting hierarchies have spurred further investigations in the study of random CSP instances~\cite{ChanNP24,ChanN25}.

\paragraph{Continuous linear programs.} Continuous linear programs have been used to study the limiting behaviors of their discrete counterparts. For example, Chan, Chen, Wu, and Zhao \cite{chan2018ranking} studied a linear program arising from the analysis of the ranking algorithm for online matching using the continuous limit, and Chan, Chen, and Jiang \cite{chan2015revealing} used a continuous LP to study the Generalized Secretary Problem. For flow problems, Strang \cite{strang1983maximal} studied maximum continuous flows in a plane, whose analysis also uses the divergence theorem. 

\subsection{Organization of the Paper}
The rest of the paper is organized as follows. We present the strengthened, Sherali--Adams style LP relaxations for non-opposite cuts in Section~\ref{sec:SA}. In Section~\ref{sec:flow_lp}, we present the discrete and continuous formulations of the auxiliary Flow LP, and show how they yield gap instances through the vertex-consistency relaxation. In Section~\ref{sec:continuous_feasible_solution}, we present a feasible solution to the continuous Flow LP for $k = 4$, thereby proving our main theorem. We end with some concluding remarks and future directions in Section~\ref{sec:concl}. 

Appendix~\ref{app:standard-cuts} extends our strengthened LP framework to ``standard'' multiway cuts. Appendix~\ref{app:k=3} demonstrates how our framework is sufficient to recover all previously known results for standard and non-opposite cuts when $k=3$. Appendix~\ref{app:determinants} includes calculations omitted from Section~\ref{sec:continuous_feasible_solution}. Appendix~\ref{app:freedom_vertex_lp} discusses degrees of freedom in \eqref{LP:dual-vertex} which are not needed for our analysis but may be of independent interest.

\paragraph{AI Disclosure.} \label{para:ai} The project grew out of a $k = 4, n = 10$ non-opposite cut integrality gap instance which we found using the minimax framework that we developed in our earlier works~\cite{BHPZ23,brakensiek2026improved}. This instance implies a lower bound of $600/499 = 1.2024048...$, better than the previous bound of 1.20016 due to~\cite{BCKM20}, but we did not have a human proof of the gap. We gave the instance to ChatGPT 5.6 Sol and asked it to produce a simple human proof, which it failed to do. Instead, it found the edge-consistency relaxation (see Section~\ref{subsec:edge}) as a computational certificate of the gap. We realized this is a powerful tool that can be used to find much stronger gap instances for Multiway Cut. With the help of ChatGPT 5.6 Sol and Claude Opus 5, we developed and numerically experimented with the other LP relaxations in Section~\ref{sec:SA}. During these experiments, we asked ChatGPT 5.6 Sol if there is a simpler formulation based on flows, and it responded with the Flow LP, and when we asked if there is a way to analyze the $n \to \infty$ limit, it produced the continuous Flow LP formulation and proved the correspondence between discrete and continuous formulations. The feasible solution for the continuous formulation which we present in Section~\ref{sec:continuous_feasible_solution} was obtained by giving ChatGPT 5.6 Sol a qualitative description of our numerical findings with the discrete Flow LP and asking it to turn it into a continuous solution with a feasibility proof.

All text in the paper was written by the human authors. ChatGPT 6 Astra and Claude Opus 5.5 provided suggestions on the exposition and references, pointed out (minor) flaws in written proofs, and proofread the paper. ChatGPT 6 Astra was also used to produce the tables and figures in the paper (using TikZ~\cite{tikz} and Matplotlib~\cite{Hunter:2007}), as well as some of the determinant calculations in Appendix~\ref{app:determinants}.

Many of the calculations used to generate Tables~\ref{table:LP_vertex}, \ref{table:LP_edge},  \ref{table:flow}, \ref{table:LP_vertex_standard}, and \ref{table:LP_edge_standard} were performed using LLM-generated code on Intel Xeon Gold 5220 and 6330 CPUs from UC Berkeley's Savio HPC cluster.

\section{Sherali--Adams Relaxations for Non-Opposite Cuts}\label{sec:SA}
In this section, we formally define the non-opposite cut framework for constructing integrality gap instances for Multiway Cut. We then present LP relaxations which strengthen the CKR relaxation for non-opposite cut, and show how to use them to find and certify integrality gap instances.

\subsection{Non-Opposite Cuts}

Let $\Delta_{k, n} \coloneqq \left\{\bx \in \left(\frac{1}{n}\mathbb{Z}_{\geq 0}\right)^k\mid \sum_{i = 1}^kx_i = 1\right\}$ and 
\[
E_{k, n} \coloneqq \left\{\{\bx, \by\} \mid \bx, \by \in \Delta_{k, n}\;,\; \exists i\neq j \in [k]\;,\; \bx = \by + \frac{1}{n}\be_i - \frac{1}{n}\be_j\right\}.
\] 
For every $\bx \in \Delta_{k, n}$, define the non-opposite cuts feasible label set
\[
\Lno(\bx) \coloneqq \begin{cases}
\{i\} & \text{if } \bx = \be_i, i \in [k]\\
\supp(\bx) \cup \{k+1\} & \text{otherwise}.
\end{cases}
\]
Here we let $\supp(\bx) = \{i \in [k] \mid x_i > 0\}$.
\begin{definition}
    A function $P: \Delta_{k, n} \to [k + 1]$ is called a \emph{non-opposite cut} if $P(\bx) \in \Lno(\bx)$ for every $\bx \in \Delta_{k, n}$.
\end{definition}

To define a non-opposite cut instance on $(\Delta_{k, n}, E_{k, n})$, it is sufficient to define a weight function $w: E_{k, n} \to \mathbb{R}_{\geq 0}$. Given a non-opposite cut $P: \Delta_{k, n} \to [k+1]$, we say that an edge $e = \{\bu, \bv\} \in E_{k, n}$ is \emph{cut} by $P$ if $P(\bu) \neq P(\bv)$.
The cost incurred by the non-opposite cut $P$ with respect to the edge weights $w$ is
\[
  \cost(w, P) \coloneqq \sum_{e =\{\bu, \bv\} \in E_{k,n}} w_e \cdot \bone[P(\bu) \neq P(\bv)].
\]
We further define 
\[
\OPTno(w) \coloneqq \min_{\substack{P: \Delta_{k, n} \to [k+1] \\ P \text{ non-opposite}}} \cost(w, P)
\]
to be the minimum total weight of edges cut by a non-opposite cut. We say that our weight function $w$ is \emph{normalized} if
\[
    \sum_{e \in E_{k,n}} w_e = n.
\]
As discussed in the introduction, for any normalized weight function $w$, we have that $\LPCKRno(w) \le 1$. A non-opposite cut gap instance implies UG-hardness of inapproximability for Multiway Cut, due to the following result:

\begin{theorem}[\cite{AMM17}]\label{thm:amm}
    Let $w: E_{k, n} \to \mathbb{R}_{\geq 0}$ be normalized. Assuming the Unique Games Conjecture, for every $\epsilon > 0$, it is NP-hard to approximate Multiway Cut, with sufficiently many terminals, within a multiplicative factor of~$\OPTno(w) - \epsilon$. 
\end{theorem}

\subsection{Strengthened LP Relaxations for Non-opposite Cuts}\label{subsec:vertex_lp}
We now define a strengthened LP relaxation for non-opposite cuts. For any $\bv \in \Delta_{k, n+1}$, define the \emph{downward cell} $D(\bv)$ of $\Delta_{k,n}$ as follows:
\[
    D(\bv) \EQ \left\{\bu \in \Delta_{k, n} \mid \exists i \in [k]\;,\; \bu = \frac{n+1}{n}\bv - \frac{1}{n}\be_i\right\}.
\]

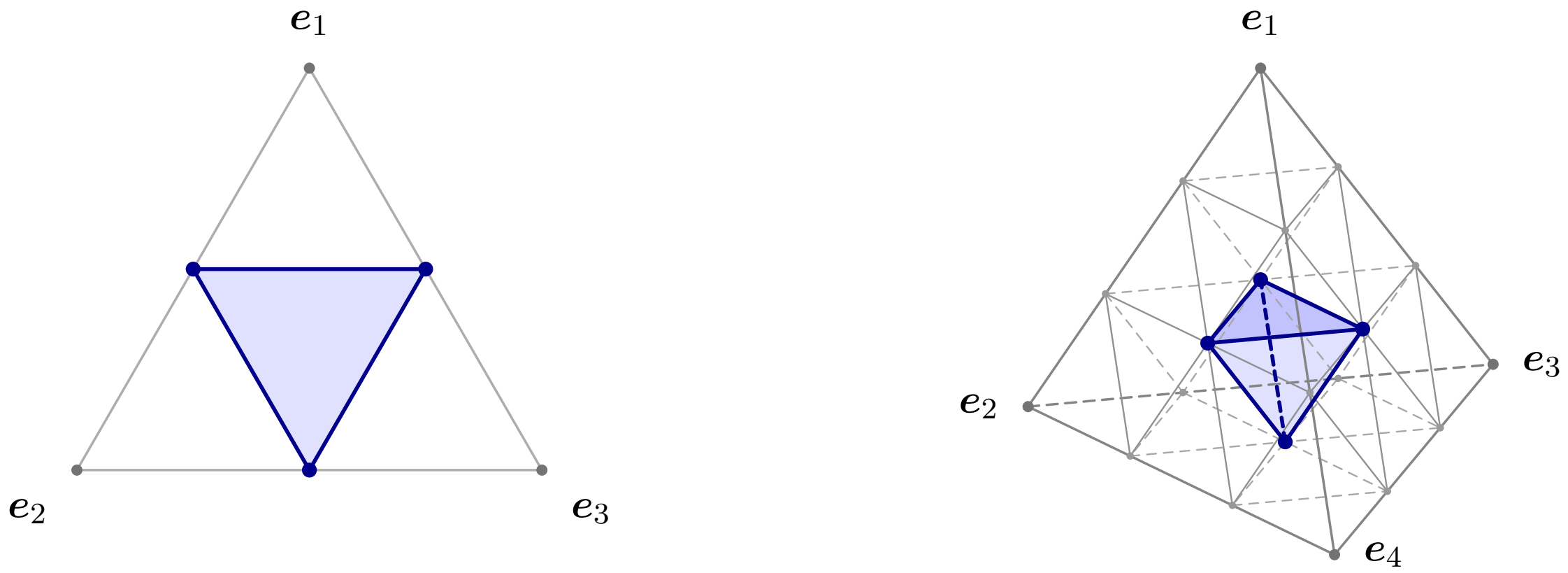
\begin{figure}[htbp]
  \centering
  \begin{tikzpicture}[
    x=1cm,y=1cm,
    line cap=round,line join=round,
    ambient/.style={draw=black!32,line width=.65pt},
    ambient hidden/.style={ambient,dash pattern=on 3pt off 2pt},
    cell edge/.style={draw=blue!55!black,line width=1.05pt},
    cell hidden/.style={cell edge,dash pattern=on 2.4pt off 1.8pt}
  ]
    \begin{scope}[shift={(2.6,0)}]
      \coordinate (A1) at (0,3.2);
      \coordinate (A2) at (-2.2,-.6);
      \coordinate (A3) at (2.2,-.6);
      \coordinate (X1) at ($(A2)!.5!(A3)$);
      \coordinate (X2) at ($(A1)!.5!(A3)$);
      \coordinate (X3) at ($(A1)!.5!(A2)$);

      \fill[blue!12] (X1)--(X2)--(X3)--cycle;
      \draw[ambient] (A1)--(A2)--(A3)--cycle;
      \draw[cell edge] (X1)--(X2)--(X3)--cycle;
      \foreach \i in {1,2,3} {
        \fill[black!55] (A\i) circle[radius=1.5pt];
        \fill[blue!55!black] (X\i) circle[radius=2pt];
      }

      \node[font=\small,align=center] at (0,-2.65)
        {\textbf{(a)} $k=3$, $n=2$\\[3pt]Downward triangle\\[3pt]$\bv = (\frac{1}{3}, \frac{1}{3}, \frac{1}{3})$\\[3pt] $D(\bv)=\{(\frac12,\frac12,0),(\frac12,0,\frac12),(0,\frac12,\frac12)\}$};

    \node[above=5pt]       at (A1) {$\be_1$};
    \node[below left=4pt]  at (A2) {$\be_2$};
    \node[below right=4pt] at (A3) {$\be_3$};
    \end{scope}

    \begin{scope}[shift={(11.6,0)}]
      \tikzset{
        ambient/.style={draw=black!48,line width=.65pt},
        grid visible/.style={draw=black!43,line width=.45pt},
        grid hidden/.style={draw=black!34,line width=.45pt,
                            dash pattern=on 2.5pt off 2pt}
      }
      \coordinate (A1) at (0,3.2);
      \coordinate (A2) at (-2.2,0);
      \coordinate (A3) at (2.2,.4);
      \coordinate (A4) at (.7,-1.4);

    \node[above=5pt] at (A1) {$\be_1$};
    \node[left=4pt]  at (A2) {$\be_2$};
    \node[right=4pt] at (A3) {$\be_3$};
    \node[right=4pt] at (A4) {$\be_4$};
      
      \coordinate (G) at (.175,.55);
      \foreach \i in {1,2,3,4}
        \coordinate (X\i) at ($(G)!-0.333333333!(A\i)$);

      \fill[blue!12] (X1)--(X2)--(X3)--cycle;
      \fill[blue!24] (X4)--(X2)--(X3)--cycle;

      \foreach \a/\b/\c in {1/2/3,2/3/4} {
        \foreach \t in {0.333333333,0.666666667} {
          \draw[grid hidden]
            ($(A\a)!\t!(A\b)$)--($(A\a)!\t!(A\c)$)
            ($(A\b)!\t!(A\a)$)--($(A\b)!\t!(A\c)$)
            ($(A\c)!\t!(A\a)$)--($(A\c)!\t!(A\b)$);
        }
      }
      \draw[ambient hidden] (A2)--(A3);
      \foreach \a/\b/\c in {1/2/4,1/3/4} {
        \foreach \t in {0.333333333,0.666666667} {
          \draw[grid visible]
            ($(A\a)!\t!(A\b)$)--($(A\a)!\t!(A\c)$)
            ($(A\b)!\t!(A\a)$)--($(A\b)!\t!(A\c)$)
            ($(A\c)!\t!(A\a)$)--($(A\c)!\t!(A\b)$);
        }
      }
      \draw[ambient] (A1)--(A2)--(A4)--(A3)--cycle;
      \draw[ambient] (A1)--(A4);
      \foreach \a/\b in {1/2,1/3,1/4,2/3,2/4,3/4}
        \foreach \t in {0.333333333,0.666666667}
          \fill[black!40] ($(A\a)!\t!(A\b)$) circle[radius=1pt];

      \draw[cell hidden] (X1)--(X4);
      \draw[cell edge] (X1)--(X2)--(X4)--(X3)--cycle;
      \draw[cell edge] (X2)--(X3);
      \foreach \i in {1,2,3,4} {
        \fill[black!55] (A\i) circle[radius=1.5pt];
        \fill[blue!55!black] (X\i) circle[radius=2pt];
      }

      \node[font=\small,align=center] at (0,-2.65)
        {\textbf{(b)} $k=4$, $n=3$\\[3pt]Downward tetrahedron\\[3pt]$\bv = (\frac14, \frac14, \frac14, \frac14)$\\[3pt]$D(\bv) = \{(\frac 13, \frac 13, \frac 13, 0), (\frac 13, \frac 13, 0, \frac 13), (\frac 13, 0, \frac 13, \frac 13), (0, \frac 13, \frac 13, \frac 13)\}$};
    \end{scope}
  \end{tikzpicture}
  \caption{Examples of downward cells.}
  \label{fig:downward-cells}
\end{figure}

See Figure~\ref{fig:downward-cells} for an illustration of downward cells in the case of $k = 3$ and $k = 4$. We call these cells downward since they are inverted compared to the original simplex. Note that some of these cells may be \emph{degenerate} in the sense that they contain fewer than $k$ vertices. For example, $D(\be_i) = \{\be_i\}$ contains only one vertex. Let $\cD_{k, n}$ be the set of downward cells in $\Delta_{k, n}$. By definition, there is a 1-to-1 correspondence between $\cD_{k, n}$ and $\Delta_{k, n+1}$. 

\begin{lemma}\label{lem:downward_cell_edge_unique}
    For every $e \in E_{k, n}$, there exists a unique $D \in \cD_{k, n}$ such that $e \subseteq D$.
\end{lemma}
\begin{proof}
    Let $e = \{\bu, \bv\}$. Then by definition there exist distinct $i, j \in [k]$ such that $\bu = \bv + \frac{1}{n}\be_i - \frac{1}{n}\be_j$, so we have $\bu, \bv \in D((n\bu + \be_j)/(n+1))$. For uniqueness, suppose $\bu, \bv \in D(\bw)$ for some $\bw \in \Delta_{k, n+1}$, then there exist $r, s \in [k]$ such that 
    \[
        (n+1)\bw \EQ n\bu+\be_r \EQ n\bv+\be_s.
    \]
    This forces $r = j$ and $s = i$, proving uniqueness.
\end{proof}

Lemma~\ref{lem:downward_cell_edge_unique} authorizes us to slightly abuse notation. Given $e \in E_{k, n}$, we let $D(e) \in \cD_{k, n}$ denote the unique downward cell containing it. For every $C \in \cD_{k, n}$, define
\[\Omegano_C = \{\sigma \in [k+1]^C \mid \forall \bu \in C \;,\; \sigma(\bu) \in \Lno(\bu)\}\] to be the set of all valid labelings of the vertices of~$C$, i.e., labelings that are consistent with the non-opposite cut requirements. Given a weight function $w: E_{k, n} \to \mathbb{R}_{\geq 0}$, we define the following \emph{vertex-consistency} LP relaxation. 
\begin{equation}
\tag{$\LPvertexno$}\label{LP:vertex}
\begin{aligned}
\text{min}\quad & \sum_{e\in E_{k, n}} w_e p_e \\
\text{s.t.}\quad & \sum_{\sigma \in \Omegano_C} \mu_{C, \sigma} \EQ 1\,, && \qquad C \in \cD_{k, n}\,, \\
    & \sum_{\substack{\sigma \in \Omegano_{C} \\ \sigma(\bu) = a}} \mu_{C, \sigma} \EQ \lambda_{\bu, a}\,, &&  \qquad C \in \cD_{k, n}, \, \bu \in C, \, a \in \Lno(\bu)\,,\\
    & p_e \EQ \sum_{\substack{\sigma\in \Omegano_{D(e)}\\ \sigma(\bu)\ne\sigma(\bv)}} \mu_{D(e),\sigma}\,, && \qquad e = \{\bu, \bv\}\in E_{k, n}\,, \\
    & \mu_{C, \sigma} \GE 0 \,, && \qquad C \in \cD_{k, n}, \, \sigma \in \Omegano_C\,.
\end{aligned}
\end{equation}

In words, \eqref{LP:vertex} assigns to each downward cell~$C$ a probability distribution~$\mu_C$ over its valid labelings, and to each vertex~$\bu$, a probability distribution~$\lambda_{\bu}$ over its valid labelings. It then enforces that for any cell and a vertex in that cell, the marginal of the cell's distribution on that vertex must agree with the vertex's distribution. For each edge $e$, \eqref{LP:vertex} defines $p_e$ to be the probability that $e$ is cut in the local distribution for $D(e)$, the unique downward cell containing~$e$. Finally, the objective function is defined to be $\sum_{e\in E_{k, n}} w_e p_e$. Note that our vertex-consistency LP is a weaker version of some level of the Sherali--Adams LP hierarchy, which maintains local distributions for all subsets of vertices up to some fixed size. 

Let $\LPvertexno(w)$ denote the optimal objective value of \eqref{LP:vertex} for a given weight function $w$. 

\begin{lemma}\label{lem:vertex_LP}
    For any $w: E_{k, n} \to \mathbb{R}_{\geq 0}$, we have
    \[
        \LPvertexno(w) \LE \OPTno(w).
    \]
\end{lemma}
\begin{proof}
    The lemma holds since any non-opposite cut $P$ gives a feasible solution to \eqref{LP:vertex} whose value is $\cost(w, P)$. In more detail, for every $C \in \cD_{k,n}$ and $\sigma \in \Omegano_C$, let $\mu_{C, \sigma} = \bone[P|_C = \sigma]$. Furthermore, for every $\bu \in \Delta_{k,n}$ and $a \in \Lno(\bu)$, let $\lambda_{\bu, a} = \bone[P(\bu) = a]$. Then, $p_{e} = \bone[\text{$e$ is cut by $P$}]$, so the objective of \eqref{LP:vertex} is equal to $\cost(w, P)$. Minimizing this cost over all non-opposite cuts~$P$ proves that $\LPvertexno(w) \leq \OPTno(w).$
\end{proof}

By Lemma~\ref{lem:vertex_LP}, if we construct a normalized weight function $w$ with $\LPvertexno(w) > 1$, then $\OPTno(w)$ exceeds $1$ by at least the same amount, allowing us to apply Theorem~\ref{thm:amm} and obtain hardness of approximation for Multiway Cut. 

To prove a lower bound on $\LPvertexno(w)$, the natural approach is to consider the dual LP to \eqref{LP:vertex}, which can be formulated as follows. Note that the $p_e$ variables and the equality constraints associated with them in \eqref{LP:vertex} can be eliminated if we plug in the definition of $p_e$ into the objective function. We then have a dual variable~$y_C$ for each normalization constraint $\sum_{\sigma} \mu_{C, \sigma} \EQ 1$, and a dual variable $z_{C, \bu, a}$ for each marginal consistency constraint $\sum_{\sigma: \sigma(\bu) = a} \mu_{C, \sigma} \EQ \lambda_{\bu, a}$. These dual variables are unrestricted since the corresponding primal constraints are equalities. The primal variables $\lambda_{\bv, a}$ give the dual constraints $\sum_{C: \bv \in C} z_{C, \bv, a} = 0$, and the primal variables $\mu_{C,\sigma}$ give a dual inequality requiring $y_C+\sum_{\bv\in C}z_{C,\bv,\sigma(\bv)}$ to be at most the total weight of the edges in $C$ cut by~$\sigma$. Finally, in the primal, only the normalization constraints have nonzero right-hand sides, which leads to the dual objective function $\sum_{C \in \cD_{k, n}} y_C$. In summary, we have:

\begin{equation}
\label{LP:dual-vertex}
\begin{aligned}
\text{max}\quad & \sum_{C \in \cD_{k, n}} y_C \\
\text{s.t.}\quad & y_C + \sum_{\bv \in C} z_{C, \bv, \sigma(\bv)} \LE \sum_{\substack{\{\bu, \bv\} \subseteq C\\ \bu\ne\bv}}w_{\{\bu, \bv\}} \cdot \bone[\sigma(\bu) \neq \sigma(\bv)], &&\qquad C \in \cD_{k, n},  \sigma \in \Omegano_C, \\
    & \sum_{C: \bv \in C} z_{C, \bv, a} \EQ 0, && \qquad \bv \in \Delta_{k, n}, a \in \Lno(\bv),\\
    & y_C, z_{C, \bv, a} \in \mathbb{R}. && 
\end{aligned}
\tag{dual-$\LPvertexno$}
\end{equation}

Here for summation $\sum_{\{\bu, \bv\} \subseteq C}$ we used the fact that there is an edge between any two distinct vertices in the same cell. Using \eqref{LP:dual-vertex}, we can already try to certify the integrality ratios of non-opposite gap instances. In Appendix~\ref{app:k=3}, we show how to recover the $k=3$ non-opposite cut gap of $\frac{6}{5} = 1.2$ by~\cite{AMM17} using \eqref{LP:dual-vertex}. We also show how our framework can be used to recover the $k=3$ gap result for standard cuts by~\cite{CCT06}. In Appendix~\ref{app:freedom_vertex_lp}, we document some degrees of freedom in the solution space of \eqref{LP:dual-vertex} which might be of interest, though they are not needed in our analysis.

Observe that in~\eqref{LP:dual-vertex}, the constraints are also linear in the $w_e$ variables. This allows us to treat them as LP variables and simultaneously optimize over them. More specifically, we consider the following LP, obtained by letting $w_e$ be non-negative variables and adding the constraint that $w$ is normalized to \eqref{LP:dual-vertex}.

\begin{equation}
\label{LP:dual-vertex-w}
\begin{aligned}
\text{max}\quad & \sum_{C \in \cD_{k, n}} y_C \\
\text{s.t.}\quad & y_C + \sum_{\bv \in C} z_{C, \bv, \sigma(\bv)} \LE \sum_{\{\bu, \bv\} \subseteq C}w_{\{\bu, \bv\}} \cdot \bone[\sigma(\bu) \neq \sigma(\bv)], &&\qquad C \in \cD_{k, n},  \sigma \in \Omegano_C, \\
    & \sum_{C: \bv \in C} z_{C, \bv, a} \EQ 0, && \qquad \bv \in \Delta_{k, n}, a \in \Lno(\bv),\\
    & \sum_{e \in E_{k, n}} w_e \EQ n, \\
    & y_C, z_{C, \bv, a} \in \mathbb{R}, \quad w_e \geq 0. && 
\end{aligned}
\tag{dual-$\LPvertexno$-$w$}
\end{equation}

\begin{lemma}\label{lem:vertex_LP_hardness}
    Let $(y_C)_C, (z_{C, \bx, a})_{C, \bx, a}, (w_e)_e$ be a feasible solution to \eqref{LP:dual-vertex-w} for some $k, n \in \mathbb{N}$, with $\sum_C y_C = r > 0$. Then assuming the Unique Games Conjecture,  Multiway Cut is NP-hard to approximate within a factor of $r - \epsilon$, for any $\epsilon > 0$.
\end{lemma}
\begin{proof}
    By LP duality, we have $\LPvertexno(w) \geq \sum_C y_C = r$. Since \eqref{LP:dual-vertex-w} enforces $w$ is normalized, the lemma then follows from Theorem~\ref{thm:amm} and Lemma~\ref{lem:vertex_LP}.
\end{proof}

Let $\LPvertexno(k, n)$ be the optimum objective value of \eqref{LP:dual-vertex-w} for any given $k$ and $n$. We used Gurobi~\cite{gurobi} to numerically compute $\LPvertexno(k, n)$, for some moderate values of~$k$ and~$n$, and the results are summarized in Table~\ref{table:LP_vertex}. By Lemma~\ref{lem:vertex_LP_hardness}, each ratio in Table~\ref{table:LP_vertex} implies (after being rigorously verified using exact arithmetic) a corresponding UG-hardness ratio for Multiway Cut, up to an arbitrarily small additive loss.

\begin{table}[ht]
\centering
\begin{tabular}{|c|c|c|c|c|}
    \hline
\backslashbox{$n$}{$k$} & 4 & 5 & 6 &7\\\hline
5 & 1.18733509 & 1.19047619 & 1.19128658 & 1.19128658 \\
10 & 1.20240480 & 1.20905902 & 1.21072958 & 1.21097816 \\
15 & 1.20799659 & 1.21488688 & 1.21712108 & 1.21755682 \\
20 & 1.21150860 & 1.21857935 & 1.22050921 & 1.22104874 \\
25 & 1.21386007 & 1.22059229 & 1.22262741 & -- \\
30 & 1.21531379 & 1.22211296 & 1.22405936 & -- \\
40 & 1.21704683 & 1.22402773 & -- & -- \\
50 & 1.21807634 & 1.22510362 & -- & -- \\
\hline
\end{tabular}
\caption{Numerically computed values of $\LPvertexno(k, n)$, rounded to eight decimal places. A dash (--) indicates that the computation required (well) over 128GB of memory and was not completed.}
\label{table:LP_vertex}
\end{table}

Finally, we note that if we take the dual again for \eqref{LP:dual-vertex-w}, then we obtain a ``primal'' LP with a particularly nice form:
\begin{equation}
\tag{$\LPvertexno$-$w$-free}\label{LP:vertex-w-free}
\begin{aligned}
\text{min}\quad & np \\
\text{s.t.}\quad & \sum_{\sigma \in \Omegano_C} \mu_{C, \sigma} \EQ 1 && \qquad C \in \cD_{k, n}, \\
    & \sum_{\substack{\sigma \in \Omegano_{C} \\ \sigma(\bu) = a}} \mu_{C, \sigma} \EQ \lambda_{\bu, a} &&  \qquad C \in \cD_{k, n}, \, \bu \in C, \, a \in \Lno(\bu),\\
    & p \GE \sum_{\sigma\in \Omegano_{D(e)}} \mu_{D(e),\sigma}\, \bone[\sigma(\bu)\ne\sigma(\bv)] && \qquad e = \{\bu, \bv\}\in E_{k, n}, \\
    & \mu_{C, \sigma} \GE 0 && \qquad C \in \cD_{k, n}, \, \sigma \in \Omegano_C.
\end{aligned}
\end{equation}

This primal formulation does not use the edge weights. It can be seen as trying to find a ``fractional'' non-opposite cut, which is a collection of local distributions over feasible non-opposite cut labelings in each downward cell satisfying vertex consistency, such that the maximum cut probability of any edge $e$ (which is defined using the local distribution for $D(e)$) is minimized.

\subsection{An Even Stronger Edge-Consistency Relaxation}\label{subsec:edge}

So far, we have discussed a strengthened LP relaxation based on the downward cells of the discrete simplex $\Delta_{k,n}$. One can obtain an even stronger relaxation by also introducing the \emph{upward cells} of~$\Delta_{k,n}.$ Assume $k \ge 3$ and $n \ge 2$ in this section. For any $\bv \in \Delta_{k,n-1}$, define the upward cell
\[
    U(\bv) \;:=\; \left\{\bu \in \Delta_{k, n} \mid \exists i \in [k]\;,\; \bu = \frac{n-1}{n}\bv + \frac{1}{n}\be_i\right\}.
\]
\begin{figure}[t]
\centering
\begingroup
\definecolor{upone}{HTML}{407FA9}
\definecolor{uptwo}{HTML}{C7873C}
\definecolor{upthree}{HTML}{438C76}
\definecolor{upfour}{HTML}{9573AA}

\begin{tikzpicture}[
  x=1cm,y=1cm,
  line cap=round,line join=round,
  every node/.style={font=\small},
  cell edge/.style={line width=.75pt},
  outer edge/.style={draw=black!65,line width=.85pt},
  vertex label/.style={inner sep=2pt}
]
  \begin{scope}
    \coordinate (A1) at (0,4.5);
    \coordinate (A2) at (-2.6,0);
    \coordinate (A3) at (2.6,0);
    \coordinate (A12) at ($(A1)!.5!(A2)$);
    \coordinate (A13) at ($(A1)!.5!(A3)$);
    \coordinate (A23) at ($(A2)!.5!(A3)$);

    \path[fill=black!2] (A1)--(A2)--(A3)--cycle;
    \path[cell edge,draw=upone,fill=upone!23]
      (A1)--(A12)--(A13)--cycle;
    \path[cell edge,draw=uptwo,fill=uptwo!23]
      (A2)--(A12)--(A23)--cycle;
    \path[cell edge,draw=upthree,fill=upthree!23]
      (A3)--(A13)--(A23)--cycle;
    \draw[outer edge] (A1)--(A2)--(A3)--cycle;

    \foreach \p in {1,2,3,12,13,23}
      \fill[black!70] (A\p) circle[radius=1.25pt];

    \node[vertex label,above=4pt] at (A1) {$\be_1$};
    \node[vertex label,below left=3pt] at (A2) {$\be_2$};
    \node[vertex label,below right=3pt] at (A3) {$\be_3$};

    \node[font=\small\bfseries,anchor=north] at (0,-.75)
      {(a) Three upward triangles in $\Delta_{3,2}$};
  \end{scope}

  \begin{scope}[shift={(7.7,1.3)}]
    \coordinate (B1) at (0,3.2);
    \coordinate (B2) at (-2.2,0);
    \coordinate (B3) at (2.2,.4);
    \coordinate (B4) at (.7,-1.4);

    \foreach \i/\j in {1/2,1/3,1/4,2/3,2/4,3/4}
      \coordinate (B\i\j) at ($(B\i)!.5!(B\j)$);

    \foreach \a/\b/\c/\tone in {1/2/4/2,1/3/4/5}{
      \path[fill=black!\tone] (B\a)--(B\b)--(B\c)--cycle;
    }

    \foreach \a/\b/\c in {1/2/3,2/3/4}{
      \foreach \u/\v/\w in {\a/\b/\c,\b/\c/\a,\c/\a/\b}{
        \ifcase\u\relax
          \or\def\cellcolor{upone}
          \or\def\cellcolor{uptwo}
          \or\def\cellcolor{upthree}
          \or\def\cellcolor{upfour}
        \fi
        \path[fill=\cellcolor,fill opacity=.08]
          (B\u)--($(B\u)!.5!(B\v)$)
          --($(B\u)!.5!(B\w)$)--cycle;
      }
    }

    \foreach \a/\b/\c/\tone in {1/2/4/2,1/3/4/5}{
      \foreach \u/\v/\w in {\a/\b/\c,\b/\c/\a,\c/\a/\b}{
        \ifcase\u\relax
          \or\def\cellcolor{upone}
          \or\def\cellcolor{uptwo}
          \or\def\cellcolor{upthree}
          \or\def\cellcolor{upfour}
        \fi
        \pgfmathsetmacro{\frontopacity}{(23+2*\tone)/100}
        \path[fill=\cellcolor,fill opacity=\frontopacity]
          (B\u)--($(B\u)!.5!(B\v)$)
          --($(B\u)!.5!(B\w)$)--cycle;
      }
    }

    \foreach \a/\b/\c in {1/2/3,2/3/4}{
      \foreach \u/\v/\w in {\a/\b/\c,\b/\c/\a,\c/\a/\b}{
        \ifcase\u\relax
          \or\def\cellcolor{upone}
          \or\def\cellcolor{uptwo}
          \or\def\cellcolor{upthree}
          \or\def\cellcolor{upfour}
        \fi
        \draw[
          draw=\cellcolor!65,
          line width=.65pt,
          dash pattern=on 2.5pt off 2pt
        ]
          ($(B\u)!.5!(B\v)$)--($(B\u)!.5!(B\w)$);
      }
    }

    \draw[
      outer edge,
      draw=black!45,
      dash pattern=on 3pt off 2pt
    ] (B2)--(B3);

    \foreach \a/\b/\c in {1/2/4,1/3/4}{
      \foreach \u/\v/\w in {\a/\b/\c,\b/\c/\a,\c/\a/\b}{
        \ifcase\u\relax
          \or\def\cellcolor{upone}
          \or\def\cellcolor{uptwo}
          \or\def\cellcolor{upthree}
          \or\def\cellcolor{upfour}
        \fi
        \draw[cell edge,draw=\cellcolor]
          ($(B\u)!.5!(B\v)$)--($(B\u)!.5!(B\w)$);
      }
    }

    \draw[outer edge] (B1)--(B2)--(B4)--(B3)--cycle;
    \draw[outer edge] (B1)--(B4);

    \foreach \p in {1,2,3,4,12,13,14,24,34}
      \fill[black!70] (B\p) circle[radius=1.25pt];
    \fill[black!40] (B23) circle[radius=1.25pt];

    \node[vertex label,above=4pt] at (B1) {$\be_1$};
    \node[vertex label,left=4pt] at (B2) {$\be_2$};
    \node[vertex label,right=4pt] at (B3) {$\be_3$};
    \node[vertex label,right=4pt] at (B4) {$\be_4$};

    \node[font=\small\bfseries,anchor=north] at (0,-2.05)
      {(b) Four upward tetrahedra in $\Delta_{4,2}$};
  \end{scope}
\end{tikzpicture}
\endgroup
\caption{An illustration of upward cells.}\label{fig:upward_cells}
\end{figure}
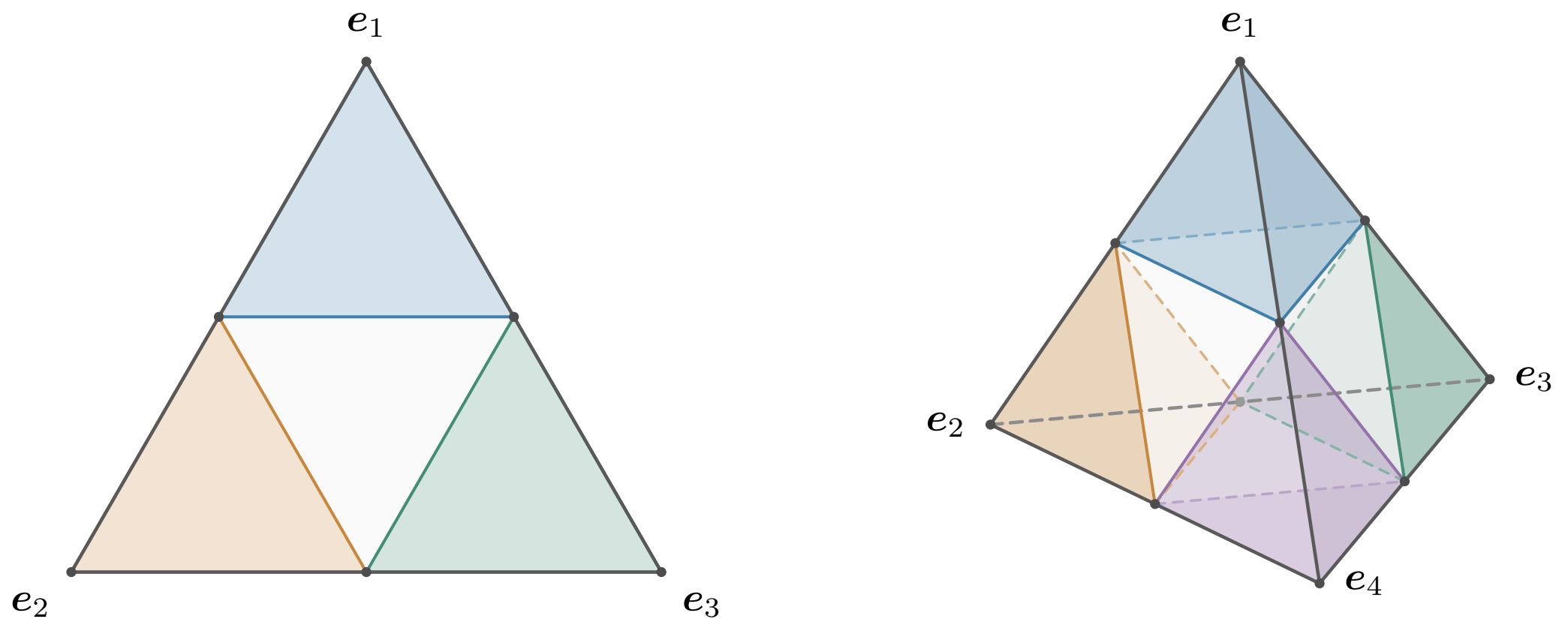

See Figure~\ref{fig:upward_cells} for an illustration of upward cells. We let $\cU_{k,n}$ denote the set of all upward cells in~$\Delta_{k,n}$. Unlike for downward cells, every upward cell has exactly $k$ vertices since $\frac{n-1}{n}\bv + \frac{1}{n}\be_i \in \Delta_{k,n}$ for every $\bv \in \Delta_{k,n-1}$. Analogous to Lemma~\ref{lem:downward_cell_edge_unique}, every $e \in E_{k,n}$ is contained in exactly one upward cell.

\begin{lemma}\label{lem:upward_cell_edge_unique}
    For every $e \in E_{k, n}$, there exists a unique $U \in \cU_{k, n}$ such that $e \subseteq U$.
\end{lemma}
\begin{proof}
    Let $e = \{\bu, \bv\}$. Then by definition there exist distinct $i, j \in [k]$ such that $\bu = \bv + \frac{1}{n}\be_i - \frac{1}{n}\be_j$, so we have $\bu, \bv \in U((n\bu - \be_i)/(n-1))$. For uniqueness, suppose $\bu, \bv \in U(\bw)$ for some $\bw \in \Delta_{k, n-1}$, then there exist $r, s \in [k]$ such that 
    \[
        (n-1)\bw \EQ n\bu-\be_r \EQ n\bv-\be_s.
    \]
    This forces $r = i$ and $s = j$, proving uniqueness.
\end{proof}
Thus, for any $e \in E_{k,n}$, we let $\cU(e)$ denote the unique upward cell $U \in \cU_{k,n}$ for which $e \subseteq U$. Analogous to downward cells, for every $C \in \cU_{k, n}$, define $\Omegano_C = \{\sigma \in [k+1]^C \mid \forall \bu \in C, \sigma(\bu) \in \Lno(\bu)\}$ to be the set of all feasible labelings of the vertices of~$C$. Combining Lemma~\ref{lem:downward_cell_edge_unique} and Lemma~\ref{lem:upward_cell_edge_unique}, we have that each $e \in E_{k,n}$ is in exactly one upward cell and one downward cell. Thus, given a weight function $w: E_{k, n} \to \mathbb{R}_{\geq 0}$, we can generalize $\LPvertexno(w)$ to account for consistency of the cell assignments on the common edge between upward and downward cells. Formally, this relaxation is $\LPedgeno(w)$, which we define as follows.

\begin{equation}
\tag{$\LPedgeno$}\label{LP:edge}
\begin{aligned}
\text{min}\quad & \sum_{e\in E_{k, n}} w_e p_e \\
\text{s.t.}\quad & \sum_{\sigma \in \Omegano_C} \mu_{C, \sigma} \EQ 1 && \qquad C \in \cD_{k, n} \cup \cU_{k,n}, \\
    & \sum_{\substack{\sigma \in \Omegano_{C} \\ \sigma(\bu) = a\\\sigma(\bv) = b}} \mu_{C, \sigma} \EQ \lambda_{e, a, b} &&  \!\!\!\begin{tabular}{l}$\qquad C \in  \cD_{k, n} \cup \cU_{k, n}, \, e = \{\bu, \bv\} \in E_{k, n}, e \subseteq C,$\\$\qquad a \in \Lno(\bu), \, b \in \Lno(\bv)$\end{tabular}\\
    & p_e \EQ \sum_{\substack{a \in \Lno(\bu)\\b \in \Lno(\bv) \\ a \neq b}} \lambda_{e, a, b} && \qquad e = \{\bu, \bv\}\in E_{k, n}, \\
    & \mu_{C, \sigma} \GE 0 && \qquad C \in \cD_{k, n} \cup \cU_{k,n}, \, \sigma \in \Omegano_C.
\end{aligned}
\end{equation}

In words, \eqref{LP:edge} assigns to each downward and upward cell~$C$ a probability distribution~$\mu_C$ over its feasible labelings, and to each edge $e = \{\bu, \bv\}$, a probability distribution~$\lambda_{e}$ over its feasible labelings \footnote{We may fix an arbitrary orientation of $e$ so that the notation $\lambda_{e, a, b}$ is well-defined.}. It then enforces that for any cell and any edge in that cell, the marginal of the cell's distribution on that edge must agree with the edge's distribution. For each edge $e$, \eqref{LP:edge} defines~$p_e$ to be the probability that $e$ is cut according to $\lambda_e$. Finally, the objective function is defined to be $\sum_{e\in E_{k, n}} w_e p_e$. Let $\LPedgeno(w)$ denote the minimum objective value of \eqref{LP:edge} for a given weight function $w$. 

Similar to Lemma~\ref{lem:vertex_LP}, we proceed to compare $\LPedgeno(w)$ to $\OPTno(w)$. In fact, we show that $\LPedgeno(w)$ sandwiches between $\LPvertexno(w)$ and $\OPTno(w)$.

\begin{lemma}\label{lem:edge_LP}
    For $k\ge 3$, $n \ge 2$, and any $w: E_{k, n} \to \mathbb{R}_{\geq 0}$, we have
    \[
        \LPvertexno(w) \LE \LPedgeno(w) \LE \OPTno(w).
    \]
\end{lemma}

\begin{proof}
    Like in the proof of Lemma~\ref{lem:vertex_LP}, let $P$ be any non-opposite cut. For any $C \in \cD_{k,n} \cup \cU_{k,n}$ and $\sigma \in \Omegano_C$, define $\mu_{C, \sigma} = \bone[P|_C = \sigma]$. Furthermore, for any $e = \{\bu, \bv\} \in E_{k,n}$ and $a \in \Lno(\bu)$ and $b \in \Lno(\bv)$, define $\lambda_{e, a, b} = \bone[P(\bu) = a \wedge P(\bv) = b]$. This is a valid assignment to \eqref{LP:edge}, with $p_{e} = \bone[P(\bu) \neq P(\bv)].$ For this assignment, the objective to \eqref{LP:edge} is $\cost(w, P)$. Minimizing this over all non-opposite cuts $P$ proves $\LPedgeno(w) \leq \OPTno(w)$.

To prove that $\LPvertexno(w) \leq \LPedgeno(w)$, it suffices to show any feasible solution to \eqref{LP:edge} corresponds to a feasible solution to \eqref{LP:vertex} with the same objective value. To do this, for each $C \in \cD_{k,n}$ and $\sigma \in \Omegano_C$, we let $\mu_{C,\sigma}$ be the same value in both \eqref{LP:edge} and \eqref{LP:vertex}. As such, for all $e \in E_{k,n}$, $p_e$ is the same in both LPs.

For $\bu \in \Delta_{k,n}$ and $a \in \Lno(\bu)$, we need to also define $\lambda_{\bu, a}$. To do this, pick an arbitrary edge $e = \{\bu, \bv\} \in E_{k,n}$ and let $\lambda_{\bu, a} = \sum_{b \in \Lno(\bv)} \lambda_{e, a, b}$. To complete the proof, it suffices to show that for any $C \in \cD_{k,n}$, $\bu \in C$, and $a \in \Lno(\bu)$ that
\[
  \sum_{\substack{\sigma \in \Omegano_C\\\sigma(\bu) = a}} \mu_{C,\sigma} \EQ \lambda_{\bu, a}.
\]
If $\bu = \be_i$ for some $i \in [k]$ and $C = \{\be_i\}$, then this equation holds by inspection as $\lambda_{\be_i, i} = 1$. Thus, we may assume that $C$ has at least two elements.

Let $e = \{\bu, \bv\} \in E_{k,n}$ be the edge such that $\lambda_{\bu, a} = \sum_{b \in \Lno(\bv)}\lambda_{e, a, b}$. Also pick an arbitrary $\bv' \in C \setminus \{\bu\}$ so that for $e' = \{\bu, \bv'\}$. We then have that
\[
\sum_{\substack{\sigma \in \Omegano_C\\\sigma(\bu) = a}} \mu_{C,\sigma} \EQ \sum_{b' \in \Lno(\bv')} \lambda_{e', a, b'}.
\]
Now, assume that $i, j, i', j' \in [k]$ are chosen such that $\bv = \bu + \frac{1}{n}\be_i - \frac{1}{n}\be_j$ and $\bv' = \bu + \frac{1}{n}\be_{i'} - \frac{1}{n}\be_{j'}$. Further define $\bv'' = \bu + \frac{1}{n}\be_i - \frac{1}{n}\be_{j'}$ and let $e'' = \{\bu, \bv''\}$.\footnote{Note that $e''$ is not a valid edge when $\bv'' = \bu$; that is, $i = j'$. If $i' \neq j$, then we can perform an analogous argument with $\bv'' = \bu + \frac{1}{n}\be_{i'} - \frac{1}{n}\be_{j}$. Otherwise, we have that $\bv$ and $\bv'$ are ``antipodes'' where $\bv + \bv'= 2\bu$. Completing the proof in this case require $k \ge 3$ and $n \ge 2$, where we know there is third vertex $\bv_1$ adjacent to $\bu$ from which we can use our existing argument twice to see that $\lambda_{\bu,a} = \sum_{b \in \Lno(\bv_1)}\lambda_{\{\bu,\bv_1\},a,b} = \sum_{b' \in \Lno(\bv')}\lambda_{\{\bu,\bv'\},a,b'}.$} From the proofs of Lemma~\ref{lem:downward_cell_edge_unique} and Lemma~\ref{lem:upward_cell_edge_unique}, we can deduce that $\{\bu, \bv, \bv''\} \subseteq D(\frac{n\bu + \be_i}{n+1})$ and $\{\bu, \bv', \bv''\} \subseteq U(\frac{n\bu - \be_{j'}}{n-1})$. From these containments, we can deduce that (writing $D$ for $D(\frac{n\bu + \be_i}{n+1})$ and $U$ for $U(\frac{n\bu - \be_{j'}}{n-1})$)

\begin{align*}
  \sum_{b \in \Lno(\bv)} \lambda_{e, a, b} &= \sum_{\substack{\sigma \in \Omegano_{D}\\\sigma(\bu) = a}} \mu_{D,\sigma} = \sum_{b'' \in \Lno(\bv'')} \lambda_{e'', a, b''}\\
  &=  \sum_{\substack{\sigma \in \Omegano_{U}\\\sigma(\bu) = a}} \mu_{U,\sigma} = \sum_{b' \in \Lno(\bv')} \lambda_{e', a, b'},
\end{align*}
as desired.
\end{proof}

Like in \eqref{LP:dual-vertex}, we can define a dual to \eqref{LP:edge} as follows. Like in \eqref{LP:vertex}, we can eliminate the equations defining each $p_e$ via substitution. For every $C \in \cD_{k,n} \cup \cU_{k,n}$, we have a dual variable $y_C$ for the equation $\sum_{\sigma \in \Omegano_C} \mu_{C,\sigma} = 1$. For each $e = \{\bu, \bv\} \in E_{k,n}$, $a \in \Lno(\bu)$, and $b \in \Lno(\bv)$, there are exactly two equations involving $\lambda_{e, a, b}$, corresponding to the downward cell~$D(e)$ and the upward cell $U(e)$. Setting these two equations equal, we get a \emph{consistency equation}
\[
  \sum_{\substack{\sigma \in \Omegano_{D(e)}\\\sigma(\bu) = a\\\sigma(\bv) = b}} \mu_{D(e), \sigma} \EQ \sum_{\substack{\sigma \in \Omegano_{U(e)}\\\sigma(\bu) = a\\\sigma(\bv) = b}} \mu_{U(e), \sigma}.
\]
Let $z_{e,a,b}$ be the dual variable for this consistency equation. Using these consistency equations, we can eliminate $\lambda_{e, a, b}$ entirely from \eqref{LP:edge}, leading to a new objective of
\[
  \min\ \ \ \sum_{e =\{\bu, \bv\}\in E_{k,n}} w_e \cdot \sum_{\substack{\sigma \in \Omegano_{D(e)}\\\sigma(\bu) \neq \sigma(\bv)}} \mu_{D(e),\sigma}.
\]
From this, we can define dual LP \eqref{LP:dual-edge} as follows.
\begin{equation}
\label{LP:dual-edge}
\begin{aligned}
\text{max}\quad & \sum_{C \in \cD_{k, n} \cup\,\cU_{k,n}} y_C \\
\text{s.t.}\quad & y_C \LE \sum_{e = \{\bu, \bv\} \subseteq C} (z_{e, \sigma(u), \sigma(v)} + w_e \cdot \bone[\sigma(u) \neq \sigma(v)]) &&\qquad C \in \cD_{k, n},  \sigma \in \Omegano_C, \\
& y_C \LE -\sum_{e = \{\bu, \bv\} \subseteq C} z_{e, \sigma(u), \sigma(v)} &&\qquad C \in \cU_{k, n},  \sigma \in \Omegano_C,\\
    & y_C, z_{e, a, b} \in \mathbb{R}. && 
\end{aligned}
\tag{dual-$\LPedgeno$}
\end{equation}
Like in~\eqref{LP:dual-vertex}, the constraints are also linear in the $w_e$ variables. This allows us to treat them as LP variables and simultaneously optimize over them. More specifically, we consider the following augmentation to \eqref{LP:dual-edge}, obtained by letting $w_e$ be non-negative variables corresponding to a normalized weight function.

\begin{equation}
\tag{dual-$\LPedgeno$-$w$}
\label{LP:dual-edge-w}
\begin{aligned}
\text{max}\quad & \sum_{C \in \cD_{k, n} \cup\,\cU_{k,n}} y_C \\
\text{s.t.}\quad & y_C \LE \sum_{e = \{\bu, \bv\} \subseteq C} (z_{e, \sigma(u), \sigma(v)} + w_e \cdot \bone[\sigma(u) \neq \sigma(v)]) &&\qquad C \in \cD_{k, n},  \sigma \in \Omegano_C, \\
& y_C \LE -\sum_{e = \{\bu, \bv\} \subseteq C} z_{e, \sigma(u), \sigma(v)} &&\qquad C \in \cU_{k, n},  \sigma \in \Omegano_C,\\
    & \sum_{e \in E_{k, n}} w_e \EQ n, \\
    & y_C, z_{e, a, b} \in \mathbb{R}, \quad w_e \geq 0. && 
\end{aligned}
\end{equation}

Using an argument analogous to that of Lemma~\ref{lem:vertex_LP_hardness}, one can show that the objective value of any feasible solution to \eqref{LP:dual-edge-w} implies a corresponding UG-hardness for the approximability of Multiway Cut.

\begin{lemma}\label{lem:edge_LP_hardness}
    Let $(y_C)_C, (z_{e, a, b})_{e, a, b}, (w_e)_e$ be a feasible solution to \eqref{LP:dual-edge-w} for some $k, n \in \mathbb{N}$, with $\sum_C y_C = r > 0$. Then assuming the Unique Games Conjecture,  Multiway Cut is NP-hard to approximate within a factor of $r - \epsilon$, for any $\epsilon > 0$.
\end{lemma}
\begin{proof}
    By LP duality, we have $\LPedge(w) \geq \sum_C y_C = r$. Since \eqref{LP:dual-edge-w} enforces $w$ is normalized, the lemma then follows from Theorem~\ref{thm:amm} and Lemma~\ref{lem:edge_LP}.
\end{proof}
We let $\LPedgeno(k, n)$ denote the objective value of \eqref{LP:dual-edge-w}.

\paragraph{Numerical comparison with \eqref{LP:dual-vertex-w}.}\text{}
In Table~\ref{table:LP_edge}, we compute the objective value of \eqref{LP:dual-edge-w} for various choices of $k$ and $n$. We note that in general, the values in Table~\ref{table:LP_edge} are modestly larger than the values in Table~\ref{table:LP_vertex}, although more running-time is needed due to the larger number of variables and equations in \eqref{LP:dual-edge-w}. We leave it as an open question whether \eqref{LP:dual-edge-w} offers any advantage in the $k, n \to \infty$ limit. In this work, we base our hardness result on \eqref{LP:dual-vertex-w} since it's much easier to analyze. 

\begin{table}[ht]
\centering
\begin{tabular}{|c|c|c|c|c|}
    \hline
\backslashbox{$n$}{$k$} & 4 & 5 & 6 &7\\\hline
5 & 1.18733509 & 1.19076065 & 1.19185373 & 1.19222303 \\
10 & 1.20240480 & 1.20962287 & 1.21133701 & 1.21166680 \\
15 & 1.20799659 & 1.21556578 & 1.21812215 & 1.21856465 \\
20 & 1.21162016 & 1.21901643 & 1.22148591 & -- \\
25 & 1.21386813 & 1.22112193 & 1.22351775 & -- \\
30 & 1.21531555 & 1.22257225 & -- & -- \\
40 & 1.21705044 & 1.22440445 & -- & -- \\
50 & 1.21809815 & 1.22547115 & -- & --\\
\hline
\end{tabular}
\caption{Numerically computed values of $\LPedgeno(k, n)$, rounded to eight decimal places. A dash (--) indicates that the computation requires over 128GB of memory and was not completed.}
\label{table:LP_edge}
\end{table}

\section{The Flow LP}\label{sec:flow_lp}

In this section, we construct an auxiliary LP called the \emph{Flow LP}, and prove that its feasible solutions can be used to construct feasible solutions for \eqref{LP:dual-vertex-w}. We then show a continuous formulation of the Flow LP and prove that it is equivalent to the discrete formulation as $n$ goes to infinity.

\subsection{The discrete formulation of the Flow LP}

Consider the following flow problem. On the discretized simplex $\Delta_{k, n}$, we send one unit of flow from each terminal $a \in [k]$, called the \emph{commodity-$a$ flow}, to its opposite face $\Gamma_a = \{\bx \in \Delta_{k, n} \mid x_a = 0\}$. We require that these flows never travel backwards, that is, at any vertex $\bx \in \Delta_{k, n}$, the commodity-$a$ flow can only go to the vertices $\{\bx - \frac{\be_a}n + \frac{\be_i}n \mid i \neq a\}$. We use $f_{a, \bx, i}$ to denote the amount of commodity-$a$ flow that travels from $\bx$ to $\bx - \frac{\be_a}n + \frac{\be_i}n$. These variables must satisfy flow conservation. To move the flow, we need to pay a cost in \emph{capacity}: for any $\bx \in \Delta_{k, n}$ and direction $i \in [k]$, we use~$b_{\bx, i}$ to denote the capacity to send flow at $\bx$ toward the direction that increases the $i$-th coordinate. More specifically, the flows must satisfy
\begin{equation}
f_{a, \bx, i} \LE b_{\bx, i}, \quad \forall a \neq i\;,\; a \in \supp(\bx).
\end{equation}
Note that the capacity is defined in a non-traditional sense: several commodities may use the capacity simultaneously, but we are paying for their max (we may take $b_{\bx, i} = \max_{a \neq i}f_{a, \bx, i}$), not their sum. We are interested in constructing flows and capacities such that the total sum of the capacities $\sum_{\bx \in \Delta_{k, n}, i \in [k]} b_{\bx, i}$ is minimized. This may be expressed as the following \emph{Flow LP}.

\begin{equation}
\begin{aligned}
 \min \quad & q_{k,n} \coloneqq \sum_{\bx \in \Delta_{k,n}}\sum_{i=1}^{k} b_{\bx,i} \\
\text{s.t.}\quad & f_{a,\bx,i} \LE b_{\bx,i} && \qquad \bx\in \Delta_{k,n}\;,\; i \in [k]\;,\; a \in \supp(\bx) \setminus \{i\}, \\
& \sum_{i\ne a} f_{a,\bx,i}-\sum_{\substack{i\ne a\\ i \in \supp(\bx)}} f_{a,\,\bx + \frac{\be_a}n - \frac{\be_i}n,\,i}\EQ \bone[\bx = \be_a] && \qquad \bx\in \Delta_{k,n}\;,\; a \in \supp(\bx)\\
& f_{a,\bx,i} \GE 0,\qquad b_{\bx,i} \GE 0 .
\end{aligned}
\label{eq:flow_lp}\tag{Flow-LP}
\end{equation}
The second constraint enforces flow conservation when $\bx$ is not the source, and it specifies one unit of starting flow when $\bx = \be_a$ is the source.

\begin{lemma}\label{lem:total_flow_dist}
    In any feasible solution to \eqref{eq:flow_lp} we have $\sum_{a, \bx, i}f_{a, \bx, i} = kn$.
\end{lemma}
\begin{proof}
    For any commodity $a$, we have one unit of flow originating from $\be_a$, and each step decreases the $a$-th coordinate by $1/n$. Hence $\sum_{\bx, i}f_{a, \bx, i} = n$. Summing over all $k$ commodities proves the lemma. 
\end{proof}

\subsection{Lifting the Flow LP to the Vertex-Consistency LP}\label{sub-lift}

We now show that any solution to \eqref{eq:flow_lp} may be naturally translated into a solution to \eqref{LP:dual-vertex-w}.

\begin{theorem}\label{thm:flow_to_vertex_LP}
    Let $k \geq 2$ and $n \geq 1$ be integers. For any feasible solution $(f_{a, \bx, i})_{a, \bx, i}$, $(b_{\bx, i})_{\bx, i}$ to \eqref{eq:flow_lp}, there exists a feasible solution $(y_C)_C, (z_{C, \bx, a})_{C, \bx, a}, (w_e)_e$ to \eqref{LP:dual-vertex-w} such that 
    \[
    \sum_{C \in \cD_{k, n}}  y_C \GE \frac{2kn}{kn +  \sum_{\bx \in \Delta_{k,n}}\sum_{i=1}^{k} b_{\bx,i}}.
    \]
\end{theorem}
\begin{proof}
    We explicitly construct a solution to \eqref{LP:dual-vertex} satisfying
    \[
    \sum_{C \in \cD_{k, n}}  y_C \EQ k, \quad \sum_{e \in E_{k, n}}w_e \LE \frac{kn + \sum_{\bx \in \Delta_{k,n}}\sum_{i=1}^{k} b_{\bx,i}}{2}.
    \]
    Then rescaling the solution appropriately gives a feasible solution to \eqref{LP:dual-vertex-w} and proves the statement.

    In the following, for any $\bx \in \Delta_{k, n}$ and $i \in [k]$, let $C_i(\bx) = D((n\bx + \be_i) / (n+1)) \in \cD_{k, n}$. We start with the $z_{C, \bx, a}$ variables. We set $z_{C, \bx, k+1} = 0$ for every cell $C$ and vertex $\bx \in \Delta_{k, n}$. For $z_{C, \bx, a}$  variables with $a \in [k]$, the overall idea is to satisfy the following:
    \begin{itemize}
        \item $z_{C_i(\bx), \bx, a} \ge 0$ if $i \neq a$ and $z_{C_i(\bx), \bx, a} \le 0$ if $i = a$.
        \item $|z_{C_a(\bx), \bx, a}|$ is the amount of commodity-$a$ flow arriving at the vertex $\bx$.
        \item For $i \neq a$, $z_{C_i(\bx), \bx, a}$ is the amount of commodity-$a$ flow going from the vertex $\bx$ to the vertex $\bx - \frac{\be_a}n + \frac{\be_i}n$.
    \end{itemize}
    More specifically, we define $z_{C_a(\bx), \bx, a} = - \sum_{i \neq a}f_{a, \bx, i}$ and $z_{C_i(\bx), \bx, a} = f_{a, \bx, i}$ for $i \neq a$. Note that this immediately satisfies the constraint $\sum_{C: \bx \in C} z_{C, \bx, a} = 0$ in \eqref{LP:dual-vertex}. For the $y_C$ variables, we set 
    \[
        y_C \EQ \begin{cases}
        1 & \text{if } \exists a\in[k], C = C_a(\be_a), \\
        0 & \text{otherwise}.
        \end{cases}
    \]
    This choice combined with the flow conservation constraint in~\eqref{eq:flow_lp} implies that for every $\bx \in \Delta_{k, n}$, $a \in \supp(\bx)$ we have
    \begin{equation}\label{eq:flow_conserv}
        \sum_{i \neq a} f_{a, \bx, i} \EQ y_{C_a(\bx)} + \sum_{\substack{ i \in \supp(\bx) \setminus \{a\}}} f_{a, \bx + \frac{\be_a}n - \frac{\be_i}n , i}.
    \end{equation}
    Finally, for the edge weights $w_e$, we define them using the following procedure.
    \begin{itemize}
        \item Initialize all weights to zero.
        \item For every $\bx \in \Delta_{k, n}$, $i \in [k]$, do the following:
        \begin{itemize}
            \item Let $S = \{a_1, \ldots, a_m\} \subseteq [k] \setminus \{i\}$ be the set of commodities $a$ for which $f_{a, \bx, i} > 0$, ordered so that $f_{a_1, \bx, i} \ge f_{a_2, \bx, i} \ge \cdots \ge f_{a_m, \bx, i}$ with ties broken arbitrarily.
            \item For $j = 1$ to $m$, let $S_j = \big\{\bx, \bx - \frac{\be_{a_1}}n + \frac{\be_i}n, \ldots, \bx - \frac{\be_{a_j}}n + \frac{\be_i}n\big\}$ and $w_j = f_{a_j, \bx, i} - f_{a_{j+1}, \bx, i}$ (with $f_{a_{j+1}, \bx, i} = 0$ if $j = m$), then $S_j$ is a clique in $(\Delta_{k, n}, E_{k, n})$ of size $j+1$, and we add $w_j/j$ to the weights of all edges in $S_j$. 
        \end{itemize}
    \end{itemize}
    Note that the total edge weights added for $\bx \in \Delta_{k, n}$, $i \in [k]$ is equal to 
    \begin{align*} 
        \sum_{j = 1}^m\frac{(f_{a_j, \bx, i} - f_{a_{j+1}, \bx, i})}{j}  \cdot \binom{j+1}{2}  \EQ \sum_{j = 1}^m\frac{(f_{a_j, \bx, i} - f_{a_{j+1}, \bx, i})\cdot(j+1)}{2} \EQ \frac{f_{a_1, \bx, i} + \sum_{j = 1}^m f_{a_j, \bx, i}}{2},
    \end{align*}
    so the total sum for the edge weights is 
    \begin{align*} 
        \sum_{\bx \in \Delta_{k,n}}\sum_{i=1}^{k} \frac{f_{a_1, \bx, i} + \sum_{j = 1}^m f_{a_j, \bx, i}}{2} & \LE \sum_{\bx \in \Delta_{k,n}}\sum_{i=1}^{k} \frac{b_{\bx, i} + \sum_{j = 1}^m f_{a_j, \bx, i}}{2} \\
        & \EQ \frac{\sum_{\bx \in \Delta_{k,n}}\sum_{i=1}^{k}b_{\bx, i} + kn}{2},
    \end{align*}
    where the inequality comes from the flow LP constraint $f_{a, \bx, i} \leq b_{\bx, i}$, and the equality is due to Lemma~\ref{lem:total_flow_dist}.

    We now verify the feasibility of our construction. It remains to show that 
    \begin{equation}\label{eq:y_z_ineq}
    y_C + \sum_{\bx \in C} z_{C, \bx, \sigma(\bx)} \LE \sum_{e = \{\bx, \by\} \subseteq C}w_e \cdot \bone[\sigma(\bx) \neq \sigma(\by)],
    \end{equation}
    for $C \in \cD_{k, n}$ and $\sigma \in \Omegano_C$. Fix some arbitrary $C$, indexed by $\bv \in \Delta_{k, n+1}$, and some $\sigma \in \Omegano_C$, we first show that 
    \begin{equation}\label{eq:y+z}
    y_C + \sum_{\bx \in C} z_{C, \bx, \sigma(\bx)} \EQ \sum_{i \in \supp(\bv)} \sum_{\substack{a \in \supp(\bv) \\ a \neq i}} f_{a, \bv^{(-i)}, i} \left(\bone[\sigma(\bv^{(-i)}) = a] - \bone[\sigma(\bv^{(-a)}) = a]\right),
    \end{equation}
    where for any $i\in\supp(\bv)$ we use $\bv^{(-i)}$ to denote $((n+1)\bv - \be_i)/n \in \Delta_{k, n}$. Indeed, any $\bx \in C$ is equal to $\bv^{(-i)}$ for some $i \in \supp(\bv)$, so we have 
    \[\sum_{\bx \in C} z_{C, \bx, \sigma(\bx)} \EQ \sum_{i \in \supp(\bv)} z_{C, \bv^{(-i)}, \sigma(\bv^{(-i)})}.
    \]
    Furthermore, by definition we have 
    \[
    z_{C, \bv^{(-i)}, \sigma(\bv^{(-i)})} \EQ \begin{cases}
    f_{a, \bv^{(-i)}, i} & \text{if } \sigma(\bv^{(-i)}) = a \neq i, \\
    - \sum_{j \neq i}f_{i, \bv^{(-i)}, j} & \text{if }\sigma(\bv^{(-i)}) = i.
    \end{cases}
    \]
    By \eqref{eq:flow_conserv}, if $i \in \supp(\bv^{(-i)})$ then we have
    \[\sum_{j \neq i}f_{i, \bv^{(-i)}, j} \EQ y_{C} + \sum_{j \in \supp(\bv) \setminus \{i\}}f_{i, \bv^{(-j)}, j},\] 
    so we obtain
    \begin{align*}
        & y_C+\sum_{\bx\in C}z_{C,\bx,\sigma(\bx)} \\
        \EQ & y_C + \sum_{i \in \supp(\bv)} z_{C, \bv^{(-i)}, \sigma(\bv^{(-i)})} \\
        \EQ & y_C + \sum_{i \in \supp(\bv)}\left(\sum_{a \in \supp(\bv) \setminus \{i\}} f_{a, \bv^{(-i)}, i} \bone[\sigma(\bv^{(-i)}) = a] - \sum_{j \neq i}f_{i, \bv^{(-i)}, j}\bone[\sigma(\bv^{(-i)}) = i]\right)\\
        \EQ & y_C + \sum_{i \in \supp(\bv)}\left(\sum_{\substack{a \in \supp(\bv) \\ a \neq i}} f_{a, \bv^{(-i)}, i} \bone[\sigma(\bv^{(-i)}) = a] - \left(y_{C} + \sum_{\substack{j \in \supp(\bv) \\ j \neq i}}f_{i, \bv^{(-j)}, j}\right)\bone[\sigma(\bv^{(-i)}) = i]\right) \\
        \EQ & y_C \left(1 - \sum_{i \in \supp(\bv) }\bone[\sigma(\bv^{(-i)}) = i]\right) + \sum_{i\neq a \in \supp(\bv)}  f_{a, \bv^{(-i)}, i} \left(\bone[\sigma(\bv^{(-i)}) = a] - \bone[\sigma(\bv^{(-a)}) = a]\right).
    \end{align*}
    Here in the last equality we reindexed the sum. The term $y_C \left(1 - \sum_{i \in \supp(\bv) }\bone[\sigma(\bv^{(-i)}) = i]\right)$ vanishes, since either $y_C = 0$, or $C = \{\be_a\}$ for some $a$ in which case $\sigma(\be_a) = a$ is forced and therefore $1 - \sum_{i \in \supp(\bv) }\bone[\sigma(\bv^{(-i)}) = i] = 0$. This proves \eqref{eq:y+z}.

    Now we are ready to prove  \eqref{eq:y_z_ineq}, with the help of \eqref{eq:y+z}. Fix some $i \in \supp(\bv)$, and let $S = \{a_1, \ldots, a_m\}$, $S_j$, $w_j$ be as defined in the edge weights definition for $\bv^{(-i)}$, $i$. The contribution from~$\bv^{(-i)}$ to the left hand side of \eqref{eq:y_z_ineq}, using the decomposition in \eqref{eq:y+z}, is equal to 
    \begin{align*}
        & \sum_{r = 1}^m f_{a_r, \bv^{(-i)}, i} \left(\bone[\sigma(\bv^{(-i)}) = a_r] - \bone[\sigma(\bv^{(-a_r)})= a_r]\right) \\
        \EQ & \sum_{r = 1}^m \sum_{j = r}^m w_j \left(\bone[\sigma(\bv^{(-i)}) = a_r] - \bone[\sigma(\bv^{(-a_r)}) = a_r]\right) \\
        \EQ &  \sum_{j = 1}^m \sum_{r = 1}^j w_j \left(\bone[\sigma(\bv^{(-i)}) = a_r] - \bone[\sigma(\bv^{(-a_r)}) = a_r]\right) 
    \end{align*}

    For each $j$, we examine the term $\sum_{r = 1}^j w_j (\bone[\sigma(\bv^{(-i)}) = a_r] - \bone[\sigma(\bv^{(-a_r)}) = a_r])$. The only case where this is positive, in which case it evaluates to $w_j$, is when $\sigma(\bv^{(-i)}) \in \{a_1, \ldots, a_j\}$ and $\sigma(\bv^{(-a_r)}) \neq a_r$ for every $r \in [j]$. In particular, this implies that for some $s \in [j]$ we have $\sigma(\bv^{(-i)}) = a_s \neq \sigma(\bv^{(-a_s)})$, and therefore $\sigma$ is not monochromatic on $S_j$. Since $S_j$ is a clique of $j+1$ vertices, it follows that $\sigma$ cuts at least $j$ edges in $S_j$ so
    \[
    \sum_{e = \{\bx, \by\} \subseteq S_j} \frac{w_j}{j}\, \bone[\sigma(\bx) \neq \sigma(\by)] \GE w_j.
    \]
    Summing over $j$ and $i$ proves \eqref{eq:y_z_ineq}.
\end{proof}

By Theorem~\ref{thm:flow_to_vertex_LP}, we have 
\[
\frac{2kn}{kn + q_{k, n}} \LE \LPvertex(k, n).
\]
We list the numerically computed values of $\frac{2kn}{kn + q_{k, n}}$ in Table~\ref{table:flow}, which serve as lower bound certificates for $\LPvertexno(k, n)$. Comparing it to Table~\ref{table:LP_vertex}, we can see that as $n$ increases, $\frac{2kn}{kn + q_{k, n}}$ is generally less than $\LPvertexno(k, n)$, and the loss seems to increase with $k$. That said, it is much cheaper to compute~$q_{k, n}$, and for $k = 4$, \eqref{eq:flow_lp} closely matches the computed $\LPvertexno(k, n)$ values. 

In addition to the values appearing in Table~\ref{table:flow}, we managed to obtain a feasible, though not optimal, solution of the flow LP for $k=7$ and $n=110$ that implies a lower bound of $\bestNum$ for the Multiway Cut problem. The solution was found using heuristic techniques and was verified using exact rational arithmetic. 

\begin{table}
\centering
\begin{tabular}{|c|c|c|c|c|c|}
    \hline
\backslashbox{$n$}{$k$} & 4 & 5 & 6 &7 & 8\\\hline
5 & 1.18733509 & 1.19047619 & 1.19128658 & 1.19128658 & 1.19128658 \\
10 & 1.20240480 & 1.20905902 & 1.21072958 & 1.21097816 & 1.21106744 \\
15 & 1.20799659 & 1.21488688 & 1.21696895 & 1.21736239 & 1.21747760 \\
20 & 1.21150860 & 1.21844218 & 1.22048504 & 1.22099302 & 1.22111950 \\
25 & 1.21386007 & 1.22053107 & 1.22246803 & 1.22308906 & 1.22322663 \\
30 & 1.21531379 & 1.22200902 & 1.22393904 & 1.22453399 & 1.22468883 \\
40 & 1.21704683 & 1.22394570 & 1.22595375 & 1.22653797 & 1.22668856 \\
50 & 1.21807631 & 1.22502167 & 1.22708443 & 1.22769517 & 1.22784877 \\
60 & 1.21883264 & 1.22573754 & 1.22781765 & 1.22845312 & 1.22860968\\
70 & 1.21931441 & 1.22626371 & 1.22834008 & -- & --\\
80 & 1.21968554 & 1.22667240 & 1.22876922 & -- &--\\
90 & 1.21998639 & 1.22697379 & 1.22908765 & -- &--\\
100 & 1.22020850 & 1.22720845 & -- & -- &--\\
\hline
\end{tabular}
\caption{Numerically computed values of $\frac{2kn}{kn + q_{k, n}}$, rounded to eight decimal places.  A dash (--) indicates that the computation could not be completed using only 128GB of memory.}
\label{table:flow}
\end{table}

\subsection{A continuous formulation for the flow LP}

In this section, we present a continuous formulation for the flow LP. Fix some $k \geq 2$ and let $\Delta_k = \{\bt \in \mathbb{R}^{k}_{\geq 0} \mid \sum_{i = 1}^kt_i = 1\}$, $\Gamma_a = \{\bt \in \Delta_k \mid t_a = 0\}$ for $a \in [k]$. The variables in our continuous formulation are:
\begin{itemize}
    \item Nonnegative measures $F_{a, i}, B_i$ on $\Delta_k$ for every $a \neq i \in [k]$. These are intended to be the continuous version of $f_{a, \bx, i}, b_{\bx, i}$ variables in \eqref{eq:flow_lp}.
    \item Probability measures $\nu_a$ on $\Gamma_a$ for every $a \in [k]$. This is intended to be the exit measure for the commodity-$a$ flow on $\Gamma_a$. These will in fact be determined by $F_{a, i}$ and the LP constraints but we add them to make the presentation simpler.
\end{itemize}
The continuous program is formulated as follows.
\begin{equation}
\begin{aligned}
 \inf \quad & Q_{k} \coloneqq \sum_{i=1}^{k} B_i(\Delta_k) \\
\text{s.t.}\quad & 0 \LE F_{a,i} \LE B_{i} && \qquad a\ne i,\\
& \sum_{i\ne a} \int_{\Delta_k} \nabla \varphi \cdot (\be_a - \be_i) \,dF_{a, i} \EQ \varphi(\be_a) - \int_{\Gamma_a}\varphi \,d\nu_a && \qquad a \in [k]\;, \;\varphi \in C^{(1)}(\Delta_k).
\end{aligned}
\label{eq:continuous_lp}\tag{Continuous-Flow-LP}
\end{equation}

Here $F_{a,i}, B_i$ are nonnegative Borel measures, and $C^{(1)}(\Delta_k)$ denotes the set of functions that are continuously differentiable on $\Delta_k$. 

The main difference is in the flow conservation constraint. In the discrete LP the constraint is formulated in terms of incoming and outgoing flow at the same vertex. Equivalently, one may multiply the conservation equations by an arbitrary function $\varphi$ on the vertices and sum over all vertices. After rearranging, each flow arc contributes the difference of $\varphi$ at its two endpoints. Summing the discrete conservation equations against $\varphi$, we obtain that the contributions at intermediate vertices cancel, leaving only the source and sink terms. Here in the continuous version we use the continuous analogue of this, where the flow arc difference is replaced by the integral of the directional derivative.

\begin{lemma}
    $Q_k$ is well-defined.
\end{lemma}
\begin{proof}
    \eqref{eq:continuous_lp} has a feasible solution. For example, we may take $F_{a, i}$ to be $1/(k-1)$ times the uniform probability measure over $\{\lambda \cdot \be_a + (1 - \lambda) \cdot \be_i \mid \lambda \in [0, 1]\}$, and $B_i = \sum_{a \neq i} F_{a, i}$. Then this is a feasible solution with $\sum_{i = 1}^k B_i(\Delta_k) = k$. On the other hand, for any feasible solution we always have $\sum_{i = 1}^k B_i(\Delta_k) \geq 0$. Thus, the infimum $Q_{k} = \inf_{B_i}\sum_{i=1}^{k} B_i(\Delta_k)$ is well-defined.
\end{proof}

The following theorem shows that the continuous version indeed captures the discrete limit.
\begin{theorem}\label{thm:flow_lp_limit}
    For every $k \geq 2$ and $n \geq 1$, we have 
    \[
    \frac{q_{k, n}}{n+1} \LE Q_k \LE \frac{q_{k, n}}{n}.
    \]
    Consequently, for every $k \geq 2$ we have
    \[
    \lim_{n \to \infty} \frac{q_{k,n}}n \EQ Q_k.
    \]
\end{theorem}
\begin{proof}
We first show that $\frac{q_{k, n}}{n+1} \leq Q_k$. To prove this part we need to ``round'' a continuous solution to a discrete solution, and we do so using the following polynomial: for any $m \in \mathbb{Z}_{\geq 0}$, $\by \in (\mathbb{Z}_{\geq 0})^k$ with $\sum_{i=1}^k y_i = m$, and $\bt \in \Delta_k$, we define
    \[
        P^{m}_{\by}(\bt) \EQ \frac{m!}{y_1!y_2! \ldots y_k!} t_1^{y_1}t_2^{y_2} \cdots t_k^{y_k}.
    \]
This is the multinomial probability of getting the outcome $\by$ in $m$ trials with the probability vector $\bt$. It is straightforward to verify that 
\[
    \frac{\partial}{\partial t_i}P^{m}_{\by}(\bt) \EQ m \, P^{m-1}_{\by - \be_i}(\bt).
\]
For any feasible solution $F_{a, i}, B_i$, to \eqref{eq:continuous_lp}, we discretize it via
    \[
        f^{(n)}_{a, \bx, i} \EQ (n+1) \int_{\Delta_k} P^{n}_{n\bx}(\bt)\, dF_{a, i}(\bt), \quad b^{(n)}_{\bx, i} \EQ (n+1) \int_{\Delta_k} P^{n}_{n\bx}(\bt)\, dB_{i}(\bt).
    \]
Then the constraint $f^{(n)}_{a, \bx, i} \leq b^{(n)}_{\bx, i}$ follows directly since the measure $B_{i}$ dominates $F_{a, i}$. To prove discrete flow conservation for commodity $a$ at $\bx$, we take $\varphi(\bt) = P^{n+1}_{n\bx + \be_a}(\bt)$ in the continuous flow conservation equation, which gives
\begin{align*}
\varphi(\be_a) - \int_{\Gamma_a} \varphi \,d\nu_a & \EQ \sum_{i \neq a} \int \nabla \varphi \cdot (\be_a - \be_i) \,dF_{a, i}(t) \\ 
& \EQ \sum_{i \neq a} \int (n+1) \left(P^{n}_{n\bx} - P^{n}_{n\bx + \be_a - \be_i}\right)\,dF_{a, i}(t) \\ 
& \EQ \sum_{i \neq a} \left(f_{a, \bx, i}^{(n)} - f_{a, \bx + (\be_a - \be_i)/n, i}^{(n)}\right)
\end{align*}
Here we interpret the terms $f_{a, \bx + (\be_a - \be_i)/n, i}^{(n)}$ to be 0 if $i \notin \supp(\bx)$, which is consistent with the derivative. For the left hand side above, we have $\int_{\Gamma_a} \varphi \,d\nu_a = 0$ since $\varphi(\bt) = P^{n+1}_{n\bx + \be_a}(\bt)$ vanishes on $\Gamma_a$, and $\varphi(\be_a) = \bone[\bx = \be_a]$, so we obtain exactly the discrete flow conservation equation. Finally, we have
\begin{align*}
    \sum_{i \in [k]} \sum_{\bx \in \Delta_{k, n}}b_{\bx, i}^{(n)} & \EQ \sum_{i \in [k]} \sum_{\bx \in \Delta_{k, n}}(n+1) \int_{\Delta_k} P^{n}_{n\bx}(\bt)\, dB_{i}(\bt) \\
    & \EQ \sum_{i \in [k]} (n+1) \int_{\Delta_k} 1\, dB_{i}(\bt) \\
    & \EQ (n+1)\sum_{i \in [k]} B_i(\Delta_k).
\end{align*}
In the second equality we used the fact that $\sum_{\bx \in \Delta_{k, n}}P^{n}_{n\bx}(\bt) = 1$. By choosing the solution $F_{a, i}, B_i$ optimally, we obtain
\[
q_{k, n} \LE \sum_{i \in [k]} \sum_{\bx \in \Delta_{k, n}}b_{\bx, i}^{(n)} \EQ (n+1)\sum_{i \in [k]} B_i(\Delta_k) \EQ (n+1)Q_k.
\]
Next, we show $Q_k \leq \frac{q_{k, n}}{n}$. For this part, we need to construct continuous solutions from discrete solutions. We use the Dirichlet distribution $\dir_{\bz}$ over $\Delta_k$ where $\bz \in (\mathbb{R}^{\geq 0})^k \setminus \{(0, \ldots, 0)\}$ is some parameter vector. When $\bz$ consists of strictly positive entries, $\dir_{\bz}$ can be defined using the probability density function
\[
g(\bt; \bz) = \frac{\Gamma(\sum_{i = 1}^k z_i)}{\prod_{i=1}^k\Gamma(z_i)} \prod_{i = 1}^k t_i^{z_i - 1},
\]
where $\Gamma$ is the Gamma function. When $\bz$ has some (but not all) zero entries, it can be defined by setting the corresponding coordinates to zero and using the Dirichlet distribution with the nonzero parameters on the resulting simplex face. See e.g.~\cite{ferguson1973bayesian} for more details. Using the identity $\Gamma(z + 1) = z \Gamma(z)$ for any $z > 0$, for any $\bz$ with $z_i > 1$ the partial derivative of $g$ with respect to $t_i$ can be expressed as:
\[
\frac{\partial}{\partial t_i}g(\bt; \bz) \EQ \left(\sum_{i = 1}^k z_i - 1\right)g(\bt; \bz - \be_i).
\]
For any $\varphi \in C^{(1)}(\Delta_k)$, $a \neq i \in [k]$, and $\bz \in (\mathbb{R}^+)^k$ with $z_a > 1$, we have the following identity:
\begin{equation}\label{eq:dirichlet_flow_partial}
    \int_{\Delta_k} \nabla \varphi \cdot (\be_a - \be_i) g(\bt; \bz + \be_i) \,d\bt \EQ \left(\sum_{i = 1}^k z_i\right) \cdot \left(\int_{\Delta_k} \varphi\, g(\bt; \bz) -  \varphi\, g(\bt; \bz + \be_i - \be_a) \, d\bt\right)
\end{equation}
Indeed, by fixing all coordinates other than $t_i$ and $t_a$, we can restrict $\varphi$ to the one-dimensional segment $t_i + t_a = s$, for some $s \in [0, 1]$. Parametrizing this segment by $t_a = r$ and $t_i = s - r$, we have $\frac{\partial\varphi}{\partial r} = \nabla \varphi \cdot (\be_a - \be_i)$, and using integration by parts we have
\begin{align*}
\int_0^s \frac{\partial\varphi}{\partial r} g(\bt; \bz + \be_i) \,dr & \EQ \varphi \ g(\bt; \bz + \be_i)\bigg|_0^s - \left(\sum_{i = 1}^k z_i\right)\left(\int_0^s \varphi\, g(\bt; \bz + \be_i - \be_a) -  \varphi\, g(\bt; \bz)\,dr \right) \\
& \EQ \left(\sum_{i = 1}^k z_i\right)\left(\int_0^s \varphi\, g(\bt; \bz) -  \varphi\, g(\bt; \bz + \be_i - \be_a)\,dr \right)
\end{align*}
The term $\varphi \ g(\bt; \bz + \be_i)\big|_0^s$ vanishes due to the $t_i^{z_i}t_a^{z_a - 1}$ factor in $g$ (recall $z_a > 1$). Integrating over all other coordinates proves \eqref{eq:dirichlet_flow_partial}. Taking a limit in \eqref{eq:dirichlet_flow_partial} then proves, for any $\bz \in (\mathbb{R}^{\geq 0})^k$ with $z_a \geq 1$, that 
\begin{equation}\label{eq:dirichlet_flow_partial_limit}
\int_{\Delta_k} \nabla \varphi \cdot (\be_a - \be_i) \,d\,\dir_{\bz + \be_i}(\bt) \EQ \left(\sum_{i = 1}^k z_i\right) \cdot \left(\int_{\Delta_k} \varphi\, d\,\dir_{\bz}(\bt) - \int_{\Delta_k}  \varphi\, d\,\dir_{\bz + \be_i - \be_a}(\bt)\right).
\end{equation}
Given some feasible discrete solution $f_{a, \bx, i}, b_{\bx, i}$, \eqref{eq:dirichlet_flow_partial_limit} motivates the definition
\begin{equation}\label{eq:cont_def}
F_{a, i} \EQ \frac{1}{n} \sum_{\substack{\bx \in \Delta_{k, n} \\ x_a > 0}}f_{a, \bx, i}\, \dir_{n\bx + \be_i}, \quad B_{i} \EQ \frac{1}{n} \sum_{\bx \in \Delta_{k, n}}b_{\bx, i}\, \dir_{n\bx + \be_i}.
\end{equation}
With this construction we have the capacity constraints $F_{a, i} \leq B_i$ immediately. For flow conservation, take any $\varphi \in C^{(1)}(\Delta_k)$ and $a \neq i \in [k]$, using \eqref{eq:dirichlet_flow_partial_limit} we have
\begin{align*}
    \sum_{i \neq a}\int_{\Delta_k} \nabla \varphi \cdot (\be_a - \be_i) \, dF_{a, i} & \EQ \frac{1}{n} \sum_{i \neq a}\int_{\Delta_k} \left(\nabla \varphi \cdot (\be_a - \be_i)\right) \, \sum_{\substack{\bx \in \Delta_{k, n} \\ x_a > 0}}f_{a, \bx, i}\, d\,\dir_{n\bx + \be_i}(\bt) \\
    & \EQ \sum_{i \neq a} \sum_{\substack{\bx \in \Delta_{k, n} \\ x_a > 0}}f_{a, \bx, i}\left(\int_{\Delta_k} \varphi\, d\,\dir_{n\bx}(\bt) -  \int_{\Delta_k}\varphi\, d\,\dir_{n\bx + \be_i - \be_a}(\bt)\right).
\end{align*}
Regrouping and canceling using the discrete flow conservation equation, we obtain
\begin{align*}
& \sum_{i \neq a}\int_{\Delta_k} \nabla \varphi \cdot (\be_a - \be_i) \, dF_{a, i} \\
\EQ & \sum_{i \neq a} f_{a, \be_a, i}\int_{\Delta_k} \varphi\, d\,\dir_{n\be_a}(\bt) - \sum_{i \neq a}\sum_{\substack{\bx \in \Delta_{k, n} \\ x_a = 0, x_i > 0}}f_{a, \bx + (\be_a - \be_i)/n, i}\int_{\Delta_k} \varphi\, d\,\dir_{n\bx}(\bt) \\
\EQ & \varphi(\be_a) - \sum_{i \neq a}\sum_{\substack{\bx \in \Delta_{k, n} \\ x_a = 0, x_i > 0}}f_{a, \bx + (\be_a - \be_i)/n, i}\int_{\Delta_k} \varphi\, d\,\dir_{n\bx}(\bt).
\end{align*}
So we satisfy the continuous flow conservation by setting 
\[
\nu_a \EQ \sum_{i \neq a}\sum_{\substack{\bx \in \Delta_{k, n} \\ x_a = 0, x_i > 0}}f_{a, \bx + (\be_a - \be_i)/n, i} \,\dir_{n\bx}
\]
which is a probability measure on $\Gamma_a$ since $\sum_{i \neq a}\sum_{\substack{\bx \in \Delta_{k, n},x_a = 0, x_i > 0}}f_{a, \bx + (\be_a - \be_i)/n, i} = 1$. This proves that \eqref{eq:cont_def} defines a feasible solution to \eqref{eq:continuous_lp}. By choosing the discrete solution optimally, we obtain that 
\begin{align*}
    Q_k \LE \sum_{i = 1}^kB_i(\Delta_k) \EQ \frac{1}{n}\sum_{i = 1}^k \sum_{\bx \in \Delta_{k, n}} b_{\bx, i} \EQ \frac{q_{k,n}}{n}.
\end{align*}
This completes the proof of the theorem.
\end{proof}

\section{A feasible solution to the continuous Flow LP when \texorpdfstring{$k = 4$}{k = 4}}\label{sec:continuous_feasible_solution}

In this section, we present an explicit feasible solution to~\eqref{eq:continuous_lp} when $k = 4$. 

\begin{theorem}\label{thm:cont_feasible_k4}
    There exists a feasible solution $(F_{a, i}, B_i)$ to~\eqref{eq:continuous_lp} when $k = 4$ such that $\sum_{i = 1}^4 B_i(\Delta_4) \leq 2.54492029$.
\end{theorem}

Our main theorem immediately follows as a corollary:
\begin{proof}[Proof of Theorem~\ref{thm:main}]
Since 
\[
\frac{8}{4 + 2.54492029} \GT 1.22232,
\]
the theorem follows from Lemma~\ref{lem:vertex_LP_hardness}, Theorem~\ref{thm:flow_to_vertex_LP}, and Theorem~\ref{thm:flow_lp_limit}.
\end{proof}

The rest of the section is devoted to constructing the measures $B_i, F_{a,i}$ on the simplex $\Delta_4$, and verifying that they satisfy the constraints in~\eqref{eq:continuous_lp}. We construct these measures in a piecewise manner, by partitioning $\Delta_4$ into the following pieces. Let $\alpha, \beta > 0$ be some parameters to be chosen later, we have:

\begin{itemize}
    \item For every terminal $a$, \emph{the corner} $K_a = \{\bt\in\Delta_4 \mid t_a\ge \alpha\}$.
    \item For every pair of terminals $a, b$, \emph{the tubes} \[T_{ab} \EQ \left\{\bt\in\Delta_4 \mid \max_{\ell \in [4]} t_\ell\le \alpha \text{ and } t_i+t_j\le\beta\;,\; \{i, j\} = [4] \setminus \{a, b\}\right\}.\]
    \item \emph{The central region} 
    \[Z \EQ \{\bt\in\Delta_4 \mid \forall\ell\in[4], t_\ell\le \alpha \text{ and }   \forall i\neq j \in [4]\;,\; t_i+t_j\ge\beta\}.\]
\end{itemize}
We also use 
\[\by_{ab}(w) \EQ w \be_a + (1 - w) \be_b, \quad 0 \leq w \leq 1\] 
to parametrize the \emph{simplex edge} $E_{ab} = \{\by_{ab}(w) \mid 0 \leq w \leq 1\}$ connecting terminals $a$ and $b$.
See Figure~\ref{fig:partition} for a picture of our partition. The parameters $\alpha, \beta$ will be chosen so that these regions have disjoint interiors; in particular, they satisfy:
\begin{equation}
    \frac{1}{2} < \alpha < 1, \quad 0 < \beta < \frac{1 - \alpha}{2}.
\end{equation}

Informally, the flows in our construction travel as follows. Take the commodity-$a$ flow (other commodities are defined in a symmetric way):
\begin{itemize}
    \item It originates at $\be_a$, and then travels alone within the corner $K_a$. No flow from the other terminals is present in this corner, so $F_{b, i} = 0$ for every $b \neq a$.
    \item It reaches the cross section $t_a = \alpha$ in three different types of pieces: 
    \begin{itemize}
        \item The simplex-edge pieces. These are atomic measures supported on $\by_{ab}(\alpha)$ for each $b \in [4] \setminus \{a\}$. 
        \item The tube pieces. These measures are supported on the boundary between the corner $K_a$ and the three tubes $T_{ab}$, one for each $b \in [4] \setminus \{a\}$.
        \item The central piece. This measure is supported on the boundary between the corner $K_a$ and the central region $Z$.
    \end{itemize}
    \item The three pieces continue in the following way:
    \begin{itemize}
        \item Each simplex-edge piece travels on the simplex edge, shedding into the corresponding tube $T_{ab}$ at a constant rate, until it reaches $\by_{ab}(1 - \alpha)$ where it exhausts.
        \item The tube flows continue in the tubes, along with the flows shed from the simplex-edge flows. The (commodity-$a$) flow in $T_{ab}$ travels only towards directions $i, j \in [4] \setminus \{a, b\}$, together with the commodity-$b$ flow. That is, $F_{a, b} = F_{b, a} = 0$ in $T_{ab}$. These flows eventually enter the central region $Z$ through the boundary between $T_{ab}$ and $Z$.
        \item The flow in the central region stays in the central region, where it shares the capacity fully with all other commodities. That is, $F_{a, i} = F_{b, i}$ for every $b \in [4] \setminus \{i\}$.
    \end{itemize}
    \item Eventually, all commodity-$a$ flow enters the central region $Z$ and in the end it exits through $Z \cap \Gamma_a$.
\end{itemize}

We now define these flows formally. We state the definition for an arbitrary terminal $a$, as all other flows are defined in a symmetric way. For each region $R \in \{K_a \mid a \in [4]\} \cup \{T_{ab} \mid a \neq b \in [4]\} \cup \{E_{ab} \mid a \neq b \in [4]\} \cup \{Z\}$, we define the flow measures $\{F_{a, i}^R \mid i \neq a\}$ supported on the region, and for every $\varphi \in C^{(1)}(\Delta_4)$, we will compute 
\[
I_a^R(\varphi) \EQ \sum_{i \neq a} \int_{\Delta_4} \nabla \varphi \cdot (\be_a - \be_i) dF^R_{a, i}
\]
which corresponds to the left hand side of the continuous flow conservation equation. These quantities will then be assembled together in the feasibility proof. Throughout the construction, we will use an additional normalization parameter $C > 0$ whose value is to be chosen later. For concreteness, we use $d\bt = dt_1\,dt_2\,dt_3$ with $t_4 = 1 - t_1 - t_2 - t_3$ to compute the integral over $\Delta_4$. When it is more convenient we may also use any other three distinct coordinates, which gives the same measure. Finally, we use the following standard form of the divergence theorem.

\begin{theorem}[See, e.g.,~\cite{Apostol1969}]\label{thm:divergence}
    Let $R \subset \mathbb{R}^3$ be a closed and bounded polyhedron with nonempty interior, and let $\bg = (g_1, g_2, g_3)$ be a continuously differentiable vector field on an open neighborhood of $R$. Then we have
    \[
    \int_R \Div \bg\, d\bx \EQ \sum_{F: \text{ facet of }R} \int_F \bn_F \cdot \bg\,dS.
    \]
    Here $\Div \bg = \sum_{i \in [3]} \frac{\partial g_i}{\partial x_i}$ is the \emph{divergence} of $\bg$, $\bn_F$ is the outward unit normal vector for the facet~$F$, $d\bx$ is the volume element, and $dS$ is the surface area element.
\end{theorem}

\subsection{The corner flows}
We first define the desired measure on the cross section $\{\bt \in \Delta_4 \mid t_a = \alpha\}$. As we discussed earlier, we define the following pieces:
\begin{align}
    d \mu_a^Z & \EQ C\, dt_i dt_j && \text{ on } K_a \cap Z, \quad i \neq j \in [4] \setminus \{a\}, \label{eq:mu_Z_def}
\end{align}
and for every $b \in [4] \setminus \{a\}$,
\begin{align}
    d \mu_{ab}^T & \EQ  \frac{C\beta}{t_i + t_j}\, dt_i dt_j && \text{ on } K_a \cap T_{ab}, \quad i \neq j \in [4] \setminus \{a, b\}, \label{eq:mu_T_ab_def} \\
    \mu_{ab}^E & \EQ  C\beta(2\alpha - 1) \delta_{\by_{ab}(\alpha)}. \label{eq:mu_E_ab_def}
\end{align}
Here $\delta_{\bx}$ denotes the unit point mass at $\bx$. These measures are chosen to match the incoming flows required by the central region, the tubes, and the simplex edges, respectively. In words, the measure on $K_a \cap Z$ is uniform, while the measure on $K_a \cap T_{ab}$ is inversely proportional to the distance from the simplex edge. 

To achieve these desired measures, we simply send the flow from the terminal $\be_a$ via straight paths. More precisely, for every $\bx$ such that $x_a = \alpha$, define the path
\[
    \gamma_{a, \bx}(\tau) \EQ \left(1 - \frac{\tau}{1 - \alpha}\right)\be_a + \frac{\tau}{1 - \alpha} \bx, \quad 0 \leq \tau \leq 1 - \alpha.
\]
For every $i \neq a$, we define the corner flow measures $F_{a, i}^{K_a}$ via:
\[
    \int \varphi dF_{a, i}^{K_a} \EQ \int x_i\left( \frac{1}{1-\alpha}\cdot\int_0^{1 - \alpha} \varphi(\gamma_{a, \bx}(\tau)) d\tau\right) d\mu_a(\bx),
\]
where 
\begin{equation}\label{eq:mu_a}
    \mu_a \EQ \mu_a^Z + \sum_{b \neq a}(\mu_{ab}^T + \mu_{ab}^E).
\end{equation} 
Also, set $B_i^{K_a} = F_{a, i}^{K_a}$ for every $i \neq a$, and all other measures are set to be 0 in this corner.

\begin{lemma}\label{lem:corner_conserv}
    For every $a \in [4]$ and $\varphi \in C^{(1)}(\Delta_4)$, we have
    \[
        I_a^{K_a}(\varphi) \EQ \mu_a(\Delta_4)\varphi(\be_a) - \int \varphi \, d\mu_a,
    \]
    and $I_b^{K_a}(\varphi) = 0$ for any $b \neq a$.
\end{lemma}
\begin{proof}
    For every $\bx \in \supp(\mu_a)$, we have
    \[
        \gamma'_{a, \bx}(\tau) \EQ -\frac{1}{1 - \alpha}\be_a + \frac{1}{1 - \alpha}\bx \EQ \sum_{i \neq a} \frac{x_i}{1 - \alpha} \,(\be_i - \be_a).
    \]
    It then follows from the chain rule that
    \[
        \frac{d}{d\tau}\varphi(\gamma_{a, \bx}(\tau)) \EQ \nabla \varphi \cdot \gamma'_{a, \bx}(\tau) \EQ \nabla \varphi \cdot \left(\sum_{i \neq a} \frac{x_i}{1 - \alpha} \,(\be_i - \be_a)\right).
    \]
    By the definition of $I_a^{K_a}(\varphi)$, we have 
    \begin{align*}
        I_a^{K_a}(\varphi) & \EQ \sum_{i \neq a} \int_{\Delta_4} \nabla \varphi \cdot (\be_a - \be_i) \, dF_{a, i}^{K_a} \\
        & \EQ \sum_{i \neq a} \int x_i\left( \frac{1}{1-\alpha}\cdot\int_0^{1 - \alpha} \nabla\varphi(\gamma_{a, \bx}(\tau)) \cdot (\be_a - \be_i)\,d\tau\right) d\mu_a(\bx) \\
        & \EQ -\int \int_0^{1 - \alpha} \frac{d}{d\tau}\varphi(\gamma_{a, \bx}(\tau)) \, d\tau \,d\mu_a(\bx) \\
        & \EQ \int (\varphi(\be_a) - \varphi(\bx))\,d\mu_a(\bx) \\
        & \EQ \mu_a(\Delta_4)\varphi(\be_a) - \int \varphi \, d\mu_a.
    \end{align*}
    Finally, we have $I_b^{K_a}(\varphi) = 0$ for any $b \neq a$ since $F_{b, i}^{K_a} = 0$.
\end{proof}

\subsection{The simplex-edge flows}
On the simplex edge $E_{ab}$, we have the simplex-edge flow $F_{a, b}^{E_{ab}}$ supported on $\{\by_{ab}(w) \mid 1-\alpha \leq w \leq \alpha\}$, defined by
\[
dB_b^{E_{ab}} \EQ dF_{a, b}^{E_{ab}} \EQ C \beta (w - 1 + \alpha)\, dw.
\]
The term $(w - 1 + \alpha)$ corresponds to the constant shedding rate of the simplex-edge flow into the tube. 

\begin{lemma}\label{lem:simplex_edge_conserv}
    For every distinct $a, b \in [4]$ and $\varphi \in C^{(1)}(\Delta_4)$, we have
    \[
        I_a^{E_{ab}}(\varphi) \EQ \int \varphi \, d\mu_{ab}^E - C \beta \int_{1-\alpha}^\alpha \varphi(\by_{ab}(w))\, dw,
    \]
    where $\mu_{ab}^E$ is defined as in Equation~\eqref{eq:mu_E_ab_def}. Furthermore, $I_i^{E_{ab}}(\varphi) = 0$ for any $i \in [4] \setminus \{a, b\}$.
\end{lemma}
\begin{proof}
    By the chain rule, we have 
    \[
        \frac{d}{dw} \varphi(\by_{ab}(w)) \EQ \nabla \varphi \cdot (\be_a - \be_b).
    \]
    It follows that 
    \begin{align*}
        I_a^{E_{ab}}(\varphi) & \EQ \int \nabla \varphi \cdot (\be_a - \be_b) dF_{a,b}^{E_{ab}} \\
        & \EQ C \beta \int_{1-\alpha}^\alpha  \frac{d}{dw} \varphi(\by_{ab}(w)) (w - 1 + \alpha)\, dw \\
        & \EQ C \beta \left( \varphi(\by_{ab}(w))(w - 1 + \alpha)\Big|_{1-\alpha}^\alpha - \int_{1-\alpha}^\alpha \varphi(\by_{ab}(w))\, dw \right) \\
        & \EQ C \beta (2\alpha - 1) \varphi(\by_{ab}(\alpha)) - C\beta \int_{1-\alpha}^\alpha \varphi(\by_{ab}(w))\, dw .
    \end{align*}
    The lemma follows since $\mu_{ab}^E = C\beta(2\alpha - 1) \delta_{\by_{ab}(\alpha)}$.
\end{proof}

\subsection{The tube flows}
In the tube $T_{ab}$, let $\{i, j\} = [4] \setminus \{a, b\}$. We define 
\begin{align*}
dB_i & \EQ dF_{a, i}^{T_{ab}} \EQ dF_{b, i}^{T_{ab}} \EQ C \beta \frac{t_i}{(t_i + t_j)^2}\, d\bt, \\
dB_j & \EQ dF_{a, j}^{T_{ab}} \EQ dF_{b, j}^{T_{ab}} \EQ C \beta \frac{t_j}{(t_i + t_j)^2}\, d\bt.
\end{align*}

All other measures are set to 0. To analyze this tube region (and later the central region), we define the measure $\eta_{ab}$ representing the flow going from $T_{ab}$ to $Z$ as follows. To make the calculations precise, let us use the coordinates 
\begin{equation}\label{eq:tube_coords}
    \bt_{ab}(w, u, v) \EQ w \be_a + (1 - w - u - v) \be_b + u \be_i + v \be_j,
\end{equation}
so $d\bt = dw\,du\,dv$. Define $\eta_{ab}$ by 
\begin{equation}\label{eq:eta_ab_def}
    \int \varphi \,d\eta_{ab} \EQ C \int_{1 - \alpha - \beta}^\alpha \int_0^\beta \varphi(\bt_{ab}(w, u, \beta - u)) \,du\,dw
\end{equation}
for continuous $\varphi$. 

\begin{lemma}\label{lem:tube_conserv}
    For every distinct $a, b \in [4]$ and $\varphi \in C^{(1)}(\Delta_4)$, we have
    \[
        I_a^{T_{ab}}(\varphi) = \int \varphi \, d\mu_{ab}^T + C \beta \int_{1-\alpha}^\alpha \varphi(\by_{ab}(w))\, dw - \int \varphi \, d \eta_{ab},
    \]
    where $\mu_{ab}^T$ is defined as in Equation~\eqref{eq:mu_T_ab_def}. Furthermore, $I_i^{T_{ab}}(\varphi) = 0$ for any $i \in [4] \setminus \{a, b\}$.
\end{lemma}
\begin{proof}
    Using the coordinates in~\eqref{eq:tube_coords}, by the chain rule we have
    \begin{align*}
        \frac{\partial}{\partial w} \varphi(\bt(w, u, v)) & \EQ \nabla \varphi \cdot (\be_a - \be_b), \\
        \frac{\partial}{\partial u} \varphi(\bt(w, u, v)) & \EQ \nabla \varphi \cdot (\be_i - \be_b), \\
        \frac{\partial}{\partial v} \varphi(\bt(w, u, v)) & \EQ \nabla \varphi \cdot (\be_j - \be_b).
    \end{align*}
    Define $\bg = (g_w, g_u, g_v) = \left(\frac{1}{u+v}, -\frac{u}{(u+v)^2}, -\frac{v}{(u+v)^2}\right)$, then we have 
    \[
        \Div \bg \EQ -\frac{1}{(u+v)^2} + \frac{2u}{(u+v)^3} - \frac{1}{(u+v)^2} + \frac{2v}{(u+v)^3} \EQ 0.
    \]
    We then have 
    \begin{align*}
        I_a^{T_{ab}}(\varphi) & \EQ C \beta \int_0^\beta \int_0^{\beta - v} \int_{1 - \alpha - u - v}^\alpha \frac{u (\nabla \varphi \cdot (\be_a - \be_i)) + v (\nabla \varphi \cdot (\be_a - \be_j))}{(u + v)^2}\, dw\,du\,dv,\\
        & \EQ C \beta \int_0^\beta \int_0^{\beta - v} \int_{1 - \alpha - u - v}^\alpha \frac{(u+v) \frac{\partial\varphi(\bt)}{\partial w} - u \frac{\partial\varphi(\bt)}{\partial u} - v \frac{\partial\varphi(\bt)}{\partial v}}{(u + v)^2}\, dw\,du\,dv \\
        & \EQ C \beta \int_0^\beta \int_0^{\beta - v} \int_{1 - \alpha - u - v}^\alpha \Div (\varphi\bg)\, dw\,du\,dv.
    \end{align*}
    We would like to apply Theorem~\ref{thm:divergence}, but $\varphi\bg$ is not defined when $u = v = 0$. To deal with this, for $\delta \in (0, \beta)$ we consider the truncated tube 
    \[
        T_{ab}^\delta \EQ \{\bt_{ab}(w, u, v) \mid u, v \geq 0, \delta \leq u + v \leq \beta, 1 - \alpha - u - v \leq w \leq \alpha\},
    \]
    and let $I_\delta$ be the contribution to $I_a^{T_{ab}}(\varphi)$ from $T_{ab}^\delta$, i.e.,
    \[
        I_\delta \EQ C\beta \int_{T_{ab}^\delta} \Div(\varphi \bg) \,dw\,du\,dv.
    \]
    Now we can apply Theorem~\ref{thm:divergence} to $I_\delta$ and obtain
    \[
        I_\delta \EQ C\beta \sum_{F: \text{facet of } T_{ab}^\delta} \int_{F} (\bn_F \cdot \bg) \varphi \,dS.
    \]
    $T_{ab}^\delta$ has six facets. On the facet with $u = 0$, we have $\bn_F = (0, -1, 0)$ and $g_u = 0$, so $\bn_F \cdot \bg = 0$. This similarly holds on the facet with $v = 0$. On the facet with $w = 1 - \alpha - u - v$, we have $\bn_F = -(1, 1, 1)/\sqrt{3}$, so again $\bn_F \cdot \bg = -(g_w + g_u + g_v)/\sqrt{3} = 0$. The nonzero contributions come from the remaining three facets. On the facet $u + v = \beta$, we have $\bn_F = (0, 1/\sqrt{2}, 1/\sqrt{2})$, and the parametrization $(w,u)\mapsto(w,u,\beta-u)$ has area element $dS=\sqrt{2}\,dw\,du$, so the factor $\sqrt{2}$ cancels and we have
    \begin{align*}
    C\beta \int_{F} (\bn_F \cdot \bg) \varphi \,dS &\EQ C\beta\int_{0}^\beta\int_{1-\alpha-\beta}^\alpha -\frac{u+v}{(u+v)^2}\varphi(\bt_{ab}(w, u, \beta - u)) \,dw\,du \\
    &\EQ -C\int_{0}^\beta\int_{1-\alpha-\beta}^\alpha \varphi(\bt_{ab}(w, u, \beta - u)) \,dw\,du \\
    &\EQ -\int \varphi \, d \eta_{ab}.
    \end{align*}
    A similar computation shows that on the facet $u + v = \delta$, we have
    \[
    C\beta \int_{F} (\bn_F \cdot \bg) \varphi \,dS \EQ \frac{C\beta}{\delta}\int_{0}^\delta\int_{1-\alpha-\delta}^\alpha \varphi(\bt_{ab}(w, u, \delta - u)) \,dw\,du.
    \]
    Finally, on the facet $w = \alpha$ we have $\bn_F = (1, 0, 0)$ and $dS = du\,dv$ so
    \[
    C\beta \int_{F} (\bn_F \cdot \bg) \varphi \,dS \EQ C \beta \int_{\substack{u, v \geq 0 \\ \delta \leq u + v \leq \beta}} \frac{\varphi(\bt_{ab}(\alpha,u,v))}{u+v} \,du\,dv.
    \]
    Combining, we obtain
    \[
    I_\delta \EQ C \beta \int_{\substack{u, v \geq 0 \\ \delta \leq u + v \leq \beta}} \frac{\varphi(\bt_{ab}(\alpha,u,v))}{u+v} \,du\,dv  +\frac{C\beta}{\delta}\int_{0}^\delta\int_{1-\alpha-\delta}^\alpha \varphi(\bt_{ab}(w, u, \delta - u)) \,dw\,du - \int \varphi \, d \eta_{ab}.
    \]
    We now take $\delta \to 0$. The first term on the right hand side converges to $ \int \varphi \, d\mu_{ab}^T$, while the second term converges to $C \beta \int_{1-\alpha}^\alpha \varphi(\by_{ab}(w))\, dw$. On the left hand side, $I_\delta$ converges to $I_a^{T_{ab}}(\varphi)$, so the lemma follows.
\end{proof}

\subsection{The central region}
The central region will be further divided into sub-regions, and we define our flow on these sub-regions so that it has constant density on each sub-region. 

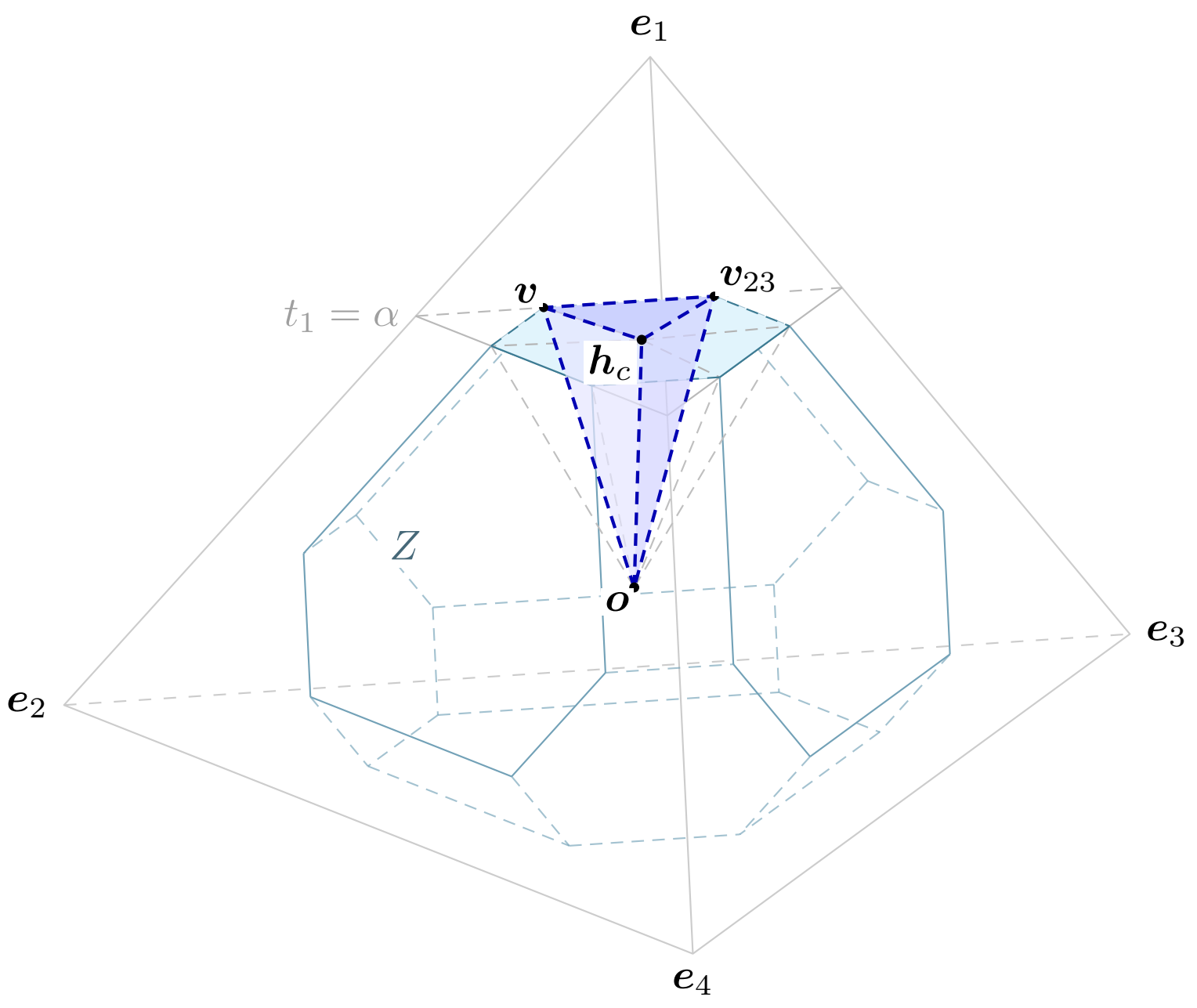
\begin{figure}[ht]
\centering
\begin{tikzpicture}[
    x=.9cm,y=.75cm,
    line join=round,
    font=\small
]
\coordinate (e1) at (.5,7.5);
\coordinate (e2) at (-5,.2);
\coordinate (e3) at (5,1);
\coordinate (e4) at (.9,-2.6);

\def\sectionDepth{.4}
\def\cornerCut{.3}

\coordinate (s2) at ($(e1)!\sectionDepth!(e2)$);
\coordinate (s3) at ($(e1)!\sectionDepth!(e3)$);
\coordinate (s4) at ($(e1)!\sectionDepth!(e4)$);

\coordinate (v1) at ($(s2)!\cornerCut!(s3)$);
\coordinate (v2) at ($(s3)!\cornerCut!(s2)$);
\coordinate (v3) at ($(s3)!\cornerCut!(s4)$);
\coordinate (v4) at ($(s4)!\cornerCut!(s3)$);
\coordinate (v5) at ($(s4)!\cornerCut!(s2)$);
\coordinate (v6) at ($(s2)!\cornerCut!(s4)$);

\coordinate (h) at (barycentric cs:s2=1,s3=1,s4=1);
\coordinate (o) at
    (barycentric cs:e1=1,e2=1,e3=1,e4=1);

\draw[gray!40,thin,dashed] (e2)--(e3);
\draw[gray!40,thin]
    (e1)--(e2)--(e4)--(e3)--cycle;
\draw[gray!40,thin] (e1)--(e4);

\draw[gray!55,thin,densely dashed] (s2)--(s3);
\draw[gray!55,thin] (s2)--(s4)--(s3);

\pgfmathsetmacro{\largeWeight}{1-\sectionDepth}
\pgfmathsetmacro{\middleWeight}{\sectionDepth*(1-\cornerCut)}
\pgfmathsetmacro{\smallWeight}{\sectionDepth*\cornerCut}

\def\permutations{%
    1/2/3/4,1/2/4/3,1/3/2/4,1/3/4/2,1/4/2/3,1/4/3/2,%
    2/1/3/4,2/1/4/3,2/3/1/4,2/3/4/1,2/4/1/3,2/4/3/1,%
    3/1/2/4,3/1/4/2,3/2/1/4,3/2/4/1,3/4/1/2,3/4/2/1,%
    4/1/2/3,4/1/3/2,4/2/1/3,4/2/3/1,4/3/1/2,4/3/2/1%
}
\foreach \a/\b/\c/\d in \permutations {
    \coordinate (z\a\b\c\d) at
      (barycentric cs:e\a=\largeWeight,e\b=\middleWeight,e\c=\smallWeight);
}

\foreach \p/\q in {%
    1234/2134,1234/1324,1234/1243,1324/3124,%
    1324/1342,1423/1432,2134/2314,2134/2143,%
    2314/3214,2314/2341,2341/3241,2341/2431,%
    2413/2431,2431/4231,3124/3214,3124/3142,%
    3214/3241,3241/3421,3412/3421,3421/4321,%
    4123/4132,4213/4231,4231/4321,4312/4321%
} {
    \draw[cyan!35!gray,thin,densely dashed,opacity=.65]
        (z\p)--(z\q);
}

\foreach \p/\q in {%
    1243/2143,1243/1423,1342/3142,1342/1432,%
    1423/4123,1432/4132,2143/2413,2413/4213,%
    3142/3412,3412/4312,4123/4213,4132/4312%
} {
    \draw[cyan!35!gray,thin] (z\p)--(z\q);
}

\fill[cyan!10]
    (v1)--(v2)--(v3)--(v4)--(v5)--(v6)--cycle;
\foreach \j in {3,...,6} {
    \draw[gray!50,thin,densely dashed] (o)--(v\j);
}
\foreach \j in {1,...,6} {
    \draw[gray!60,thin,densely dashed] (h)--(v\j);
}
\draw[cyan!50!black,thin,densely dashed]
    (v6)--(v1)--(v2)--(v3)
    (v4)--(v5);
\draw[cyan!50!black,thin]
    (v3)--(v4)
    (v5)--(v6);

\fill[blue!10,opacity=.7] (o)--(v1)--(h)--cycle;
\fill[blue!17,opacity=.7] (o)--(h)--(v2)--cycle;
\fill[blue!23,opacity=.7] (h)--(v1)--(v2)--cycle;
\draw[blue!70!black,thick,densely dashed]
    (o)--(v1)--(v2)--cycle
    (o)--(h)
    (v1)--(h)--(v2);

\fill (h) circle (1.2pt)
    node[below left,xshift=-1pt,fill=white,inner sep=1pt]
    {$\bh_c$};
\fill (o) circle (1.2pt)
    node[below left,fill=white,inner sep=1pt] {$\bo$};
\fill (v1) circle (1.1pt)
    node[above left,fill=white,inner sep=1pt] {$\bv$};
\fill (v2) circle (1.1pt)
    node[above right,fill=white,inner sep=1pt] {$\bv_{23}$};

\node[cyan!35!black,fill=white,inner sep=2pt]
    at (-1.8,2) {$Z$};
\node[above] at (e1) {$\be_1$};
\node[left]  at (e2) {$\be_2$};
\node[right] at (e3) {$\be_3$};
\node[below] at (e4) {$\be_4$};
\node[left,gray!75] at (s2) {$t_1=\alpha$};
\end{tikzpicture}
\caption{An illustration of one of these tetrahedra inside the central region. The highlighted tetrahedron is $\conv(\bo,\bh_c,\bv,\bv_{23})$.}\label{fig:central_coning}
\end{figure}

The sub-regions are defined as follows. Take the boundary facets of the central region. These facets consist of the four ``corner hexagons'' $Z \cap K_a$, four ``sink hexagons'' $Z \cap \Gamma_a$, and six ``tube rectangles'' $Z \cap T_{ab}$. We triangulate each facet by taking its center and then joining it to its edges. We then create tetrahedra by taking each of these triangles and ``coning'' it to the center 
\[\bo = \left(\frac14, \frac14, \frac14, \frac14\right)\] of the central region. In total, this produces $4 \times 6 + 4 \times 6 + 6 \times 4 = 72$ tetrahedra. See Figure~\ref{fig:central_coning} for an illustration.

To make references to these tetrahedra easier, we give names to its vertices. Let
\[
\bv \EQ (\alpha, 1 -\alpha - \beta, \beta, 0).
\]
Then all vertices of the central region can be obtained by permuting the coordinates of $\bv$. Let $\bv_{ij}$ be obtained from $\bv$ by swapping coordinates $i$ and $j$. Define
\begin{align*}
     \bh_c & \EQ \left(\alpha, \frac{1-\alpha}{3}, \frac{1-\alpha}{3}, \frac{1-\alpha}{3}\right),\\
     \bh_s & \EQ \left(\frac13, \frac13, \frac13, 0\right), \\
     \bh_t & \EQ \left(\frac{1 - \beta}{2}, \frac{1 - \beta}{2}, \frac{\beta}{2}, \frac{\beta}{2}\right).
\end{align*}
These are the centers of $Z \cap K_1$, $Z \cap \Gamma_4$ and $Z \cap T_{12}$ respectively. 

For each tetrahedron $T$, we assign a constant vector $\bc^T$ intended to be the densities for the flow measures in $T$ (up to the normalization factor $C$), that is, 
\[
dB_i^T \EQ dF_{a, i}^T \EQ C c_i \,d\bt, \quad \forall i \neq a.
\] We require that 
\begin{equation}
\sum_{i \in [4]} c_i^T \EQ 1.
\end{equation}

There are six types of tetrahedra, up to coordinate permutations. We list their representative density vectors in Table~\ref{tab:density_vectors}, and the density vectors for other tetrahedra may be obtained by permuting the coordinates.

\begin{table}
    \centering
\renewcommand{\arraystretch}{1.3}
\begin{tabular}{lll}
\hline
Tetrahedron $T$ & Symmetry & Density vector $\bc^T$\\
\hline
$\conv(\bo,\bh_c,\bv,\bv_{23})$ & $2\leftrightarrow3$
    & $(0,p_c,p_c,1-2p_c)$\\
$\conv(\bo,\bh_c,\bv,\bv_{34})$ & $3\leftrightarrow4$
    & $(0,1-2q_c,q_c,q_c)$\\
$\conv(\bo,\bh_s,\bv,\bv_{12})$ & $1\leftrightarrow2$
    & $(p_s,p_s,1-2p_s,0)$\\
$\conv(\bo,\bh_s,\bv,\bv_{23})$ & $2\leftrightarrow3$
    & $(1-2q_s,q_s,q_s,0)$\\
$\conv(\bo,\bh_t,\bv,\bv_{12})$ & $1\leftrightarrow2$
    & $(0,0,p_t,1-p_t)$\\
$\conv(\bo,\bh_t,\bv,\bv_{34})$ & $3\leftrightarrow4$
    & $(0,0,1/2,1/2)$\\
\hline
\end{tabular}

    \caption{The density vectors for six representative tetrahedra.}
    \label{tab:density_vectors}
\end{table}

Note that some entries in these vectors must be 0, due to the following reasoning:
\begin{itemize}
    \item Any tetrahedron bordering $K_a$ (whose bordering surface has nonzero area) must have $c_a = 0$, since other commodities cannot cross this boundary to enter $K_a$.
    \item Similarly, any tetrahedron bordering $\Gamma_a$ must have $c_a = 0$, since only commodity $a$ exits through $\Gamma_a$.
    \item On a tube face $Z \cap T_{ab} \subset \{\bt \mid t_a + t_b = 1 - \beta\}$, only commodities $a$ and $b$ cross the boundary, so we have $c_a = c_b = 0$.
\end{itemize}
Furthermore, for any region that stays invariant if we swap coordinates $i$ and $j$, we impose the natural constraint $c_i = c_j$. The remaining undetermined variables are left as unknowns in Table~\ref{tab:density_vectors}. The subscripts $c, s, t$ are chosen for corner, sink, and tube respectively.

We now derive the conditions that these densities must satisfy in order to satisfy flow conservation on faces between these tetrahedra. For any continuously differentiable $\varphi = \varphi(t_1, t_2, t_3, 1 - t_1 - t_2 - t_3)$ we have 
\[
    \frac{\partial \varphi}{\partial t_i} \EQ \nabla\varphi \cdot (\be_i - \be_4), \quad i = 1, 2, 3.
\]
Since we have $\sum_{i = 1}^4 c_i^T = 1$, for any $a \in [4]$ we have
\[
    \sum_{i\neq a}c_i^T (\be_a - \be_i) \EQ \be_a - \bc^T.
\]
It follows that 
\begin{align*}
    \sum_{i \neq a} c_i^T \nabla \varphi \cdot (\be_a - \be_i) & \EQ \nabla \varphi \cdot \left(\be_a - \bc^T\right) \\
    & \EQ \nabla \varphi \cdot \left(\sum_{i = 1}^3(\bone[a = i] - c_i) (\be_i - \be_4)\right) \\
    & \EQ \sum_{i = 1}^3 (\bone[a = i] - c_i) \cdot \frac{\partial \varphi}{\partial t_i}.
\end{align*}
The second equality holds since both vectors agree on the first three coordinates and they both have coordinate sum 0. Applying Theorem~\ref{thm:divergence} gives us 
\begin{align}
    \sum_{i \neq a} \int \nabla \varphi \cdot (\be_a - \be_i) \, dF_{a, i}^T & \EQ C\int \sum_{i = 1}^3 (\bone[a = i] - c_i) \cdot \frac{\partial \varphi}{\partial t_i} \,d\bt \\
    & \EQ C\sum_{F: \text{ facet of } T} \int_F \bn_F \cdot \left(\bone[a = i] - c_i^T\right)_{i = 1, 2, 3} \varphi\,dS. \label{eq:central_faces}
\end{align}

If two tetrahedra $T_1$ and $T_2$ share a face $F$, then their contributions from $F$ cancel if we have 
\begin{equation}\label{eq:face_condition}
(c_i^{T_1} - c_i^{T_2})_{i = 1, 2, 3} \cdot \bn_F \EQ 0,
\end{equation}
where either orientation for $\bn_F$ may be used. Every such face $F$ is of the form $\conv(\bo, \bu, \bw)$ and Equation~\eqref{eq:face_condition} is equivalent to 
\begin{equation}\label{eq:face_condition_det}
\det\left(\begin{array}{c}(u_i - o_i)_{i = 1, 2 ,3} \\(w_i - o_i)_{i = 1, 2 ,3} \\(c_i^{T_1} - c_i^{T_2})_{i = 1, 2 ,3}\end{array}\right) \EQ 0.
\end{equation}
We evaluate these determinants to obtain concrete conditions on the density vectors $\bc^T$. There are six types of internal faces up to coordinate permutations, three of which given by joining $\bo$ to one of the edges of $Z$, such as $\conv(\bo, \bv, \bv_{23})$, and three of which given by joining $\bo$ to one of the vertices of $Z$ and one of the centers of some facet of $Z$, such as $\conv(\bo, \bv, \bh_c)$. The determinants for each of these six types of internals faces are listed in Table~\ref{tab:determinants}, and their calculations can be found in Appendix~\ref{app:determinants}.

\begin{table}
\begin{center}
\begin{tabular}{|Sl|Sl|}
\hline
Internal face & Determinant value \\
\hline
$\conv(\bo,\bv,\bv_{34})$
    & $\displaystyle
        \beta \left(\alpha - \frac{1}{4}\right)(1-2q_c)$ \\ \hline

$\conv(\bo,\bh_t,\bv)$
    & $\displaystyle
       \frac{1-2\beta}{4}(1-2\alpha-\beta) \left(p_t-\frac12\right)$ \\\hline

$\conv(\bo,\bh_c,\bv)$
    & $\begin{aligned}[t]
       -\frac{4\alpha-1}{12}\bigl[
          &3(\alpha-1+2\beta)p_c+1-\alpha-3\beta+3\beta q_c\bigr]
       \end{aligned}$ \\\hline

$\conv(\bo,\bv,\bv_{12})$
    & $\displaystyle
       \frac{2\alpha+\beta-1}{2}\left((1- 2\beta) - (1-2\beta)p_t - p_s\right)$ \\\hline

$\conv(\bo,\bv,\bv_{23})$
    & $\displaystyle
       \frac{1-\alpha-2\beta}{2}
        \bigl[(1-4\alpha)p_c+2\alpha-1+q_s\bigr]$ \\\hline

$\conv(\bo,\bh_s,\bv)$
    & $\begin{aligned}[t]
       \frac1{12}\bigl[
          &(6\alpha-3+3\beta)p_s+(3\alpha-3+6\beta)q_s
             +2-3\alpha-3\beta\bigr]
       \end{aligned}$ \\
\hline
\end{tabular}
\end{center}
\caption{Determinant conditions derived from Equation~\eqref{eq:face_condition_det}}
\label{tab:determinants}
\end{table}

\begin{lemma}\label{lem:parameters}
    Assume that $\beta > 0$, $\frac12 < \alpha \leq 1 - 3\beta$, and
    \begin{equation}\label{eq:alpha_beta}
        4(1-\alpha)^2 - 24(1 - \alpha)\beta + 6\beta^2 + 12\beta - 1 \EQ 0.
    \end{equation}
    Then, the following choice of parameters
    \begin{align*}
        q_c & \EQ p_t \EQ \frac{1}{2}, \\
        p_c & \EQ \frac{2\alpha - 2 + 3\beta}{6(\alpha - 1 + 2\beta)}, \\
        p_s & \EQ \frac{1}{2} - \beta, \\
        q_s & \EQ 1 - 2\alpha + (4\alpha - 1)p_c,
    \end{align*}
    defines density vectors with nonnegative entries and all determinants in Table~\ref{tab:determinants} are 0.
\end{lemma}
\begin{proof}
    Setting the first five determinants in Table~\ref{tab:determinants} to 0 gives the choice of parameters, and finally plugging $p_s$ and $q_s$ back in the sixth determinant gives~\eqref{eq:alpha_beta}. With $\beta > 0$, $\frac12 < \alpha \leq 1 - 3\beta$, we have $\alpha - 1 + 2\beta \leq -\beta < 0$, so $p_c = \frac13 - \frac{\beta}{6(\alpha - 1 + 2\beta)} \in (1/3, 1/2]$, 
    \[
    \frac{2}{3} (1 - \alpha) \LE q_s \EQ (1 - 2\alpha) + (4\alpha - 1)p_c \LE \frac{1}{2},
    \]
    and $p_s \in (0, 1/2)$. So all entries in these density vectors are nonnegative.
\end{proof}

This completes the choice of the density vectors. We define the central region measures by setting 
\[
F_{a, i}^Z \EQ \sum_{T} F_{a, i}^T, \quad B_i^Z \EQ \sum_{T} B_i^T.
\]

\begin{lemma}\label{lem:central_conserv}
    For every $a \in [4]$ and $\varphi \in C^{(1)}(\Delta_4)$, we have
    \[
        I_a^Z(\varphi) \EQ \int \varphi \,d\mu_a^Z + \sum_{b \neq a} \int \varphi \, d\eta_{ab} - \int \varphi \,d\nu_a,
    \]
    where $d\mu_a^Z$ is defined as in~\eqref{eq:mu_Z_def}, $d\eta_{ab}$ is defined as in~\eqref{eq:eta_ab_def}, and $d\nu_a$ is defined on $Z \cap \Gamma_a$ via
\[
    d\nu_a \EQ C \,dt_i\,dt_j, \quad i \neq j \in [4] \setminus \{a\}.
\]
\end{lemma}
\begin{proof}
    By the choice of our density vectors, all internal-face contributions cancel, so it remains to compute the contributions from the boundary of $Z$. By \eqref{eq:central_faces}, we need to evaluate contributions of the form
    \[
    C\int_F \bn_F \cdot \left(\bone[a = i] - c_i^T\right)_{i = 1, 2, 3} \varphi\,dS.
    \]
    where $F$ lies on the boundary of $Z$.

    If $F$ is on the corner face $Z \cap K_1$, then $\bn_F = (1, 0, 0)$ and $dS = dt_2\,dt_3$, and also $c_1^T = 0$, so 
    \[
    C\int_F \bn_F \cdot \left(\bone[a = i] - c_i^T\right)_{i = 1, 2, 3} \varphi\,dS \EQ C \int_F \bone[a = 1] \varphi\,dt_2\,dt_3 \EQ \bone[a = 1]\int_F \varphi \,d\mu_a^Z.
    \]
    Similarly, if $F$ is on the sink face $Z \cap \Gamma_1$, then $\bn_F = (-1, 0, 0)$, $dS = dt_2\,dt_3$ and $c_1^T = 0$, so 
    \[
    C\int_F \bn_F \cdot \left(\bone[a = i] - c_i^T\right)_{i = 1, 2, 3} \varphi\,dS \EQ -C \int_F \bone[a = 1]\varphi\,dt_2\,dt_3 \EQ -\bone[a = 1]\int_F \varphi\, d\nu_a.
    \]
    If $F$ is on the tube face $Z \cap T_{12}$, then $\bn_F = (1, 1, 0) / \sqrt{2}$, $dS = \sqrt{2}dt_1\,dt_3$, and $c_1 = c_2 = 0$, so we have
    \[
    C\int_F \bn_F \cdot \left(\bone[a = i] - c_i^T\right)_{i = 1, 2, 3} \varphi\,dS \EQ C \int_F \bone[a \in {1,2}]\varphi\,dt_1\,dt_3 \EQ \bone[a \in \{1, 2\}]\int_F \varphi\, d\eta_{12}.
    \]
    The other faces follow by symmetry, and summing these contributions yields the lemma.
\end{proof}

\subsection{Normalization and feasibility}

To choose the normalization parameter $C$, we compute 
\begin{align*}
    \int_{Z \cap \Gamma_1} \,d\nu_1 \EQ \int_{Z \cap \Gamma_1} C\,dt_2\,dt_3 \EQ C \, \left( \frac{(1 - 3\beta)^2}{2} - 3\,\frac{(1 - \alpha - 2\beta)^2}{2}\right).
\end{align*}
Since $\nu_1$ is a probability measure, we have 
\begin{equation}\label{eq:C_def}
C \EQ \left(\frac{(1 - 3\beta)^2}{2} - 3\,\frac{(1 - \alpha - 2\beta)^2}{2}\right)^{-1}.
\end{equation} 

We can now put all the pieces together and define
\begin{equation}\label{eq:final_def}
    F_{a, i} \EQ \sum_R F_{a, i}^R, \quad B_i = \sum_R B_i^R, \quad \forall i \neq a \in [4].
\end{equation}
where the summation variable $R$ is taken over all corners, simplex edges, tubes, and the central region.

\begin{lemma}
    Equation~\eqref{eq:final_def} defines a feasible solution to \eqref{eq:continuous_lp}.
\end{lemma}
\begin{proof}
    By construction we have $0 \leq F_{a, i} \leq B_i$ for every $i \neq a \in [4]$. The flow conservation constraint follows by summing the contributions from Lemmas~\ref{lem:corner_conserv}, \ref{lem:simplex_edge_conserv}, \ref{lem:tube_conserv}, and \ref{lem:central_conserv}.
\end{proof}

\subsection{Optimizing the parameters}

We now compute the total capacities $\sum_i B_i^R(\Delta_4)$ in each region $R$. For a corner region $K_a$, we have 
\[
    \sum_i B_i^{K_a}(\Delta_4) \EQ \sum_{i \neq a} \int \,dF_{a, i}^{K_a} \EQ 1 - \alpha.
\]
For a simplex edge $E_{ab}$, we have
\[
    \sum_i B_i^{E_{ab}}(\Delta_4) \EQ \int \,dF_{a, b}^{E_{ab}} + \int \,dF_{b, a}^{E_{ab}} \EQ \int_{1-\alpha}^\alpha C\beta(2\alpha-1)\,dw \EQ C\beta(2\alpha-1)^2.
\]
For a tube $T_{ab}$, letting $\{i, j\} = [4] \setminus \{a, b\}$, we have
\[
    \sum_\ell B_\ell^{T_{ab}}(\Delta_4) \EQ \int \,dF_{a, i}^{T_{ab}} + \int \,dF_{a, j}^{T_{ab}} \EQ \int_{T_{ab}} C\beta\frac{1}{t_i + t_j}\,d\bt \EQ C \beta^2 \left(2\alpha - 1 + \frac{\beta}{2}\right).
\]
For the central region $Z$, we have $\sum_{i}dB_i^Z = C d\bt$ since the density vectors $\bc^T$ satisfying $\sum_i c_i^T = 1$ for every tetrahedron $T$ in $Z$. It follows that 
\begin{align*}
    \sum_i B_i^{Z}(\Delta_4) \EQ \int_Z C\,d\bt & \EQ C \left(\frac{1}{6} - 4 \times \frac{(1 - \alpha)^3}{6} - 6 \times \left( \frac{(2\alpha - 1)\beta^2}{2} + \frac{\beta^3}{3}\right)\right) \\
    & \EQ C \left(\frac{1}{6} - \frac{2(1 - \alpha)^3}{3} - 3(2\alpha - 1)\beta^2 - 2\beta^3\right).
\end{align*}
Adding these together, we obtain
\begin{equation}\label{eq:total_cost}
    \sum_{i = 1}^4 B_i(\Delta_4) \EQ 4(1 - \alpha) + C\left(\frac{1}{6} - \frac{2(1 - \alpha)^3}{3} + 6\beta(2\alpha-1)^2 + 3\beta^2(2\alpha-1) + \beta^3\right).
\end{equation}

\begin{figure}[t]
    \centering
    \includegraphics[scale=0.8]{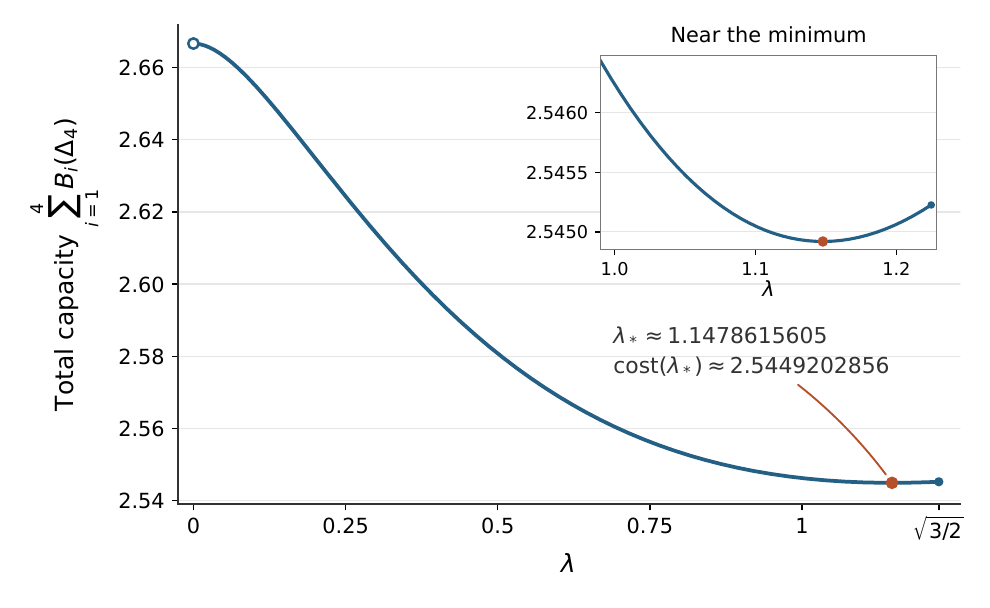}
    \caption{A plot for the cost function in \eqref{eq:continuous_cost}.}
    \label{fig:continous_flow_cost}
\end{figure}
We seek to minimize this expression, under Equation~\eqref{eq:alpha_beta} and Equation~\eqref{eq:C_def}. To satisfy Equation~\eqref{eq:alpha_beta}, we parametrize
\begin{equation}
    \alpha \EQ \frac{1}{2} + \frac{2\lambda^2}{2\lambda^2 + 12\lambda + 3}, \quad \beta \EQ \frac{2\lambda}{2\lambda^2 + 12\lambda + 3},
\end{equation}
under which Equation~\eqref{eq:total_cost} (after also plugging in Equation~\eqref{eq:C_def}) becomes 
\begin{equation}\label{eq:continuous_cost}
    \sum_{i = 1}^4 B_i(\Delta_4) \EQ \frac{8(144\lambda^5+612\lambda^4+872\lambda^3 +468\lambda^2+108\lambda+9)} {(2\lambda^2+12\lambda+3) (4\lambda^4+144\lambda^3+180\lambda^2+72\lambda+9)}.
\end{equation}
The condition in Lemma~\ref{lem:parameters} for $\alpha$ and $\beta$ gives $\lambda \in (0, \sqrt{3/2}]$. See Figure~\ref{fig:continous_flow_cost} for a plot of this function. Using numerical optimization, when
\[
    \lambda \approx 1.14786156048, \quad \alpha \approx 0.63576706375, \quad \beta \approx 0.11827825622,
\]
we get $\sum_{i = 1}^4 B_i(\Delta_4) \leq 2.54492029$. This completes the proof of Theorem~\ref{thm:cont_feasible_k4}. 

\section{Concluding remarks}\label{sec:concl}

We obtained substantially improved lower bounds on the integrality ratio of the CKR relaxation of the Multiway Cut problem, which by a result of \cite{MNRS08} translate, under UGC, to a hardness of approximation results for the problem. Our improved lower bounds are based on the non-opposite multiway cuts framework of \cite{AMM17} and on the idea of using stronger LP relaxations to obtain improved integrality ratios for weaker LP relaxations. Such an idea was used before for other problems, e.g., by Goemans \cite{Goemans95} for the Traveling Salesperson Problem (TSP), and by Vicari \cite{Vicari20} for the Steiner Tree problem. However, to the best of our knowledge, this idea was not used before to obtain lower bounds on integrality ratios of LP relaxations known to be equal to the approximability ratios of the corresponding problems. Another novel idea used to obtain our lower bounds is considering a continuous limit of stronger LP relaxation and presenting an explicit almost-optimal solution of this continuous LP relaxation.

We presented an explicit analytic solution of a continuous LP relaxation that implies a lower bound of $1.22232$ on the approximability ratio of the Multiway Cut problem with an arbitrary number of terminals. While this result was found using AI tools (see the AI disclosure), the proof of this result is self-contained and does not require computational verification. We also presented strong numerical evidence that suggests that the lower bound can be improved to at least $\bestNum$.

We conclude by mentioning a few future directions.
\begin{itemize}
    \item Our $k = 4$ solution for \eqref{eq:continuous_lp} is very likely not optimal. ChatGPT suggested that a tiny perturbation in the corner region can slightly improve the objective value. It would be interesting to find the $k = 4$ optimal solution to \eqref{eq:continuous_lp} and understand its structure.
    \item We already have numerical evidence that the Flow LP approach leads to a Multiway Cut lower bound above $1.23$. Furthermore, ChatGPT claims to have found a $k = 7$ feasible solution for \eqref{eq:continuous_lp} that leads to a bound of $\approx 1.2308$. However the construction seems complicated and ad-hoc, and we have not yet verified it. It would be interesting to understand solutions of the continuous flow LP for $k>4$.
    \item Our approach can also be used to find integrality gap instances for a fixed number of terminals, by using the standard Multiway Cut formulation in the LP instead of the non-opposite cut. This leads to hardness of approximation results for a fixed number of terminals (see Appendix~\ref{app:standard-cuts}, Table~\ref{table:LP_vertex_standard} and Table~\ref{table:LP_edge_standard}). Obtaining a rigorous understanding of the LP solutions in this scenario is also interesting.
    \item An interesting theoretical question to explore would be to understand how much do we gain if we add more local distributions and consistency requirements to the LP formulation, or more broadly speaking if we go to higher levels in the Sherali--Adams hierarchy. A concrete question would be, do we get a strictly larger ratio for \eqref{LP:edge} than for \eqref{LP:vertex} when $k, n \to \infty$? 
    \item Finally, we believe that the idea of using stronger LP to certify integrality ratios can be used for other optimization problems as well. This is especially relevant for problems where we know that the Basic LP/SDP is optimal in the worst case and therefore might have a ``psychological blind spot'' for using stronger relaxations.
    
\end{itemize}

\appendix
\section{Adaptation of Our Framework to Standard Cuts}\label{app:standard-cuts}

In this section, we show that our strengthened LP framework in Section~\ref{sec:SA} can be extended to standard multiway cuts. We state the relevant definitions and LP formulations, but we omit proofs due to them being near-identical to those in Section~\ref{sec:SA}.

\subsection{Standard Cuts}
For every $\bx \in \Delta_{k, n}$, define the standard cuts feasible label set
\[
L(\bx) \coloneqq \begin{cases}
\{i\} & \text{if } \bx = \be_i, i \in [k]\\
[k] & \text{otherwise}.
\end{cases}
\]
\begin{definition}
    A function $P: \Delta_{k, n} \to [k]$ is called a \emph{standard cut} if $P(\bx) \in L(\bx)$ for every $\bx \in \Delta_{k, n}$.
\end{definition}
To define a standard cut instance on $(\Delta_{k, n}, E_{k, n})$, it is sufficient to define a weight function $w: E_{k, n} \to \mathbb{R}_{\geq 0}$. Given a standard cut $P: \Delta_{k, n} \to [k]$, we say that an edge $e = \{\bu, \bv\} \in E_{k, n}$ is \emph{cut} by $P$ if $P(\bu) \neq P(\bv)$. The cost incurred by a standard cut $P$ with respect to the edge weights $w$ is
\[
  \cost(w, P) \coloneqq \sum_{e =\{\bu, \bv\} \in E_{k,n}} w_e \cdot \bone[P(\bu) \neq P(\bv)].
\]
We further define 
\[
\OPT(w) \coloneqq \min_{\substack{P: \Delta_{k, n} \to [k] \\ P \text{ standard}}} \cost(w, P)
\]
to be the minimum total weight of edges cut by a standard cut. We say that our weight function $w$ is \emph{normalized} if
\[
    \sum_{e \in E_{k,n}} w_e \EQ n.
\]
For every $C \in \cD_{k, n} \cup \cU_{k,n}$, define
\[\Omega_C \EQ \{\sigma \in [k]^C \mid \forall \bu \in C \;,\; \sigma(\bu) \in L(\bu)\}\]

As discussed in the introduction, for any normalized weight function $w$, we have  \mbox{$\LPCKR(w) \le 1$}. As such, by a result of Manokaran, Naor, Raghavendra, and Schwartz \cite{MNRS08}, a standard cut gap instance implies UG-hardness of inapproximability for Multiway Cut with $k$ terminals.

\begin{theorem}[\cite{MNRS08}, see also \cite{Raghavendra08}]\label{thm:mnrs}
    Let $w: E_{k, n} \to \mathbb{R}_{\geq 0}$ be a normalized weight function. Assuming the Unique Games Conjecture, for every $\varepsilon > 0$, it is NP-hard to approximate Multiway Cut with $k$ terminals within a multiplicative factor of~$\OPT(w) - \varepsilon$. 
\end{theorem}

\subsection{Vertex-consistency LP}

Analogous to \eqref{LP:vertex}, we define a strengthening of the CKR LP relaxation for standard cuts which enforces vertex consistency on downward cells.

\begin{equation}
\tag{$\LPvertex$}\label{LP:vertex-standard}
\begin{aligned}
\text{min}\quad & \sum_{e\in E_{k, n}} w_e p_e \\
\text{s.t.}\quad & \sum_{\sigma \in \Omega_C} \mu_{C, \sigma} \EQ 1\,, && \qquad C \in \cD_{k, n}\,, \\
    & \sum_{\substack{\sigma \in \Omega_{C} \\ \sigma(\bu) = a}} \mu_{C, \sigma} \EQ \lambda_{\bu, a}\,, &&  \qquad C \in \cD_{k, n}, \, \bu \in C, \, a \in L(\bu)\,,\\
    & p_e \EQ \sum_{\substack{\sigma\in \Omega_{D(e)}\\ \sigma(\bu)\ne\sigma(\bv)}} \mu_{D(e),\sigma}\,, && \qquad e = \{\bu, \bv\}\in E_{k, n}\,, \\
    & \mu_{C, \sigma} \GE 0 \,, && \qquad C \in \cD_{k, n}, \, \sigma \in \Omega_C\,.
\end{aligned}
\end{equation}

By suitably adapting the proof of Lemma~\ref{lem:vertex_LP}, one can show that the objective value $\LPvertex(w)$ is always at most $\OPT(w)$. We now define the dual of \eqref{LP:vertex-standard}.

\begin{equation}
\label{LP:dual-vertex-standard}
\begin{aligned}
\text{max}\quad & \sum_{C \in \cD_{k, n}} y_C \\
\text{s.t.}\quad & y_C + \sum_{\bv \in C} z_{C, \bv, \sigma(\bv)} \LE \sum_{\substack{\{\bu, \bv\} \subseteq C\\ \bu\ne\bv}}w_{\{\bu, \bv\}} \cdot \bone[\sigma(\bu) \neq \sigma(\bv)], &&\qquad C \in \cD_{k, n},  \sigma \in \Omega_C, \\
    & \sum_{C: \bv \in C} z_{C, \bv, a} \EQ 0, && \qquad \bv \in \Delta_{k, n}, a \in L(\bv),\\
    & y_C, z_{C, \bv, a} \in \mathbb{R}. && 
\end{aligned}
\tag{dual-$\LPvertex$}
\end{equation}

Since the dependence on $w$ in \eqref{LP:dual-vertex-standard} is linear, we can add a constraint that $w$ is normalized and optimize over $w$.

\begin{equation}
\label{LP:dual-vertex-w-standard}
\begin{aligned}
\text{max}\quad & \sum_{C \in \cD_{k, n}} y_C \\
\text{s.t.}\quad & y_C + \sum_{\bv \in C} z_{C, \bv, \sigma(\bv)} \LE \sum_{\{\bu, \bv\} \subseteq C}w_{\{\bu, \bv\}} \cdot \bone[\sigma(\bu) \neq \sigma(\bv)], &&\qquad C \in \cD_{k, n},  \sigma \in \Omega_C, \\
    & \sum_{C: \bv \in C} z_{C, \bv, a} \EQ 0, && \qquad \bv \in \Delta_{k, n}, a \in L(\bv),\\
    & \sum_{e \in E_{k, n}} w_e \EQ n, \\
    & y_C, z_{C, \bv, a} \in \mathbb{R}, \quad w_e \geq 0. && 
\end{aligned}
\tag{dual-$\LPvertex$-$w$}
\end{equation}

We let $\LPvertex(k, n)$ denote the objective value of \eqref{LP:dual-vertex-w-standard}. Similar to \eqref{LP:vertex-w-free}, we can take the dual of \eqref{LP:dual-vertex-w-standard}.

\begin{equation}
\tag{$\LPvertex$-$w$-free}\label{LP:vertex-w-free-standard}
\begin{aligned}
\text{min}\quad & np \\
\text{s.t.}\quad & \sum_{\sigma \in \Omega_C} \mu_{C, \sigma} \EQ 1 && \qquad C \in \cD_{k, n}, \\
    & \sum_{\substack{\sigma \in \Omega_{C} \\ \sigma(\bu) = a}} \mu_{C, \sigma} \EQ \lambda_{\bu, a} &&  \qquad C \in \cD_{k, n}, \, \bu \in C, \, a \in L(\bu),\\
    & p \GE \sum_{\sigma\in \Omega_{D(e)}} \mu_{D(e),\sigma}\, \bone[\sigma(\bu)\ne\sigma(\bv)] && \qquad e = \{\bu, \bv\}\in E_{k, n}, \\
    & \mu_{C, \sigma} \GE 0 && \qquad C \in \cD_{k, n}, \, \sigma \in \Omega_C.
\end{aligned}
\end{equation}

\subsection{Edge-consistency LP}

We further define a family of edge-consistency LPs for standard cuts analogous to \eqref{LP:edge}.

\begin{equation}
\tag{$\LPedge$}\label{LP:edge-standard}
\begin{aligned}
\text{min}\quad & \sum_{e\in E_{k, n}} w_e p_e \\
\text{s.t.}\quad & \sum_{\sigma \in \Omega_C} \mu_{C, \sigma} \EQ 1 && \qquad C \in \cD_{k, n} \cup \cU_{k,n}, \\
    & \sum_{\substack{\sigma \in \Omega_{C} \\ \sigma(\bu) = a\\\sigma(\bv) = b}} \mu_{C, \sigma} \EQ \lambda_{e, a, b} &&  \!\!\!\begin{tabular}{l}$\qquad C \in  \cD_{k, n} \cup \cU_{k, n}, \, e = \{\bu, \bv\} \in E_{k, n}, e \subseteq C,$\\$\qquad a \in L(\bu), \, b \in L(\bv)$\end{tabular}\\
    & p_e \EQ \sum_{\substack{a \in L(\bu)\\b \in L(\bv) \\ a \neq b}} \lambda_{e, a, b} && \qquad e = \{\bu, \bv\}\in E_{k, n}, \\
    & \mu_{C, \sigma} \GE 0 && \qquad C \in \cD_{k, n} \cup \cU_{k,n}, \, \sigma \in \Omega_C.
\end{aligned}
\end{equation}

Likewise, by adapting the proof of Lemma~\ref{lem:edge_LP}, we can show that $\LPedge(w) \in [\LPvertex(w), \OPT(w)]$. The dual of \eqref{LP:edge-standard} is as follows.
\begin{equation}
\label{LP:dual-edge-standard}
\begin{aligned}
\text{max}\quad & \sum_{C \in \cD_{k, n} \cup\,\cU_{k,n}} y_C \\
\text{s.t.}\quad & y_C \LE \sum_{e = \{\bu, \bv\} \subseteq C} (z_{e, \sigma(u), \sigma(v)} + w_e \cdot \bone[\sigma(u) \neq \sigma(v)]) &&\qquad C \in \cD_{k, n},  \sigma \in \Omega_C, \\
& y_C \LE -\sum_{e = \{\bu, \bv\} \subseteq C} z_{e, \sigma(u), \sigma(v)} &&\qquad C \in \cU_{k, n},  \sigma \in \Omega_C,\\
    & y_C, z_{e, a, b} \in \mathbb{R}. && 
\end{aligned}
\tag{dual-$\LPedge$}
\end{equation}
If optimize over $w$ and constrain that $w$ is normalized, we get the following LP.
\begin{equation}
\tag{dual-$\LPedge$-$w$}
\label{LP:dual-edge-w-standard}
\begin{aligned}
\text{max}\quad & \sum_{C \in \cD_{k, n} \cup\,\cU_{k,n}} y_C \\
\text{s.t.}\quad & y_C \LE \sum_{e = \{\bu, \bv\} \subseteq C} (z_{e, \sigma(u), \sigma(v)} + w_e \cdot \bone[\sigma(u) \neq \sigma(v)]) &&\qquad C \in \cD_{k, n},  \sigma \in \Omega_C, \\
& y_C \LE -\sum_{e = \{\bu, \bv\} \subseteq C} z_{e, \sigma(u), \sigma(v)} &&\qquad C \in \cU_{k, n},  \sigma \in \Omega_C,\\
    & \sum_{e \in E_{k, n}} w_e \EQ n, \\
    & y_C, z_{e, a, b} \in \mathbb{R}, \quad w_e \geq 0. && 
\end{aligned}
\end{equation}

We let $\LPedge(k, n)$ denote the objective value of \eqref{LP:dual-edge-w-standard}.

\subsection{Numerical results}

In Table~\ref{table:LP_vertex_standard} and Table~\ref{table:LP_edge_standard}, we present some numerical computations which estimate $\LPvertex(k, n)$ and $\LPedge(k, n)$ for various values of $k$ and $n$. In general, the computed values exceed the best-known rigorously computed integrality gap of $6/(5 + \frac{1}{k-1})$ for small $k$ due to \cite{AMM17}.

\begin{table}[ht]
\centering
\begin{tabular}{|c|c|c|c|c|}
    \hline
\backslashbox{$n$}{$k$} & 4 & 5 & 6 &7\\\hline
5  & 1.10915492 & 1.12768449 & 1.13939139 & 1.14745908 \\
10 & 1.11801242 & 1.13892723 & 1.15218171 & 1.16132008 \\
15 & 1.12215766 & 1.14315626 & 1.15683438 & 1.16623058 \\
20 & 1.12451796 & 1.14595005 & 1.15959361 & 1.16901434 \\
25 & 1.12576767 & 1.14749897 & 1.16134048 & 1.17090926 \\
30 & 1.12656553 & 1.14848994 & 1.16246282 & -- \\
40 & 1.12772743 & 1.14971504 & -- & -- \\
50 & 1.12835807 & 1.15053900 & -- & -- \\
\hline\hline
\cite{AMM17} & 1.12500000 & 1.14285714 & 1.15384615 & 1.16129032 \\
\hline
\end{tabular}
\caption{Numerically computed values of $\LPvertex(k, n)$, rounded to eight decimal places. A dash (--) indicates that the computation was not completed. The final line is the best-known lower bounds for small $k$ due to \cite{AMM17}.}
\label{table:LP_vertex_standard}
\end{table}

\begin{table}[ht]
\centering
\begin{tabular}{|c|c|c|c|c|}
    \hline
\backslashbox{$n$}{$k$} & 4 & 5 & 6 &7\\\hline
5  & 1.10915492 & 1.12768449 & 1.13939138 & 1.14745908 \\
10 & 1.11824689 & 1.13931036 & 1.15262982 & 1.16184033 \\
15 & 1.12221748 & 1.14349119 & 1.15732303 & 1.16685015 \\
20 & 1.12459511 & 1.14621877 & 1.16001245 & --\\
25 & 1.12587061 & 1.14779277 & 1.16175131 & -- \\
30 & 1.12664860 & 1.14876641 & -- & -- \\
40 & 1.12781239 & 1.14997073 & -- & -- \\
50 & 1.12843901  & -- & -- & -- \\
\hline\hline
\cite{AMM17} & 1.12500000 & 1.14285714 & 1.15384615 & 1.16129032 \\
\hline
\end{tabular}
\caption{Numerically computed values of $\LPedge(k, n)$, rounded to eight decimal places. A dash (--) indicates that the computation was not completed. The final line is the best-known lower bounds for small $k$ due to \cite{AMM17}.}
\label{table:LP_edge_standard}
\end{table}

\section{Analyzing the Vertex-Consistency LP for $k=3$}\label{app:k=3}

In this appendix, we analyze the vertex-consistency LP when $k = 3$ for both standard cuts and non-opposite cuts. This gives new proofs for the optimal $k = 3$ gap results in the literature~\cite{CCT06, AMM17}.

One way to visualize the solutions for the dual of vertex-consistency LP for $k=3$ is as follows. 
\begin{enumerate}
\item Recall that each edge $e \in E_{3, n}$ belongs to a unique downward cell $D(e) \in \cD_{3, n}$. For $k=3$, these downward cells are either downward triangles or single edges (we may ignore singleton cells since they don't contain edges). Given a solution to the vertex-consistency LP, we have the variable $p_e$ for each edge $e$ which represents the cut probability of $e$ in the local distribution of $D(e)$. The objective value for the LP is equal to $\sum_{e}{{w_e}p_e}$. 
\item We choose a \emph{valuation function} which assigns a value $\val(C)$ to each downward cell $C$ based on the local distribution probabilities of the vertices in $C$. We choose $\val(C)$ so that $\sum_{e \in E(C)}w_e{p_e} \geq \val(C)$, where $E(C)$ denotes the set of edges in $C$. This implies that $\sum_{e}{{w_e}p_e} \geq \sum_{C}{\val(C)}$. If we choose this valuation function so that $\sum_{C}{\val(C)}$ is the same regardless of  the vertex-consistency LP solution, this gives us a lower bound on the objective value of the vertex-consistency LP.
\end{enumerate}
We make this precise with the following definition, which works for both standard cuts and non-opposite cuts.
\begin{definition}
Fix an edge weight function $w: E_{3, n} \to \mathbb{R}^{\geq 0}$. We define a valuation function (with respect to $w$) as follows: for every downward cell $C$, $\val(C)$ is a linear function in the vertex marginals~$\lambda_{\bu}$ for every vertex $\bu \in C$, satisfying that for all distributions $\mu$ over $\Omega_C$, $\sum_{e \in E(C)}w_e{p_e} \geq \val(C)$ (here $p_e$ and $\val(C)$ are both induced by $\mu$).
\end{definition}
\begin{proposition}
Let $K > 0$ be some constant. If\, $\val$ is a valuation function such that \mbox{$\sum_{C}{\val(C)} \geq K$} for any vertex-consistency LP solution, then the objective value for the vertex-consistency LP is at least $K$.
\end{proposition}
\subsection{Derivation of the $k=3$ Lower Bound for Standard Cuts}
We use a slight modification of the instance used by Karger et al. \cite{KKSTY04} to show a lower bound of $\frac{12}{11}$ for $k=3$ as this modification makes our analysis cleaner and allows us to match the slightly stronger bound of \cite{CCT06}, which is not only tight asymptotically but also for instances of a given side length $n$. Throughout this subsection, we consider the scaled discretized simplex $(3N)\Delta_{3, 3N}$ for some $N \in \mathbb{N}$.
\begin{enumerate}
\item Along the sides of the simplex, if we consider the edge $e$ which is the $j$th edge from one corner and the $j'$th edge from another corner where $j' = 3N+1-j$, we take $w_e = 3N+2 - \min{\{j,j',N\}} = N + 1 + \max{\{2N+1-j,0\}} + \max{\{2N+1-j',0\}}$.
\item For all edges $e$ which are not on the side of the simplex which have an endpoint with a coordinate greater than $2N$ (i.e., an endpoint of the edge is within distance $N-1$ of the closest corner), we set $w_e = 1$ if this coordinate changes along $e$ (i.e., the edge is parallel to a side of the simplex between the closest corner and one of the other two corners) and $0$ otherwise (i.e., the edge is parallel to the side of the simplex between the other corners).
\item All other edges have weight $1$.
\end{enumerate}
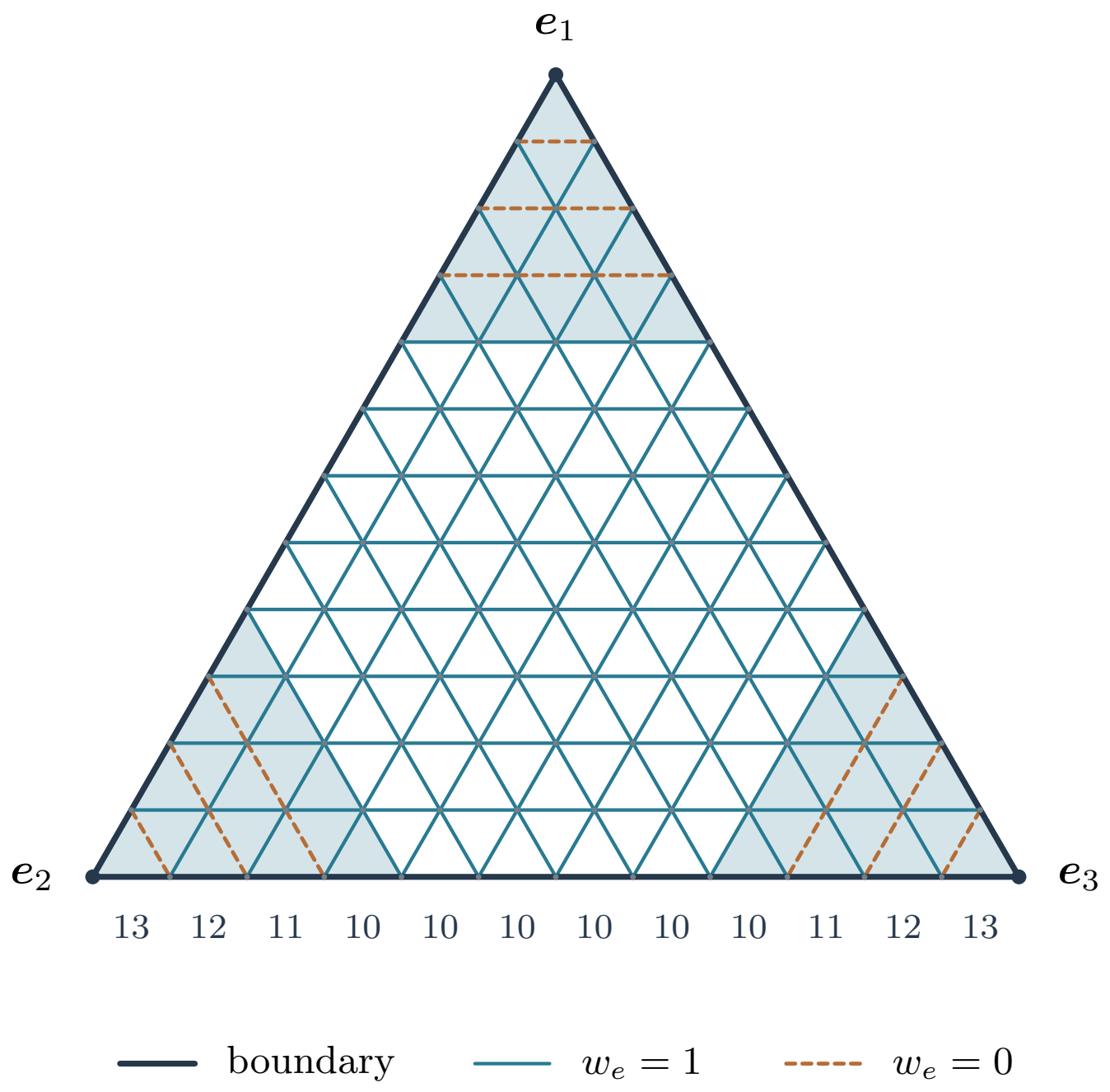
\begin{figure}
    \centering\begingroup
\definecolor{cewblue}{HTML}{287A92}
\definecolor{cewzero}{HTML}{B66D35}
\definecolor{cewboundary}{HTML}{26384B}
\begin{tikzpicture}[
  x=1cm,y=1cm,line cap=round,line join=round,
  every node/.style={font=\small},
  cew unit/.style={draw=cewblue,line width=.8pt},
  cew zero/.style={draw=cewzero,line width=.9pt,
    dash pattern=on 2.1pt off 1.6pt},
  cew boundary/.style={draw=cewboundary,line width=1.35pt},
  cew note/.style={font=\footnotesize,text=black!65,align=center}
]
\def\cewN{4}
\def\cewn{12}
\pgfmathsetmacro{\cewheight}{6*sqrt(3)*.62}

\newcommand{\cewDrawEdge}[4]{%
  \pgfmathtruncatemacro{\cewa}{\cewn-#1}%
  \pgfmathtruncatemacro{\cewb}{#1-#2}%
  \pgfmathtruncatemacro{\cewc}{#2}%
  \pgfmathtruncatemacro{\cewd}{\cewn-#3}%
  \pgfmathtruncatemacro{\cewe}{#3-#4}%
  \pgfmathtruncatemacro{\cewf}{#4}%
  \pgfmathtruncatemacro{\cewisboundary}{%
    (\cewa==0 && \cewd==0) ||
    (\cewb==0 && \cewe==0) ||
    (\cewc==0 && \cewf==0)}%
  \pgfmathtruncatemacro{\cewiszero}{%
    (\cewa==\cewd && \cewa>2*\cewN) ||
    (\cewb==\cewe && \cewb>2*\cewN) ||
    (\cewc==\cewf && \cewc>2*\cewN)}%
  \ifnum\cewisboundary=1
    \def\cewstyle{cew boundary}%
  \else
    \ifnum\cewiszero=1
      \def\cewstyle{cew zero}%
    \else
      \def\cewstyle{cew unit}%
    \fi
  \fi
  \draw[\cewstyle]
    (\cewprefix-#1-#2)--(\cewprefix-#3-#4);
}

\begin{scope}
  \def\cewprefix{A}
  \foreach \r in {0,...,12}{
    \foreach \s in {0,...,\r}{
      \coordinate (A-\r-\s) at
        ({.62*(\s-\r/2)},{\cewheight-.62*sqrt(3)*\r/2});
    }
  }
  \foreach \a/\b/\c in {0-0/4-0/4-4,12-0/8-0/12-4,12-12/8-8/12-8}{
    \path[fill=cewblue!20] (A-\a)--(A-\b)--(A-\c)--cycle;
  }
  \foreach \r in {0,...,12}{
    \foreach \s in {0,...,\r}{
      \ifnum\s<\r
        \pgfmathtruncatemacro{\sn}{\s+1}
        \cewDrawEdge{\r}{\s}{\r}{\sn}
      \fi
      \ifnum\r<12
        \pgfmathtruncatemacro{\rn}{\r+1}
        \pgfmathtruncatemacro{\sn}{\s+1}
        \cewDrawEdge{\r}{\s}{\rn}{\s}
        \cewDrawEdge{\r}{\s}{\rn}{\sn}
      \fi
    }
  }
  \foreach \r in {0,...,12}{
    \foreach \s in {0,...,\r}{
      \fill[cewboundary!65] (A-\r-\s) circle[radius=.65pt];
    }
  }
  \foreach \p in {0-0,12-0,12-12}
    \fill[cewboundary] (A-\p) circle[radius=1.7pt];
  \node[above=4pt] at (A-0-0) {$\be_1$};
  \node[left=5pt] at (A-12-0) {$\be_2$};
  \node[right=5pt] at (A-12-12) {$\be_3$};
  \foreach \j in {1,...,12}{
    \pgfmathtruncatemacro{\jm}{\j-1}
    \pgfmathtruncatemacro{\weight}{3*\cewN+2-min(\j,\cewn+1-\j,\cewN)}
    \node[font=\scriptsize,text=cewboundary,below=5pt]
      at ($(A-12-\jm)!.5!(A-12-\j)$) {$\weight$};
  }
\end{scope}

\begin{scope}[yshift=-1.5cm]
  \draw[cew boundary] (-3.5,0)--(-2.9,0);
  \node[anchor=west,font=\footnotesize] at (-2.78,0) {boundary};
  \draw[cew unit] (-.65,0)--(-.05,0);
  \node[anchor=west,font=\footnotesize] at (.07,0) {$w_e=1$};
  \draw[cew zero] (1.85,0)--(2.45,0);
  \node[anchor=west,font=\footnotesize] at (2.57,0) {$w_e=0$};
\end{scope}
\end{tikzpicture}
\endgroup
    \caption{An illustration for the edge weights when $N = 4$}\label{fig:k=3_weights_standard}
\end{figure}
See Figure~\ref{fig:k=3_weights_standard} for an illustration for the edge weights.
\begin{remark}
The only difference between this instance and the instance of \cite{KKSTY04} is that the weights of the edges along the sides of the simplex are increased by $1$ for our instance.
\end{remark}
This instance can be analyzed using the following building blocks. 
\begin{enumerate}
\item Edges near a corner: For each edge $e$ which has an endpoint whose $1$-coordinate is greater than $2N$, orienting the edge $e = \{\bu,\bv\}$ so that it points away from the corner (i.e., $u_1 > v_1$), we take 
\[
\val(e) \EQ \frac{2}{3}\left(\lambda_{\bu, 1} - \lambda_{\bv, 1}\right) + \frac{1}{3}\left(\lambda_{\bv, 2} - \lambda_{\bu, 2}\right) + \frac{1}{3}\left(\lambda_{\bv, 3} - \lambda_{\bu, 3}\right).
\]
We take analogous valuations for the edges which are close to corner $2$ or corner $3$. For a downward cell $C$ near the corner, we define $\val(C) = \sum_{e \in E(C)}w_e\val(e)$.
\item Center downward cells: For each center downward cell $C$, orient the cell $C = \{\bu,\bv,\bw\}$ so that $\bu$ is farthest from corner $3$, $\bv$ is farthest from corner $2$, and $\bw$ is farthest from corner $1$, we take 
\[
\val(C) \EQ \frac{2}{3}\left(\lambda_{\bu, 1} + \lambda_{\bv, 1} - 2\lambda_{\bw, 1}\right) + \frac{2}{3}\left(\lambda_{\bu, 2} + \lambda_{\bw, 2} - 2\lambda_{\bv, 2}\right) + \frac{2}{3}\left(\lambda_{\bv, 3} + \lambda_{\bw, 3} - 2\lambda_{\bu, 3}\right).
\]
\item For the edge $e$ on the side of the triangle between corner $1$ and corner $2$ which is the $j$th edge from corner $1$ and the $j'$th edge from corner $2$ where $j \in [N,2N+1]$ and $j' = 3N+1-j$, we view it as $2\left(2N+1 - j\right)$ edges going from corner $1$ to corner $2$ and $2\left(2N+1 - j'\right)$ edges going from corner $2$ to corner $1$. In other words, using the orientation from item 1, we take 
\begin{align*}
& \qquad {w_e}\val(e) 
\\&\EQ 2\left(2N+1 - j\right)\left(\frac{2}{3}\left(\lambda_{\bu, 1} - \lambda_{\bv, 1}\right) + \frac{1}{3}\left(\lambda_{\bv, 2} - \lambda_{\bu, 2}\right) + \frac{1}{3}\left(\lambda_{\bv, 3} - \lambda_{\bu, 3}\right)\right) \\
& \qquad +2\left(2N+1 - j'\right)\left(\frac{1}{3}\left(\lambda_{\bu, 1} - \lambda_{\bv, 1}\right) + \frac{2}{3}\left(\lambda_{\bv, 2} - \lambda_{\bu, 2}\right) + \frac{1}{3}\left(\lambda_{\bu, 3} - \lambda_{\bv, 3}\right)\right)\\
&\EQ \frac{3N+3+(j'-j)}{3}\left(\lambda_{\bu, 1} - \lambda_{\bv, 1}\right) \\
& \qquad + \frac{3N+3+(j-j')}{3}\left(\lambda_{\bv, 2} - \lambda_{\bu, 2}\right) + \frac{2(j-j')}{3}\left(\lambda_{\bu, 3} - \lambda_{\bv, 3}\right).
\end{align*}
Note that this gives a total weight of $2\left(2N+1 - j\right) + 2\left(2N+1 - j'\right) = 8N+4-2(j + j') = 8N + 4 - 2(3N+1) = 2N+2 = w_e$ for $e$.
\end{enumerate}
\begin{remark}
The correspondence between these valuation functions and the flows is as follows. For each component, for each vertex $v$ in the component and each label $i$, the coefficient for $\lambda_{v,i}$ gives the amount of flow of commodity $i$ going out from vertex $v$ for that component. For example, for the center downward cells, 
\begin{enumerate}
\item We have $\frac{2}{3}$ units of commodity $1$ flowing from $u$ and $v$ and $\frac{4}{3}$ units of commodity $1$ flowing into $w$.
\item We have $\frac{2}{3}$ units of commodity $2$ flowing from $u$ and $w$ and $\frac{4}{3}$ units of commodity $2$ flowing into $v$.
\item We have $\frac{2}{3}$ units of commodity $3$ flowing from $v$ and $w$ and $\frac{4}{3}$ units of commodity $3$ flowing into $u$.
\end{enumerate}
\end{remark}
\begin{proposition}
For each of the downward cells $C$, for any feasible solution to the vertex-consistency LP, $\sum_{e \in E(C)}{w_ep_e} \geq \val(C)$.
\end{proposition}
\begin{proof}
A solution to the vertex-consistency LP gives a local probability distribution for the labels of the vertices in $C$. Thus, it is sufficient to show that $\sum_{e \in E(C)}{w_ep_e} \geq \val(C)$ for any actual assignment of labels to the vertices.

For edges $e = \{\bu,\bv\}$ near corner $1$, $\val(e) = \frac{2}{3}\left(\lambda_{\bu, 1} - \lambda_{\bv, 1}\right) + \frac{1}{3}\left(\lambda_{\bv, 2} - \lambda_{\bu, 2}\right) + \frac{1}{3}\left(\lambda_{\bv, 3} - \lambda_{\bu, 3}\right)$. Observe that $\val(e) = 0$ if $\bu$ and $\bv$ have the same label and $\val(e) \leq 1$ regardless of the labels of $\bu$ and $\bv$. To see this, observe that the value given to $\bu$ is at most $\frac{2}{3}$ and the value given to $\bv$ is at most $\frac{1}{3}$ as $\bv$ can't have both label $2$ and label $3$. Similar arguments apply to all of the other edges.

For downward cells $C$ in the center, $\val(C) = 0$ if all of the vertices of $C$ have the same labels and $\val(C) \leq 2$ regardless of the labels of the vertices in $C$ as the value assigned to each vertex is at most $\frac{2}{3}$. Since at least two edges of $C$ are cut whenever the vertices of $C$ don't all have the same color, $\sum_{e \in E(C)}{w_ep_e} \geq \val(C)$.
\end{proof}
\begin{remark}
The gap between the objective value for the vertex-consistency LP and the objective value for the CKR relaxation comes from the center downward cells. In particular, The vertex-consistency LP captures the fact that if $\lambda_{u,3} = 0$, $\lambda_{v,2} = 0$, and $\lambda_{w,3} = 0$ then it is impossible for all three of the vertices to have the same color so there must be at least two cut edges. However, if $\lambda_{u,1} = \lambda_{u,2} = \lambda_{u,1} = \lambda_{u,2} = \lambda_{u,1} = \lambda_{u,2} = \frac{1}{2}$ then the CKR relaxation will only give a cost of $\frac{1}{2}$ to each of the three edges of this downward cell for a total cost of $\frac{3}{2}$.
\end{remark}
\begin{lemma}
For all feasible solutions to the vertex-consistency LP, $\sum_{C}{\val(C)} = 12N+4$.
\end{lemma}
\begin{proof}
Observe that corner $1$ appears as $\bu$ for two edges of weight $3N+1$ and must have label $1$ so the contribution from corner $1$ is $2(3N+1) \cdot \frac{2}{3} = \frac{12N+4}{3}$. Similarly, the other two corners also contribute $\frac{12N+4}{3}$ so the total contribution to $\sum_{C}{\val(C)}$ from the corners is $12N+4$.

The other vertices can be analyzed as follows:
\begin{enumerate}
\item Vertices in the triangle near corner $1$: Each vertex whose $1$-coordinate is greater than $2N$ which is not on a side of a triangle appears as $\bv$ in two edges of weight $1$ and appears as $\bu$ in two edges of weight $1$. 
Thus, the total contribution from this vertex is 
\[
2\left(-\frac{2}{3}\lambda_{\bu, 1} + \frac{1}{3}\lambda_{\bu, 2} + \frac{1}{3}\lambda_{\bu, 3}\right) + 2\left(\frac{2}{3}\lambda_{\bu, 1} - \frac{1}{3}\lambda_{\bu, 2} - \frac{1}{3}\lambda_{\bu, 3}\right) \EQ 0
\]
A similar argument holds for vertices in other corners.
\item Vertices in the central hexagon: Each vertex whose coordinates are all greater than $0$ and less than $2N$ appears as $\bw$ in one downward cell, appears as $\bv$ in one downward cell, and appears as $\bu$ in one downward cell. 
Thus, the total contribution from this vertex is 
\[
\left(-\frac{4}{3}\lambda_{\bu, 1} + \frac{2}{3}\lambda_{\bu, 2} + \frac{2}{3}\lambda_{\bu, 3}\right) + \left(\frac{2}{3}\lambda_{\bu, 1} - \frac{4}{3}\lambda_{\bu, 2} + \frac{2}{3}\lambda_{\bu, 3}\right) + \left(\frac{2}{3}\lambda_{\bu, 1} + \frac{2}{3}\lambda_{\bu, 2} - \frac{4}{3}\lambda_{\bu, 3}\right) = 0
\]
\item Vertices on a side of the triangle near a corner: Letting $\bv^{(j)}$ be the $j$th vertex on the path from corner $1$ to corner $2$, for all $j \in  [N - 1]$, $\bv^{(j)}$ appears as $\bv$ for an edge with weight $3N-j + 2$, appears as $\bu$ for an edge with weight $3N-j+1$, and appears as $\bu$ for an edge with weight $1$. Thus, the total contribution from $\bv^{(j)}$ is 
\begin{align*}
& (3N-j + 2)\left(-\frac{2}{3}\lambda_{\bv^{(j)}, 1} + \frac{1}{3}\lambda_{\bv^{(j)}, 2} + \frac{1}{3}\lambda_{\bv^{(j)}, 3}\right) \\
& \qquad + ((3N-j+1)+1)\left(\frac{2}{3}\lambda_{\bv^{(j)}, 1} - \frac{1}{3}\lambda_{\bv^{(j)}, 2} - \frac{1}{3}\lambda_{\bv^{(j)}, 3}\right) \EQ 0
\end{align*}
A similar argument holds for other vertices on a side of the triangle near a corner.
\item Vertices in the middle of a side of the triangle: Letting $\bv^{(j)}$ be the $j$th vertex on the path from corner $1$ to corner $2$, for all $j \in  [N,2N]$, we have the following contributions: 
\begin{enumerate}
\item[1.] For the edge $\{\bv^{(j-1)},\bv^{(j)}\}$, since $j' - j = (3N+1-j)-j = 3N+1-2j$, $\bv^{(j)}$  contributes 
\[
-\frac{3N+3+(3N+1-2j)}{3}\lambda_{\bv^{(j)}, 1} + \frac{3N+3+(2j-3N-1)}{3}\lambda_{\bv^{(j)}, 2} + \frac{2(3N+1-2j)}{3}\lambda_{\bv^{(j)}, 3}.
\]
\item[2.] For the edge $\{\bv^{(j)},\bv^{(j+1)}\}$, since $j'-1 - (j+1) = 3N-1-2j$, $\bv^{(j)}$  contributes 
\[
\frac{3N+3+(3N-1-2j)}{3}\lambda_{\bv^{(j)}, 1} - \frac{3N+3+(2j-3N+1)}{3}\lambda_{\bv^{(j)}, 2} - \frac{2(3N-1-2j)}{3}\lambda_{\bv^{(j)}, 3}.
\]
\item[3.] For the downward triangle containing $\bv^{(j)}$, $\bv^{(j)}$ contributes $\frac{2}{3}\lambda_{\bv^{(j)}, 1} + \frac{2}{3}\lambda_{\bv^{(j)}, 2}-\frac{4}{3}\lambda_{\bv^{(j)}, 3}$.
\end{enumerate}
The sum of these contributions is $0$. A similar argument holds for vertices which are in the middle of the other sides of the triangle. 
\item Vertices on the boundary between a corner triangle and the central hexagon: Each vertex whose $1$-coordinate is $2N$ and whose other coordinates are greater than $0$ appears as $\bv$ for two edges, appears as $\bv$ for one downward cell, and appears as $\bu$ for a downward cell. Thus, the total contribution from this vertex is 
\[
2\left(-\frac{2}{3}\lambda_{\bu, 1} + \frac{1}{3}\lambda_{\bu, 2} + \frac{1}{3}\lambda_{\bu, 3}\right) + \left(\frac{2}{3}\lambda_{\bu, 1} - \frac{4}{3}\lambda_{\bu, 2} + \frac{2}{3}\lambda_{\bu, 3}\right) + \left(\frac{2}{3}\lambda_{\bu, 1} + \frac{2}{3}\lambda_{\bu, 2} - \frac{4}{3}\lambda_{\bu, 3}\right) = 0
\]
A similar argument holds for other vertices which are on the boundary between a corner triangle and the central hexagon.
\end{enumerate}
\end{proof}
\begin{lemma}
The value of the CKR LP relaxation is at most $11N+4$.
\end{lemma}
\begin{proof}
Karger et al. \cite{KKSTY04} showed that the LP value for their instance is $11N+1$. Our modification increases this value by $3$ which gives an LP value of $11N+4$. 

An alternative direct argument is as follows. If we take the natural LP solution where the probability that a vertex is not assigned label $i$ is proportional to its distance from corner $i$, this gives a cost of $\frac{1}{3N}$ to each edge. The corresponding LP value will be $\frac{1}{3N}$ times the total weight of the edges.

We can count the total weight of the edges as follows:
\begin{enumerate}
\item Interior of the corner triangles: There are $2\left(\sum_{j=1}^{N-1}{j}\right) = N^2 - N$ edges in the interior of each corner triangle. Since there are $3$ corner triangles, this gives a total contribution of $3N^2 - 3N$.
\item Sides of the triangle near the corners: The sum of the weights of the $j$th edge along a side of the triangle for $j = 1,\ldots,N$ is $N(2N+1) + \sum_{j=1}^{N}{j} = N(2N+1) + \frac{N(N+1)}{2}$. There are six such segments which gives a total contribution of $15N^2 + 9N$.
\item Middle segments of the sides of the triangle: The sum of the weights of $j$th edge along a side of the triangle for $j=N+1,\ldots,2N$ is $N(2N+2)$. There are three such segments which gives a total contribution of $6N^2 + 6N$
\item Central hexagon: There are $9N^2$ edges which are in the interior of the central hexagon or are on the boundary between the central hexagon and a corner triangle. To see this, observe that for each side of the triangle, this region contains $3\left(\sum_{j=1}^{N}{j} + \sum_{j=1}^{N-1}{j}\right) = 3\left(\frac{N^2 + N}{2} + \frac{N^2 - N}{2}\right) = 3N^2$ edges which are parallel to this side of the triangle. This gives a total contribution of $3 \cdot 3N^2 = 9N^2$.
\end{enumerate}
Putting everything together, the total weight of the edges is 
\[
(3N^2 - 3N) + (15N^2 + 9N) + (6N^2 + 6N) + 9N^2 = 33N^2 + 12N
\]
which gives an LP value of $\frac{33N^2 + 12N}{3N} = 11N + 4$.
\end{proof}
Thus, we have a ratio of $\frac{12N+4}{11N+4}$ which goes to $\frac{12}{11}$ as $N \to \infty$. Note that when $n = 3N$, this matches the tight ratio of $\frac{12n+12}{11n+12}$ shown by \cite{CCT06} as $\frac{12n+12}{11n+12} = \frac{36N + 12}{33N + 12} = \frac{12N+4}{11N+4}$.
\subsection{Vertex-Based Derivation of the $k=3$ Non-Opposite Lower Bound}
We use a slight modification of the instance used by Angelidakis, Makarychev, and Manurangsi \cite{AMM17} to prove a lower bound of $\frac{6}{5}$ for $k=3$ for non-opposite cuts as this modification allows us to obtain a slightly better lower bound for an instance of a given size. Throughout this subsection, we consider the scaled discretized simplex $(3N)\Delta_{3, 3N}$ for some $N \in \mathbb{N}$.
\begin{enumerate}
\item Along the sides of the simplex, if we consider the edge $e$ which is the $j$th edge from one corner and the $j'$th edge from another corner where $j' = 3N+1-j$, we take $w_e = N+2x - \min{\{j,j',N\}}$ for some $x \in [0,2/3]$. 

As we will see from the analysis, choosing a higher $x$ gives a better lower bound but we must choose $x \leq \frac{2}{3}$. Note that $x = \frac{1}{2}$ gives the lower bound instance from \cite{AMM17}.
\item For all edges $e$ which are not on the side of the simplex which have an endpoint with a coordinate greater than $2N$ (i.e., an endpoint of the edge is within distance $N-1$ of the closest corner), we set $w_e = 1$ if this coordinate changes along $e$ (i.e., the edge is parallel to a side of the triangle between the closest corner and one of the other two corners) and $0$ otherwise (i.e., the edge is parallel to the side of the triangle between the other corners).
\item All other edges have weight $1$.
\end{enumerate}

\begin{figure}[h]
\centering
\begingroup
\definecolor{noewblue}{HTML}{287A92}
\definecolor{noewzero}{HTML}{B66D35}
\definecolor{noewboundary}{HTML}{26384B}
\begin{tikzpicture}[
  x=1cm,y=1cm,line cap=round,line join=round,
  every node/.style={font=\small},
  noew unit/.style={draw=noewblue,line width=.8pt},
  noew zero/.style={draw=noewzero,line width=.9pt,
    dash pattern=on 2.1pt off 1.6pt},
  noew boundary/.style={draw=noewboundary,line width=1.35pt},
  noew note/.style={font=\footnotesize,text=black!65,align=center}
]
\def\noewN{4}
\def\noewn{12}
\pgfmathsetmacro{\noewheight}{6*sqrt(3)*.62}

\newcommand{\noewDrawEdge}[4]{%
  \pgfmathtruncatemacro{\noewa}{\noewn-#1}%
  \pgfmathtruncatemacro{\noewb}{#1-#2}%
  \pgfmathtruncatemacro{\noewc}{#2}%
  \pgfmathtruncatemacro{\noewd}{\noewn-#3}%
  \pgfmathtruncatemacro{\noewe}{#3-#4}%
  \pgfmathtruncatemacro{\noewf}{#4}%
  \pgfmathtruncatemacro{\noewisboundary}{%
    (\noewa==0 && \noewd==0) ||
    (\noewb==0 && \noewe==0) ||
    (\noewc==0 && \noewf==0)}%
  \pgfmathtruncatemacro{\noewiszero}{%
    (\noewa==\noewd && \noewa>2*\noewN) ||
    (\noewb==\noewe && \noewb>2*\noewN) ||
    (\noewc==\noewf && \noewc>2*\noewN)}%
  \ifnum\noewisboundary=1
    \def\noewstyle{noew boundary}%
  \else
    \ifnum\noewiszero=1
      \def\noewstyle{noew zero}%
    \else
      \def\noewstyle{noew unit}%
    \fi
  \fi
  \draw[\noewstyle]
    (\noewprefix-#1-#2)--(\noewprefix-#3-#4);
}

\begin{scope}
  \def\noewprefix{A}
  \foreach \r in {0,...,12}{
    \foreach \s in {0,...,\r}{
      \coordinate (A-\r-\s) at
        ({.62*(\s-\r/2)},{\noewheight-.62*sqrt(3)*\r/2});
    }
  }
  \foreach \a/\b/\c in {0-0/4-0/4-4,12-0/8-0/12-4,12-12/8-8/12-8}{
    \path[fill=noewblue!20] (A-\a)--(A-\b)--(A-\c)--cycle;
  }
  \foreach \r in {0,...,12}{
    \foreach \s in {0,...,\r}{
      \ifnum\s<\r
        \pgfmathtruncatemacro{\sn}{\s+1}
        \noewDrawEdge{\r}{\s}{\r}{\sn}
      \fi
      \ifnum\r<12
        \pgfmathtruncatemacro{\rn}{\r+1}
        \pgfmathtruncatemacro{\sn}{\s+1}
        \noewDrawEdge{\r}{\s}{\rn}{\s}
        \noewDrawEdge{\r}{\s}{\rn}{\sn}
      \fi
    }
  }
  \foreach \r in {0,...,12}{
    \foreach \s in {0,...,\r}{
      \fill[noewboundary!65] (A-\r-\s) circle[radius=.65pt];
    }
  }
  \foreach \p in {0-0,12-0,12-12}
    \fill[noewboundary] (A-\p) circle[radius=1.7pt];
  \node[above=4pt] at (A-0-0) {$\be_1$};
  \node[left=5pt] at (A-12-0) {$\be_2$};
  \node[right=5pt] at (A-12-12) {$\be_3$};
  \foreach \j in {1,...,12}{
    \pgfmathtruncatemacro{\jm}{\j-1}
    \pgfmathtruncatemacro{\offset}{4-min(\j,13-\j,4)}
    \ifnum\offset=0
      \def\weight{2x}
    \else
      \edef\weight{\offset+2x}
    \fi
    \node[font=\scriptsize,text=noewboundary,rotate=55,anchor=east,yshift=-4pt]
      at ($(A-12-\jm)!.5!(A-12-\j)$) {$\weight$};
  }
\end{scope}

\begin{scope}[yshift=-1.75cm]
  \draw[noew boundary] (-3.5,0)--(-2.9,0);
  \node[anchor=west,font=\footnotesize] at (-2.78,0) {boundary};
  \draw[noew unit] (-.65,0)--(-.05,0);
  \node[anchor=west,font=\footnotesize] at (.07,0) {$w_e=1$};
  \draw[noew zero] (1.85,0)--(2.45,0);
  \node[anchor=west,font=\footnotesize] at (2.57,0) {$w_e=0$};
\end{scope}
\end{tikzpicture}
\endgroup
\caption{An illustration for the edge weights when $N = 4$. }
\end{figure}
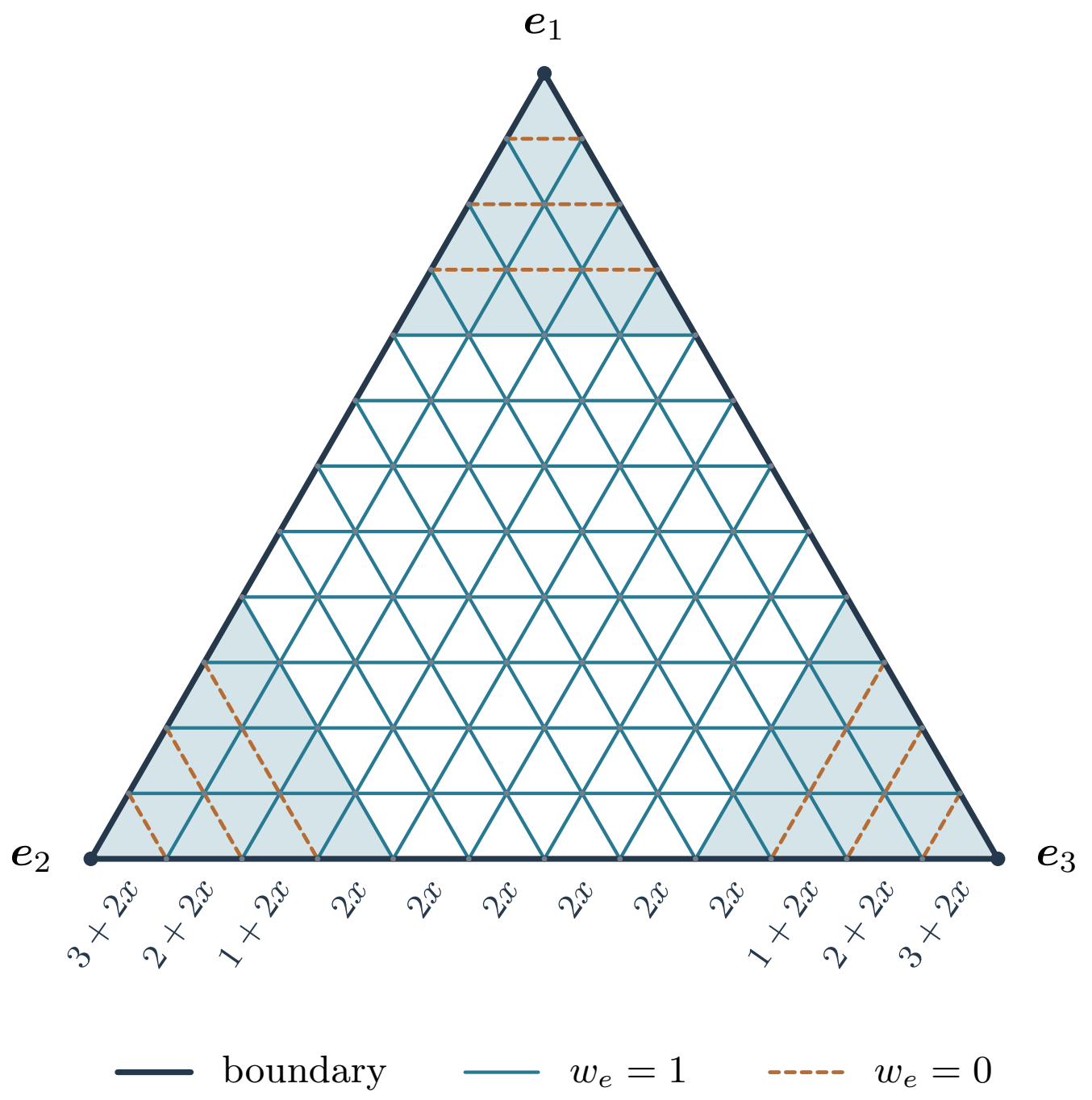

This instance can be analyzed using the following building blocks. While the edge weights in the central hexagon are uniform, instead of treating it as a single region, we split it into six triangles where three of these triangles are mirror images of the corner triangles and three of these triangles are upward non-corner triangles which have a side which is the middle segment of the path between two of the corners.
\begin{enumerate}
\item Edges near a corner: For each edge $e$ which has an endpoint whose $1$-coordinate is greater than $2N$, orienting the edge $e = \{\bu,\bv\}$ so that it points away from the corner (i.e., $u_1 > v_1$), we take $\val(e) = \lambda_{\bu, 1} - \lambda_{\bv, 1}$
We take analogous valuations for the edges which are close to corner $2$ or corner $3$.
\item Edges in a middle segment: For each edge $e$ in the middle segment of the path between corner $1$ and corner $2$, (i.e., both endpoints of $e$ have $1$-coordinate and $2$-coordinate between~$N$ and~$2N$), orienting the edge $e = \{\bu,\bv\}$ so that it points away from the corner (i.e., $\bu$ has a larger $1$-coordinate than $\bv$), we take $\val(e) = \frac{1}{2}\left(\lambda_{\bu, 1} - \lambda_{\bv, 1}\right) + \frac{1}{2}\left(\lambda_{\bv, 2} - \lambda_{\bu, 2}\right)$.
We take analogous valuations for the edges which are in the other middle segments.
\item Downward cells in a triangle which is the mirror image of a corner triangle: For each center downward cell $C$ which is in the mirror image of the corner triangle for corner $1$ (i.e., for all vertices $\bv \in C$, the $1$-coordinate of $\bv$ is at most $2N$ and the $2$-coordinate and $3$-coordinate of $\bv$ are at most $N$), orient the cell $C = \{\bu,\bv,\bw\}$ so that $\bu$ is farthest from corner $3$, $\bv$ is farthest from corner $2$, and $\bw$ is farthest from corner $1$. 
\begin{enumerate}
\item[1.] If $C$ is not on the left or right side of this triangle (i.e., $w_2, w_3 < N$) then we take 
\[
\val(C) \EQ \left(\lambda_{\bu, 1} + \lambda_{\bv, 1} - 2\lambda_{\bw, 1}\right) + \left(\lambda_{\bu, 2} - \lambda_{\bv, 2}\right) + \left(\lambda_{\bv, 3} - \lambda_{\bu, 3}\right).
\]
\item[2.] If $C$ is on the left side but not the right side of this triangle (i.e., $w_2 = N$ and $w_3 < N$) then we take 
\begin{align*}
\val(C) \EQ & \left(x\lambda_{\bu, 1} + \lambda_{\bv, 1} - (1+x)\lambda_{\bw, 1}\right) + \left(x\lambda_{\bu, 2} - \lambda_{\bv, 2} + (1-x)\lambda_{\bw, 2}\right) \\
& \qquad \qquad + \left((x-2)\lambda_{\bu, 3} + \lambda_{\bv, 3} + (1-x)\lambda_{\bw, 3}\right).
\end{align*}
Similarly, if $C$ is on the right side but not the left side of this triangle (i.e., $w_3 = N$ and $w_2 < N$) then we take 
\begin{align*}
\val(C) \EQ & \left(\lambda_{\bu, 1} + x\lambda_{\bv, 1} - (1+x)\lambda_{\bw, 1}\right) + \left(\lambda_{\bu, 2} + (x-2) \lambda_{\bv, 2} + (1-x)\lambda_{\bw, 2}\right) \\
& \qquad \qquad + \left(-\lambda_{\bu, 3} + x\lambda_{\bv, 3} + (1-x)\lambda_{\bw, 3}\right).
\end{align*}
\item[3.] If $C$ is the cell touching the center of the entire triangle (i.e., $\bw = (N,N,N)$) then we take 
\begin{align*}
\val(C) \EQ & \left(x\lambda_{\bu, 1} + x\lambda_{\bv, 1} - 2x\lambda_{\bw, 1}\right) + \left(x\lambda_{\bu, 2} + (x-2)\lambda_{\bv, 2} + (2-2x)\lambda_{\bw, 2}\right) \\
& \qquad \qquad + \left((x-2)\lambda_{\bu, 3} + x\lambda_{\bv, 3} + (2-2x)\lambda_{\bw, 3}\right).
\end{align*}
\end{enumerate}
We take analogous valuations for the downward cells which are in triangles which are mirror images of the corner triangles near corner $2$ and corner $3$.
\item Downward cells in upward non-corner triangles: For each center downward cell $C$ which is in triangle which has the middle segment of the path between corner $1$ and corner $2$ as one of its sides (i.e., for all vertices $\bv \in C$, $v_1, v_2 \geq N$), orient the cell $C = \{\bu,\bv,\bw\}$ so that $\bu$ is farthest from corner $3$, $\bv$ is farthest from corner $2$, and $\bw$ is farthest from corner $1$. We take 
\[
\val(C) \EQ \left(\lambda_{\bv, 1} - \lambda_{\bw, 1}\right) + \left(\lambda_{\bw, 2} - \lambda_{\bv, 2}\right) + \left(-2\lambda_{\bu, 3} + \lambda_{\bv, 3} + \lambda_{\bw, 3}\right).
\]
We take analogous valuations for the downward cells which are in the other upward non-corner triangles.
\end{enumerate}
\begin{proposition}
For each of these downward cells $C$, for any feasible solution to the vertex-consistency LP, $\sum_{e \in E(C)}{w_ep_e} \geq \val(C)$.
\end{proposition}
\begin{proof}[Proof sketch]
This can be shown in a similar way as before by checking that 
\begin{enumerate}
\item For each edge $e$, the maximum coefficient for $\lambda_{\bu, 1},\lambda_{\bu, 2},\lambda_{\bu, 3}$ (which is the maximum value given to $\bu$) plus the maximum coefficient for $\lambda_{\bv, 1},\lambda_{\bv, 2},\lambda_{\bv, 3}$ (which is the maximum value given to $\bv$) is at most $1$.
\item For each downward cell $C$, the maximum coefficient for $\lambda_{\bu, 1},\lambda_{\bu, 2},\lambda_{\bu, 3}$ (which is the maximum value given to $\bu$) plus the maximum coefficient for $\lambda_{\bv, 1},\lambda_{\bv, 2},\lambda_{\bv, 3}$ (which is the maximum value given to $\bv$) plus the maximum coefficient for $\lambda_{\bw, 1},\lambda_{\bw, 2},\lambda_{\bw, 3}$ (which is the maximum value given to $\bw$) is at most $2$. For example, for the downward cells containing the center vertex $(N,N,N)$, this sum is $x + x + (2-2x) = 2$.
\end{enumerate}
\end{proof}
\begin{lemma}
For all feasible solutions to the vertex-consistency LP, $\sum_{C}{\val(C)} \geq 6N+(12x-6)$. If $x = \frac{2}{3}$, this is always an equality.
\end{lemma}
\begin{proof}
Observe that corner $1$ appears as $\bu$ for two edges of weight $N+2x-1$ and must have label $1$ so the contribution from corner $1$ is $2(N+2x-1)$. Similarly, the other two corners also contribute $2(N+2x-1)$ so the total contribution to $\sum_{C}{\val(C)}$ from the corners is $6(N+2x-1) = 6N + (12x-6)$.

The other vertices can be analyzed as follows:
\begin{enumerate}
\item Vertices in the triangle near corner $1$: Each vertex whose $1$-coordinate is greater than $2N$ which is not on a side of a triangle appears as $\bv$ in two edges of weight $1$ and appears as $\bu$ in two edges of weight $1$. 
Thus, the total contribution from this vertex is 
$2(-\lambda_{\bu, 1}) + 2\lambda_{\bu, 1} = 0$.
A similar argument holds for vertices in other corners.
\item Vertices in a triangle mirroring a corner triangle: For each vertex in the triangle which is the mirror image of the corner triangle for corner $1$ (i.e., the $1$ coordinate of the vertex is less than~$2N$ and the $2$-coordinate and $3$-coordinate of the vertex are less than $N$) appears as $\bw$ in one downward cell, appears as $\bv$ in one downward cell, and appears as $\bu$ in one downward cell. 
Thus, the total contribution from this vertex is 
\[
(-2\lambda_{\bu, 1}) + \left(\lambda_{\bu, 1} - \lambda_{\bu, 2} + \lambda_{\bu, 3}\right) + \left(\lambda_{\bu, 1} + \lambda_{\bu, 2} - \lambda_{\bu, 3}\right) \EQ 0
\]
A similar argument holds for vertices in other triangles mirroring a corner triangle.
\item Vertices in an upward non-corner triangle: For each vertex in the triangle which has the middle segment of the path from corner $1$ to corner $2$ as a side (i.e., the $1$-coordinate and $2$-coordinate of the vertex are both greater than $N$ and the $3$-coordinate is greater than 0), this vertex appears as $\bw$ in one downward cell, appears as $\bv$ in one downward cell, and appears as $\bu$ in one downward cell. Thus, the total contribution from this vertex is 
\[
(-\lambda_{\bu, 1} + \lambda_{\bu, 2} + \lambda_{\bu, 3}) + \left(\lambda_{\bu, 1} - \lambda_{\bu, 2} + \lambda_{\bu, 3}\right) + \left(-2\lambda_{\bu, 3}\right) \EQ 0
\]
\item Vertices on a side of the triangle near a corner: Letting $\bv^{(j)}$ be the $j$th vertex on the path from corner $1$ to corner $2$, for all $j \in  [N-1]$, $\bv^{(j)}$ appears as $\bv$ for an edge with weight $N+2x-j$, appears as $\bu$ for an edge with weight $N+2x-j-1$, and appears as $\bu$ for an edge with weight $1$. Thus, the total contribution from $\bv^{(j)}$ is 
\[
(N+2x-j)(-\lambda_{\bv^{(j)}, 1}) + (N+2x-j-1)\lambda_{\bv^{(j)}, 1} + \lambda_{\bv^{(j)}, 1} \EQ 0
\]
A similar argument holds for other vertices on a side of the triangle near a corner.
\item Vertices in the middle of a side of the triangle: Letting $\bv^{(j)}$ be the $j$th vertex on the path from corner $1$ to corner $2$, for all $j \in  [N+1,2N-1]$, we have the following contributions: 
\begin{enumerate}
\item[1.] The edge $\{\bv^{(j-1)},\bv^{(j)}\}$ has weight $2x$. For this edge, $\bv^{(j)}$ contributes $2x\frac{\lambda_{\bv^{(j)}, 2}-\lambda_{\bv^{(j)}, 1}}{2} = x(\lambda_{\bv^{(j)}, 2}-\lambda_{\bv^{(j)}, 1})$.
\item[2.] The edge $\{\bv^{(j)},\bv^{(j+1)}\}$ has weight $2x$. For this edge, $\bv^{(j)}$ contributes $2x\frac{\lambda_{\bv^{(j)}, 1}-\lambda_{\bv^{(j)}, 2}}{2} = x(\lambda_{\bv^{(j)}, 1}-\lambda_{\bv^{(j)}, 2})$.
\item[3.] For the downward triangle containing $\bv^{(j)}$, $\bv^{(j)}$ contributes $-2\lambda_{\bv^{(j)}, 3}$ which is $0$ because~$\bv^{(j)}$ is on the side opposite corner $3$ so $\lambda_{\bv^{(j)}, 3} = 0$.
\end{enumerate}
The sum of these contributions is $0$. A similar argument holds for vertices which are in the middle of the other sides of the triangle. 
\item Vertices on the boundary between a corner triangle and the mirror image of this triangle: Each vertex whose $1$-coordinate is $2N$ and whose other coordinates are greater than $0$ appears as $\bv$ for two edges, appears as $\bv$ for one downward cell, and appears as $\bu$ for a downward cell. Thus, the total contribution from this vertex is 
\[
-2\lambda_{\bu, 1} + (\lambda_{\bu, 1} - \lambda_{\bu, 2} + \lambda_{\bu, 3}) + (\lambda_{\bu, 1} + \lambda_{\bu, 2} - \lambda_{\bu, 3}) \EQ 0
\]
A similar argument holds for vertices which are on the boundary between another corner triangle and its mirror image.
\item Vertices on the boundary between the mirror image of a corner triangle and a non-corner upwards triangle: Each vertex whose $1$-coordinate is greater than $N$ and less than $2N$ and whose $2$-coordinate is equal to $N$ appears as $\bw$ in one downward cell of the mirror triangle, appears as $\bu$ in one downward cell of the mirror triangle, and appears as $\bv$ in one downward cell of the non-corner upwards triangle (note that the coefficients for the first two downward cells are different). Thus, the total contribution from this vertex is 
\begin{align*}
& \left((-1-x)\lambda_{\bu, 1} + (1-x)\lambda_{\bu, 2} + (1-x)\lambda_{\bu, 3}\right) + \\
& \qquad \left(x\lambda_{\bu, 1} + x\lambda_{\bu, 2} + (x-2)\lambda_{\bu, 3}\right) + \left(\lambda_{\bu, 1} - \lambda_{\bu, 2} + \lambda_{\bu, 3}\right) \EQ 0
\end{align*}
A similar argument holds for other vertices which are on the boundary between a mirror triangle and a non-corner upwards triangle.
\item Vertices at the intersection of a corner triangle, mirror triangle, and non-corner upwards triangle. For the vertex with coordinates $(2N,N,0)$, this vertex appears as $\bv$ for an edge in the corner triangle with weight $2x$, appears as $\bu$ for an edge in the non-corner upwards triangle with weight $2x$, and appears as $\bu$ in a downward cell of the mirror triangle. Thus, the total contribution from this vertex is 
\[
-2x\lambda_{\bu, 1} + 2x\frac{\lambda_{\bu, 1}-\lambda_{\bu, 2}}{2} + \left(x\lambda_{\bu, 1} + x\lambda_{\bu, 2} + (x-2)\lambda_{\bu, 3}\right) = (x-2)\lambda_{\bu, 3}
\]
which is $0$ because this vertex is on the side opposite corner $3$ so $\lambda_{\bu, 3} = 0$.

A similar argument holds for other vertices which are corners of three triangles.
\item For the center vertex $(N,N,N)$, the total contribution is $(-2x + (2-2x) + (2-2x))(\lambda_{\bu, 1} + \lambda_{\bu, 2} + \lambda_{\bu, 3})$ which is $0$ if $x = \frac{2}{3}$ and is guaranteed to be greater than or equal to $0$ as long as $x \leq \frac{2}{3}$.
\end{enumerate}
\end{proof}
\begin{lemma}
The value of the CKR LP relaxation is at most $5N + 6x - 2$.
\end{lemma}
\begin{proof}
Angelidakis, Makarychev, and Manurangsi \cite{AMM17} showed that the LP value for their instance is $5N + 1$. Our modification changes this value by $(6x-3)$ which gives an LP value of $5N + 6x - 2$.

An alternative direct argument is as follows. If we take the natural LP solution where the probability that a vertex is not assigned label $i$ is proportional to its distance from corner $i$, this gives a cost of $\frac{1}{3N}$ to each edge. The corresponding LP value will be $\frac{1}{3N}$ times the total weight of the edges.

We can count the total weight of the edges as follows:
\begin{enumerate}
\item Interior of the corner triangles: There are $2\left(\sum_{j=1}^{N-1}{j}\right) = N^2 - N$ edges in the interior of each corner triangle. Since there are $3$ corner triangles, this gives a total contribution of $3N^2 - 3N$.
\item Sides of the triangle near the corners: The sum of the weights of the $j$th edge along a side of the triangle for $j = 1,\ldots,N$ is $2xN + \sum_{j=1}^{N-1}{j} = 2Nx + \frac{N(N-1)}{2}$. There are six such segments which gives a total contribution of $3N^2 + (12x - 3)N$.
\item Middle segments of the sides of the triangle: The sum of the weights of $j$th edge along a side of the triangle for $j=N+1,\ldots,2N$ is $2xN$. There are three such segments which gives a total contribution of $6xN$
\item Central hexagon: As before, there are $9N^2$ edges which are in the interior of the central hexagon or are on the boundary between the central hexagon and a corner triangle (the edge weights in the central hexagon are uniform so we can treat it as a single region here).
\end{enumerate}
Putting everything together, the total weight of the edges is 
\[
(3N^2 - 3N) + (3N^2 + 12xN - 3N) + 6xN + 9N^2 \EQ 15N^2 +18xN - 6N
\]
which gives an LP value of $\frac{15N^2 + 18xN - 6N}{3N} = 5N + 6x - 2$.
\end{proof}
Setting $x = \frac{2}{3}$ gives a vertex-consistency LP value of $6N+2$ and a CKR LP value of $5N+2$ for a ratio of $\frac{6N+2}{5N+2}$ which goes to $\frac{6}{5}$ as $N \to \infty$.
\section{Calculations for Determinants in Table~\ref{tab:determinants}}\label{app:determinants}

We reproduce the content of Table~\ref{tab:determinants} here.

\begin{center}
\begin{tabular}{|Sl|Sl|}
\hline
Internal face & Determinant value \\
\hline
$\conv(\bo,\bv,\bv_{34})$
    & $\displaystyle
        \beta \left(\alpha - \frac{1}{4}\right)(1-2q_c)$ \\ \hline

$\conv(\bo,\bh_t,\bv)$
    & $\displaystyle
       \frac{1-2\beta}{4}(1-2\alpha-\beta) \left(p_t-\frac12\right)$ \\\hline

$\conv(\bo,\bh_c,\bv)$
    & $\begin{aligned}[t]
       -\frac{4\alpha-1}{12}\bigl[
          &3(\alpha-1+2\beta)p_c+1-\alpha-3\beta+3\beta q_c\bigr]
       \end{aligned}$ \\\hline

$\conv(\bo,\bv,\bv_{12})$
    & $\displaystyle
       \frac{2\alpha+\beta-1}{2}\left((1- 2\beta) - (1-2\beta)p_t - p_s\right)$ \\\hline

$\conv(\bo,\bv,\bv_{23})$
    & $\displaystyle
       \frac{1-\alpha-2\beta}{2}
        \bigl[(1-4\alpha)p_c+2\alpha-1+q_s\bigr]$ \\\hline

$\conv(\bo,\bh_s,\bv)$
    & $\begin{aligned}[t]
       \frac1{12}\bigl[
          &(6\alpha-3+3\beta)p_s+(3\alpha-3+6\beta)q_s
             +2-3\alpha-3\beta\bigr]
       \end{aligned}$ \\
\hline
\end{tabular}
\end{center}

We start by considering the face $\conv(\bo, \bv, \bv_{34})$, shared by the tetrahedra $\conv(\bo, \bh_c, \bv, \bv_{34})$ and $\conv(\bo, \bh_t, \bv, \bv_{34})$ with density vectors 
\[(0, 1-2q_c,q_c,q_c)\quad \text{and} \quad (0,0,1/2,1/2)\] respectively. We have $\bv = (\alpha, 1-\alpha-\beta,\beta,0)$ and $\bv_{34} = (\alpha, 1-\alpha-\beta,0,\beta)$, so Equation~\eqref{eq:face_condition_det} gives 
\begin{align*}
    \det\left(\begin{array}{ccc}
    \alpha - \frac{1}{4} & 1 - \alpha - \beta - \frac{1}{4} & \beta - \frac{1}{4} \\
    \alpha - \frac{1}{4} & 1 - \alpha - \beta - \frac{1}{4} & -\frac{1}{4} \\
    0& 1 - 2q_c &q_c - \frac{1}{2}\end{array}\right) & \EQ 
    \det\left(\begin{array}{ccc}
    \alpha - \frac{1}{4} & 1 - \alpha - \beta - \frac{1}{4} & \beta - \frac{1}{4} \\
    0 & 0 & -\beta \\
    0& 1 - 2q_c &q_c - \frac{1}{2}\end{array}\right) \\
    & \EQ \beta \left(\alpha - \frac{1}{4}\right)(1-2q_c).
\end{align*}
The face $\conv(\bo,\bh_t,\bv)$ is shared by the tetrahedra $\conv(\bo,\bh_t,\bv,\bv_{12})$ and $\conv(\bo,\bh_t,\bv,\bv_{34})$, with density vectors
\[
    (0,0,p_t,1-p_t) \quad \text{and} \quad(0,0,1/2,1/2)
\]
respectively. We have $\bh_t = ((1-\beta)/2, (1-\beta)/2, \beta/2, \beta/2)$, so Equation~\eqref{eq:face_condition_det} gives
\begin{align*}
    \det\left(\begin{array}{ccc}
        \frac{1-2\beta}{4} & \frac{1-2\beta}{4} & \frac{2\beta-1}{4} \\
        \alpha-\frac14 & 1-\alpha-\beta-\frac14 & \beta-\frac14 \\
        0 & 0 & p_t-\frac12
    \end{array}\right)
    &\EQ
    \det\left(\begin{array}{ccc}
        \frac{1-2\beta}{4} & 0 & \frac{2\beta-1}{4} \\
        \alpha-\frac14 & 1-2\alpha-\beta & \beta-\frac14 \\
        0 & 0 & p_t-\frac12
    \end{array}\right)\\
    & \EQ \frac{1-2\beta}{4}(1-2\alpha-\beta) \left(p_t-\frac12\right).
\end{align*}
The face $\conv(\bo,\bh_c,\bv)$ is shared by the tetrahedra $\conv(\bo,\bh_c,\bv,\bv_{23})$ and $\conv(\bo,\bh_c,\bv,\bv_{34})$, with density vectors
\[
    (0,p_c,p_c,1-2p_c) \quad \text{and} \quad (0,1-2q_c,q_c,q_c)
\]
respectively. We have $\bh_c = (\alpha, (1-\alpha)/3, (1-\alpha)/3, (1-\alpha)/3)$, so Equation~\eqref{eq:face_condition_det} gives
\begin{align*}
    & \det\left(\begin{array}{ccc}
        \alpha-\frac14 & \frac{1-\alpha}{3} -\frac14 & \frac{1-\alpha}{3}-\frac14 \\
        \alpha-\frac14 & 1-\alpha-\beta-\frac14 & \beta-\frac14 \\
        0 & p_c-1+2q_c & p_c-q_c
    \end{array}\right) \\
    \EQ &\det\left(\begin{array}{ccc}
        \alpha-\frac14 & \frac{1-\alpha}{3}-\frac14
            & \frac{1-\alpha}{3}-\frac14 \\
        0 & \frac{2(1-\alpha)}3-\beta
            & \beta-\frac{1-\alpha}{3} \\
        0 & p_c-1+2q_c & p_c-q_c
    \end{array}\right)\\
    \EQ &\left(\alpha-\frac14\right)
    \left[
        \left(\frac{2(1-\alpha)}3-\beta\right)(p_c-q_c)
        -\left(\beta-\frac{1-\alpha}{3}\right)(p_c-1+2q_c)
    \right]\\
    \EQ & \frac{4\alpha-1}{12} \bigl[3(1-\alpha-2\beta)p_c -(1-\alpha-3\beta)-3\beta q_c\bigr].
\end{align*}

For the face $\conv(\bo,\bv,\bv_{12})$, the adjacent tetrahedra
are $\conv(\bo,\bh_s,\bv,\bv_{12})$ and
$\conv(\bo,\bh_t,\bv,\bv_{12})$, with density vectors
\[
    (p_s,p_s,1-2p_s,0) \quad \text{and} \quad (0,0,p_t,1-p_t),
\]
respectively. Since
$\bv_{12}=(1-\alpha-\beta,\alpha,\beta,0)$,
Equation~\eqref{eq:face_condition_det} gives
\begin{align*}
    &\det\left(\begin{array}{ccc}
        \alpha-\frac14 & 1-\alpha-\beta-\frac14
            & \beta-\frac14 \\
        1-\alpha-\beta-\frac14 & \alpha-\frac14
            & \beta-\frac14 \\
        p_s & p_s & 1-2p_s-p_t
    \end{array}\right)\\
    \EQ&
    \det\left(\begin{array}{ccc}
        \alpha-\frac14 & 1-\alpha-\beta-\frac14
            & \beta-\frac14 \\
        1-2\alpha-\beta & 2\alpha+\beta-1 & 0 \\
        p_s & p_s & 1-2p_s-p_t
    \end{array}\right)\\
    \EQ&
    \det\left(\begin{array}{ccc}
        \frac12-\beta & 1-\alpha-\beta-\frac14
            & \beta-\frac14 \\
        0 & 2\alpha+\beta-1 & 0 \\
        2p_s & p_s & 1-2p_s-p_t
    \end{array}\right)\\
    \EQ&
    (2\alpha+\beta-1)
    \left[
        \left(\frac12-\beta\right)(1-2p_s-p_t)
        -2\left(\beta-\frac14\right)p_s
    \right]\\
    \EQ&
    (2\alpha+\beta-1)
    \left[
        \left(\frac12-\beta\right)(1-p_t)-\frac{p_s}{2}
    \right]\\
    \EQ&\frac{2\alpha+\beta-1}{2}\left((1- 2\beta) - (1-2\beta)p_t - p_s\right).
\end{align*}

For the face $\conv(\bo,\bv,\bv_{23})$, the adjacent tetrahedra are $\conv(\bo,\bh_c,\bv,\bv_{23})$ and $\conv(\bo,\bh_s,\bv,\bv_{23})$, with density vectors
\[
    (0,p_c,p_c,1-2p_c) \quad \text{and} \quad (1-2q_s,q_s,q_s,0),
\]
respectively. Since $\bv_{23}=(\alpha,\beta,1-\alpha-\beta,0)$, Equation~\eqref{eq:face_condition_det} gives 
\begin{align*}
    &\det\left(\begin{array}{ccc}
        \alpha-\frac14 & 1-\alpha-\beta-\frac14
            & \beta-\frac14 \\
        \alpha-\frac14 & \beta-\frac14
            & 1-\alpha-\beta-\frac14 \\
        2q_s-1 & p_c-q_s & p_c-q_s
    \end{array}\right)\\
    \EQ&
    \det\left(\begin{array}{ccc}
        \alpha-\frac14 & 1-\alpha-\beta-\frac14
            & \beta-\frac14 \\
        0 & \alpha+2\beta-1 & 1-\alpha-2\beta \\
        2q_s-1 & p_c-q_s & p_c-q_s
    \end{array}\right)\\
    \EQ&
    \det\left(\begin{array}{ccc}
        \alpha-\frac14 & \frac12-\alpha & \beta-\frac14 \\
        0 & 0 & 1-\alpha-2\beta \\
        2q_s-1 & 2(p_c-q_s) & p_c-q_s
    \end{array}\right)\\
    \EQ&
    -(1-\alpha-2\beta)
    \det\left(\begin{array}{cc}
        \alpha-\frac14 & \frac12-\alpha\\
        2q_s-1 & 2(p_c-q_s)
    \end{array}\right)\\
    \EQ&
    (1-\alpha-2\beta)
    \left[
        \left(\frac12-\alpha\right)(2q_s-1)
        -2\left(\alpha-\frac14\right)(p_c-q_s)
    \right]\\
    \EQ&
    (1-\alpha-2\beta)
    \left[
        \left(\frac12-2\alpha\right)p_c
        +\alpha-\frac12+\frac{q_s}{2}
    \right]\\
    \EQ&
    \frac{1-\alpha-2\beta}{2}
    \bigl[(1-4\alpha)p_c+2\alpha-1+q_s\bigr].
\end{align*}

Finally, the face $\conv(\bo,\bh_s,\bv)$ is adjacent to $\conv(\bo,\bh_s,\bv,\bv_{12})$ and $\conv(\bo,\bh_s,\bv,\bv_{23})$, with density vectors
\[
    (p_s,p_s,1-2p_s,0)
    \quad\text{and}\quad
    (1-2q_s,q_s,q_s,0),
\]
respectively. Since $\bh_s=(1/3,1/3,1/3,0)$,
Equation~\eqref{eq:face_condition_det} gives
\begin{align*}
    &\det\left(\begin{array}{ccc}
        \frac1{12} & \frac1{12} & \frac1{12} \\
        \alpha-\frac14 & 1-\alpha-\beta-\frac14
            & \beta-\frac14 \\
        p_s-1+2q_s & p_s-q_s & 1-2p_s-q_s
    \end{array}\right)\\
    \EQ&
    \det\left(\begin{array}{ccc}
        \frac1{12} & 0 & 0 \\
        \alpha-\frac14 & 1-2\alpha-\beta & \beta-\alpha \\
        p_s-1+2q_s & 1-3q_s & 2-3p_s-3q_s
    \end{array}\right)\\
    \EQ&
    \frac1{12}\bigl[
        (1-2\alpha-\beta)(2-3p_s-3q_s)
        -(\beta-\alpha)(1-3q_s)
    \bigr]\\
    \EQ&
    \frac1{12}\bigl[
        (6\alpha-3+3\beta)p_s
        +(3\alpha-3+6\beta)q_s
        +2-3\alpha-3\beta
    \bigr].
\end{align*}

\section{Some Degrees of Freedom for the Vertex-consistency LP}\label{app:freedom_vertex_lp}

We identify here some degrees of freedom in feasible solutions to \eqref{LP:dual-vertex}. They may be used to justify some simplifying assumptions on the optimal solution. 

\begin{lemma}\label{lem:lp_shift}
    Let $(y, z)$ be any feasible solution to \eqref{LP:dual-vertex}. Let $(m_{\bv, C})_{\bv \in \Delta_{k, n}, C \in \cD_{k, n}}$ be such that $\sum_{C: \bv \in C} m_{\bv, C} = 0$ for every $\bv \in \Delta_{k, n}$. Then the solution $(y', z')$ defined by 
    \begin{align*}
        z_{C, \bv, a}' & \EQ z_{C, \bv, a} - m_{\bv, C}\,, \\
        y_C' & \EQ y_C + \sum_{\bv \in C} m_{\bv, C}\,,
    \end{align*}
    is also a feasible solution which achieves the same objective value.
\end{lemma}
\begin{proof}
    By definition we have
    \[
    y_C' + \sum_{\bv \in C} z_{C, \bv, \sigma(\bv)}' \EQ y_C + \sum_{\bv \in C} z_{C, \bv, \sigma(\bv)}
    \]
    for every $C$ and every $\sigma \in \Omegano_C$, and 
    \begin{align*}
    \sum_{C: \bv \in C} z_{C, \bv, a}' & \EQ \sum_{C: \bv \in C} (z_{C, \bv, a} - m_{\bv, C}) \EQ 0
    \end{align*}
    for every $\bv$ and $a \in \Lno(v)$, since $\sum_{C: \bv \in C} m_{\bv, C}= 0$. Thus $(y', z')$ is also a feasible solution to \eqref{LP:dual-vertex}. Furthermore, the objective value is preserved:
    \begin{align*}
    \sum_{C}y_C' \EQ \sum_{C}\left(y_C +  \sum_{\bv \in C} m_{\bv, C}\right) \EQ \sum_{C}y_C.
    \end{align*}
    This completes the proof.
\end{proof}

Note that in the transformation of Lemma~\ref{lem:lp_shift} we only needed the values $m_{\bv, C}$ for $\bv \in C$. 

\begin{lemma}\label{lem:local_mean_zero}
    There exists an optimal solution to \eqref{LP:dual-vertex} such that for every $C \in \cD_{k, n}$ and $\bv \in C$
    \[
    \sum_{\substack{a \in \Lno(\bv)}} z_{C, \bv, a} \EQ 0.
    \]
\end{lemma}
\begin{proof}
    Take some optimal solution $(y, z)$ and let $m_{\bv, C} = \frac{1}{|\Lno(\bv)|}\sum_{b \in \Lno(\bv)}z_{C, \bv, b}$ for $\bv \in C$. Then we have
    \[
    \sum_{C: \bv \in C} m_{\bv, C} \EQ \sum_{C: \bv \in C} \frac{1}{|\Lno(\bv)|}\sum_{b \in \Lno(\bv)}z_{C, \bv, b} \EQ \frac{1}{|\Lno(\bv)|} \sum_{b \in \Lno(\bv)} \sum_{C: \bv \in C} z_{C, \bv, b} \EQ 0.
    \]
    We may therefore apply Lemma~\ref{lem:lp_shift} and obtain $(y', z')$ which is also optimal and furthermore satisfies
    \[
    \sum_{\substack{a \in \Lno(\bv)}} z_{C, \bv, a}' \EQ \sum_{\substack{a \in \Lno(\bv)}} (z_{C, \bv, a} - m_{\bv, C}) \EQ 0. \qedhere
    \]
\end{proof}

\begin{lemma}\label{lem:y_nonzero}
    For every $C \in \cD_{k, n}$, there exists an optimal solution to \eqref{LP:dual-vertex} such that $y_{C'} = 0$ for every $C' \neq C$.
\end{lemma}
\begin{proof}
    It is sufficient to show that for any distinct $C, C' \in \cD_{k,n}$ and $\varepsilon \in \mathbb{R}$, we can modify a feasible solution $(y, z)$ to \eqref{LP:dual-vertex} so that $y_{C}$ increases by $\varepsilon$ and $y_{C'}$ decreases by $\varepsilon$ while all other $y$ variables are unchanged. Consider the bipartite graph $H = (\cD_{k, n}, \Delta_{k, n}, E_H)$ where $(C, \bv) \in E_H$ if and only if $\bv \in C$. Observe that $H$ is connected, so we may find a simple path from $C = C_1, \bv^{(1)}, C_2, \ldots, \bv^{(q)}, C_{q+1} = C'$ in $H$. Then for every $i \in [q]$ we define $m_{\bv^{(i)}, C_i} = \varepsilon$ and $m_{\bv^{(i)}, C_{i+1}} = -\varepsilon$, and set $m$ to be 0 everywhere else. Then this choice of $m$ satisfies the condition in Lemma~\ref{lem:lp_shift}, so we may apply the lemma, which leaves every $y$ variable in the interior of the path (as well as outside the path) unchanged and changes the two endpoints on the path as desired.
\end{proof}

\bibliography{ref_lb}
\bibliographystyle{alphaurl}

\end{document}